%% file: ms.tex
\documentclass[a4paper,11pt]{article}
\pdfoutput=1 

\usepackage{jinstpub} 

\usepackage{lineno}
\usepackage{subfig}
\usepackage{multirow}

\nolinenumbers

\title{Test beam performance of a CBC3-based mini-module for the Phase-2 CMS Outer Tracker before and after neutron irradiation}

\author{The Tracker Group of the CMS Collaboration}

\emailAdd{uplegger@fnal.gov}

\abstract
{
  The Large Hadron Collider (LHC) at CERN will undergo major upgrades to increase the instantaneous luminosity up to 
5--7.5$\times10^{34}$\,cm$^{-2}$s$^{-1}$. This High Luminosity upgrade of the LHC (HL-LHC) will deliver a total of 
3000--4000\,fb$^{-1}$ of proton-proton collisions at a center-of-mass energy of 13--14\,TeV.
To cope with these challenging environmental conditions, the strip tracker of the CMS experiment 
will be upgraded using modules with two	closely-spaced silicon sensors to provide information 
to include tracking in the Level-1 trigger selection. 
  This paper describes the performance, in a test beam experiment, of the first  prototype module based on 
  the final version of the CMS Binary Chip front-end ASIC before and after the module was irradiated with neutrons. 
Results demonstrate that the prototype module satisfies the requirements, providing efficient tracking information, after 
being irradiated with a total fluence comparable to the one expected through the lifetime of the experiment. 
}


\begin{document}
\maketitle
\flushbottom
\input{tex/introduction}
\input{tex/daq}

\input{tex/ftbf}

\input{tex/trackingMonicelli}
\input{tex/datasets}

\input{tex/calibration}
\section{Detector setup\label{sec:setup}}
\input{tex/detectorOptimization}
\input{tex/alignmentBeamParameters}
\section{Performance\label{sec:res}}
\input{tex/efficiencyAnalysis}

\input{tex/timeWalkAnalysis}

\input{tex/clusterWidthAnalysis}
\input{tex/resolutionAnalysis}

\input{tex/chargeAnalysis}

\input{tex/stubAnalysis}
\input{tex/conclusions}

\acknowledgments
\input{tex/Acknowledgements_2022}

\bibliographystyle{unsrt}
\bibliography{Bibliography}

\input{tex/TrackerAuthorList_2023}







\end{document}

%% file: tex/introduction.tex
\newcommand{\Vcth}{$\mathrm{V_{CTH}}$}
\newcommand{\pt}{$p_\text{T}$}


\section{Introduction}
\begin{figure}[t]
	\centering
	\includegraphics[width=0.8\textwidth]{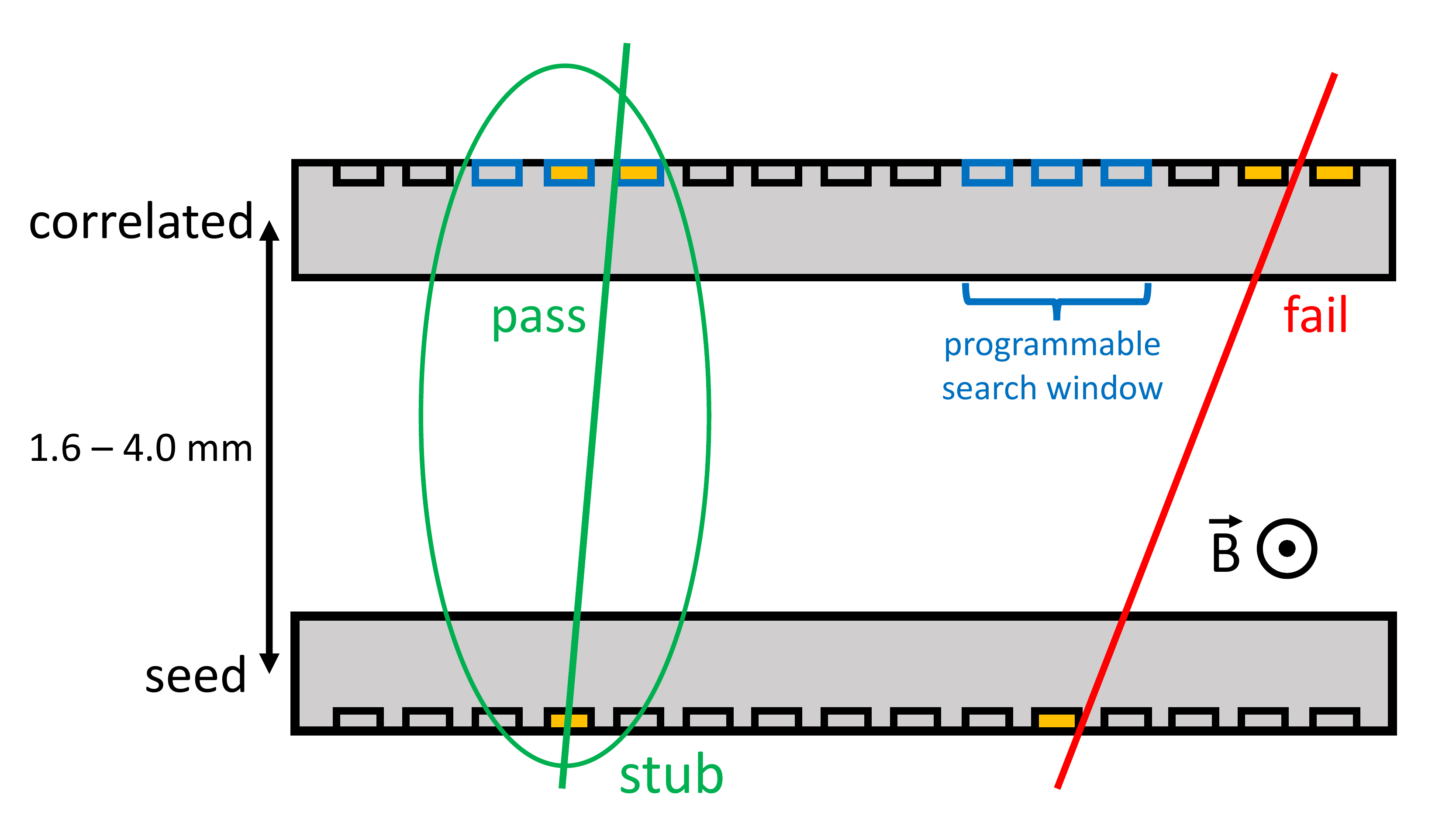}
	\caption[Module sketch]{Illustration of the identification of particles with high transverse momentum. 
	A hit in the bottom sensor opens a programmable search window in the top
	sensor such that particles with momentum larger than about 2--3 GeV should pass
	through this sensor area creating a stub, as pictured by the particle trajectory on the left. 
	Particles with momentum smaller than 2--3 GeV will instead fail to go through 
	the programmable search window not creating a stub, as pictured by the particle trajectory on the right.}
	\label{fig:PtModuleSketch}
\end{figure}

The Large Hadron Collider (LHC) at CERN will undergo major upgrades to increase the instantaneous luminosity up to 
5--7.5$\times10^{34}$\,cm$^{-2}$s$^{-1}$. This High Luminosity upgrade of the LHC (HL-LHC)~\cite{HL-LHC:TDR} will provide 
an unprecedented sample of proton-proton collision data at a center-of-mass energy of 13--14\,TeV, extending significantly 
the discovery potential of direct searches for new phenomena and the sensitivity of the precision measurement program. 
Compared to the LHC, the luminosity upgrade will increase the collision rates, detector occupancies, 
and the radiation induced damage.
To cope with these challenging experimental conditions, the entire silicon tracking system of the CMS 
experiment~\cite{CMS:Experiment} will be upgraded~\cite{CMS:TrackerTDR} with a new tracker, 
which will include an Inner Tracker based on silicon pixel modules and an Outer Tracker made from silicon modules with
strip and macro-pixel sensors.\\
The new CMS Outer Tracker will be equipped with so-called \pt-modules, which consist of pairs of radiation-tolerant silicon sensors 
parallel to each other and separated by a few millimeters. 
This new design concept will enable CMS to include tracking information in the first level of the trigger (L1) selection.
For the first time a selection on the transverse momentum of tracks can be used to select 
events of interest with high efficiency while maintaining trigger rates within the allowed bandwidth.
Figure~\ref{fig:PtModuleSketch} illustrates the concept of operation of these \pt-modules. 
A hit in the bottom (seed) sensor opens a programmable search window in the top (correlated) sensor 
such that particles with momentum larger than 2--3 GeV should pass through this area while particles with lower momentum 
should not, due to their large curvature in the magnetic field. 
The seed sensor hit and the correlated sensor hit falling in the search window form a 
hit pair called stub which is used by the L1 trigger to decide whether the data for an event are to be 
read out.
\\Modules installed at radii from 20 to 60\,cm from the beam line are named pixel-strip (PS) modules and consist of one strip sensor and one macro-pixel sensor, 
where macro refers to the unusual pixel size when compared to the Inner Tracker pixels, which for this sensor is $1.447\,\text{mm}\times100\,\mathrm{\mu m}$.
Modules at larger radii 
consist of two identical strip sensors and are named strip-strip (2S) modules. 
\begin{figure}[h]
	\centering
	\includegraphics[width=0.8\textwidth]{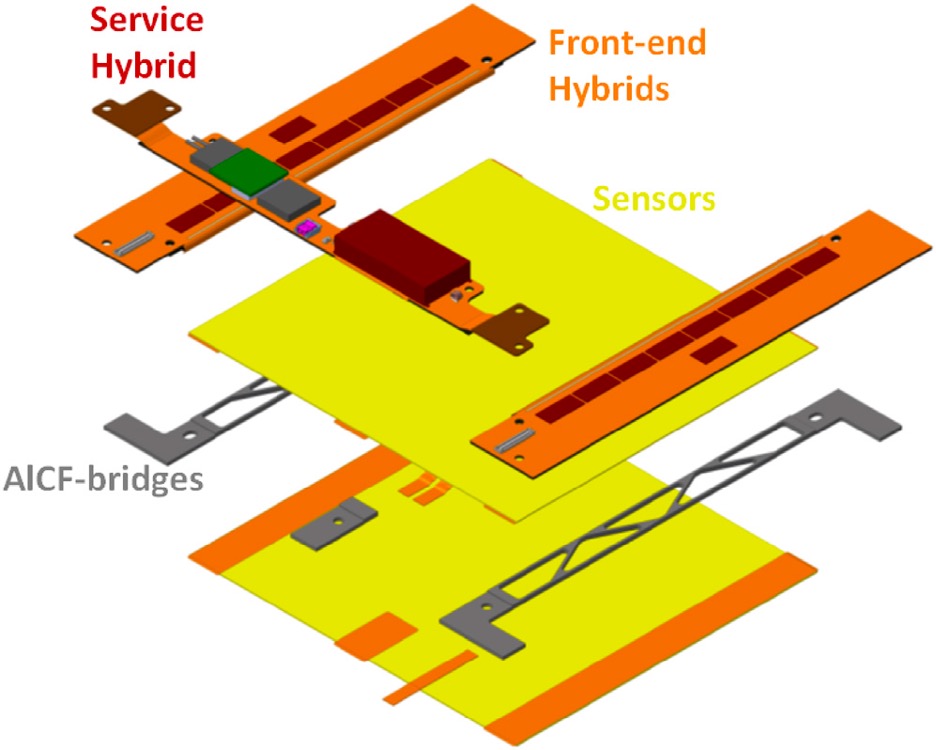}
	\caption[Module drawing]{Exploded view of a 2S module. 
	The seed and correlated sensors, in yellow, are separated by two aluminum carbon fiber bridges. 
	The two front-end hybrids, in orange, are mounted on the two sides of the sensors. 
	The service hybrid, in red, with two DC-DC converters inside an electro-magnetic shielding box 
	and the optical link to the DAQ, is mounted between the two front-end hybrids.}
	\label{fig:2SModuleDrawing}
\end{figure}
An exploded view drawing of the \mbox{2S module is shown in Fig.~\ref{fig:2SModuleDrawing}.}\\
The PS modules are read out by the Macro Pixel ASIC (MPA~\cite{MPA}) and by the Short Strip ASIC (SSA~\cite{SSA}), while  
the 2S modules are read out by the CMS Binary Chip (CBC~\cite{CBC3}). 
The MPA and CBC front-end ASICs have the logic implemented to correlate the hits in the seed and correlated sensors 
and create the stubs.
The stub information is transmitted to the L1 trigger at the LHC collision frequency of 40\,MHz,
while the strip hit information is read out on receipt of a L1 trigger accept signal.
Results from the characterization
of a prototype 2S module based on the previous version of the CBC are discussed in Ref.~\cite{CBC2Testbeam},
while this paper presents the performance of the first module 
based on a version of the CBC ASIC (CBC3) that incorporates the complete functionality, including the trigger logic.
This module, featuring smaller sensors and built with only two CBC chips instead of the sixteen that equip a tracker module, is called mini-module.\\
Although the tracker will operate in the CMS magnetic field, all measurements presented in this paper 
were done in the absence of a magnetic field since no magnet was available at the Fermilab Test Beam Facility (FTBF).
\\This paper is organized as follows:  Sec.~\ref{sec:cbc3} summarizes the main features of the CBC3 chip, 
Sec.~\ref{sec:minimod} describes the mini-module and Sec.~\ref{sec:daq}
discusses the mini-module data acquisition system (DAQ).
The Fermilab Test Beam Facility is briefly described in Sec.~\ref{sec:ftbf} and the alignment 
and tracking software, in Sec.~\ref{sec:align}, 
while the Rhode Island Nuclear Science Center irradiation facility is described in Sec.~\ref{sec:irradiation}.
The collected data sets are summarized in Sec.~\ref{sec:datasets}.
The final part of the paper presents the performance of the mini-module before and after irradiation, starting in
Sec.~\ref{sec:calib} with the module calibration, followed by the steps
to set up the necessary parameters to take data, Sec.~\ref{sec:setup}. 
Finally, Sec.~\ref{sec:res} presents the results of the module performance in the beam.


\section{The CBC3 front-end ASIC\label{sec:cbc3}}
\begin{figure}[t]
	\centering
	\includegraphics[width=0.8\textwidth]{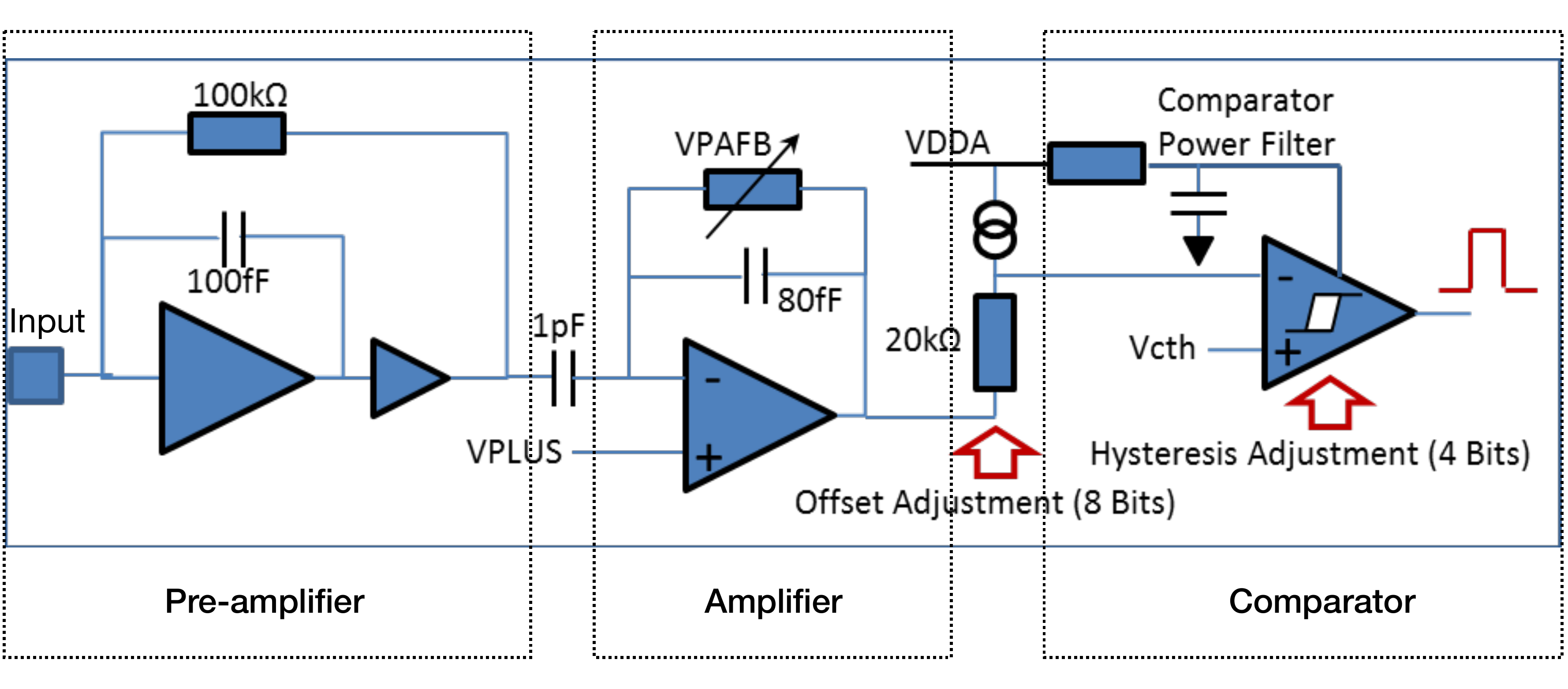}
	\caption[CBC3 Analog FrontEnd]{Block diagram of the analog front-end of the CBC3 ASIC.}
	\label{fig:CBC3AnalogueFrontEnd}
\end{figure}

The CBC series is designed for the readout of the 2S modules of the Phase-2 CMS tracker.
Manufactured in a 130\,nm CMOS technology, the CBC chip reads out the charge produced
via ionization by charged particles traversing the silicon sensors.
The CBC converts the produced charge into a hit or no-hit binary value for each of the channels. 
Compared to earlier versions of the CBC~\cite{CBC2}, the CBC3, mounted on the mini-module, is the first version of the ASIC to include 
the full trigger logic circuitry for detecting potential high momentum tracks in each bunch crossing event.\\
Each CBC3 chip has 254 channels, each with a pre-amplifier, amplifier and the possibility to individually 
trim the effective threshold.
The CBC3 operates with the 127 odd-numbered channels connected to the seed sensor and the 127 even-numbered channels connected to 
the correlated sensor of the module. One CBC3 ASIC then services 127 strips from each of two sensors.\\
The block diagram of the analog front-end of the CBC3 is shown in Fig.~\ref{fig:CBC3AnalogueFrontEnd}.
Charge generated in the strip is read out by a pre-amplifier and integrated onto a
100\,fF feedback capacitor. The feedback capacitor is discharged by a 100\,k$\Omega$~resistive feedback network.
The resulting \mbox{voltage} pulse from the pre-amplifier is further amplified by a capacitive gain amplifier, 
which is stabilized by a large feedback resistance, VPAFB.
To compensate for any channel to channel threshold mismatch, due to the fabrication process, 
each amplifier has a programmable offset adjustment, named VPLUS.
These control currents are programmed via an 8-bit register in each channel, 
providing enough resolution to obtain a channel to channel variation of the average no-particle signal at the comparator input, called pedestal,
of less than one hundred electrons.
The comparator stage detects signals which cross a defined adjustable global chip threshold, \Vcth, 
and will produce a digital 1 output for as long as the signal stays above the threshold.
The various analog biases used by the front-end circuits are derived on-chip, from the analog voltage VDDA, using bias generation
circuits powered by a voltage reference band gap. The band gap circuit uses a PMOS transistor in place of
the more traditional diode, which makes it more tolerant to radiation damage, but leaves it susceptible
to process variation.
\\AC coupled n-in-p type silicon sensors were chosen for the tracker, so the CBC3 channel will operate for the readout of electrons. 
Since the AC coupling circuit is included in the sensor, there is no leakage current compensation circuit included in the CBC3 design.
Each input is bonded to one strip of a silicon sensor.


\section{The mini-module\label{sec:minimod}}
\begin{figure}[t]
	\centering
	\subfloat[]{\includegraphics[height=7cm]{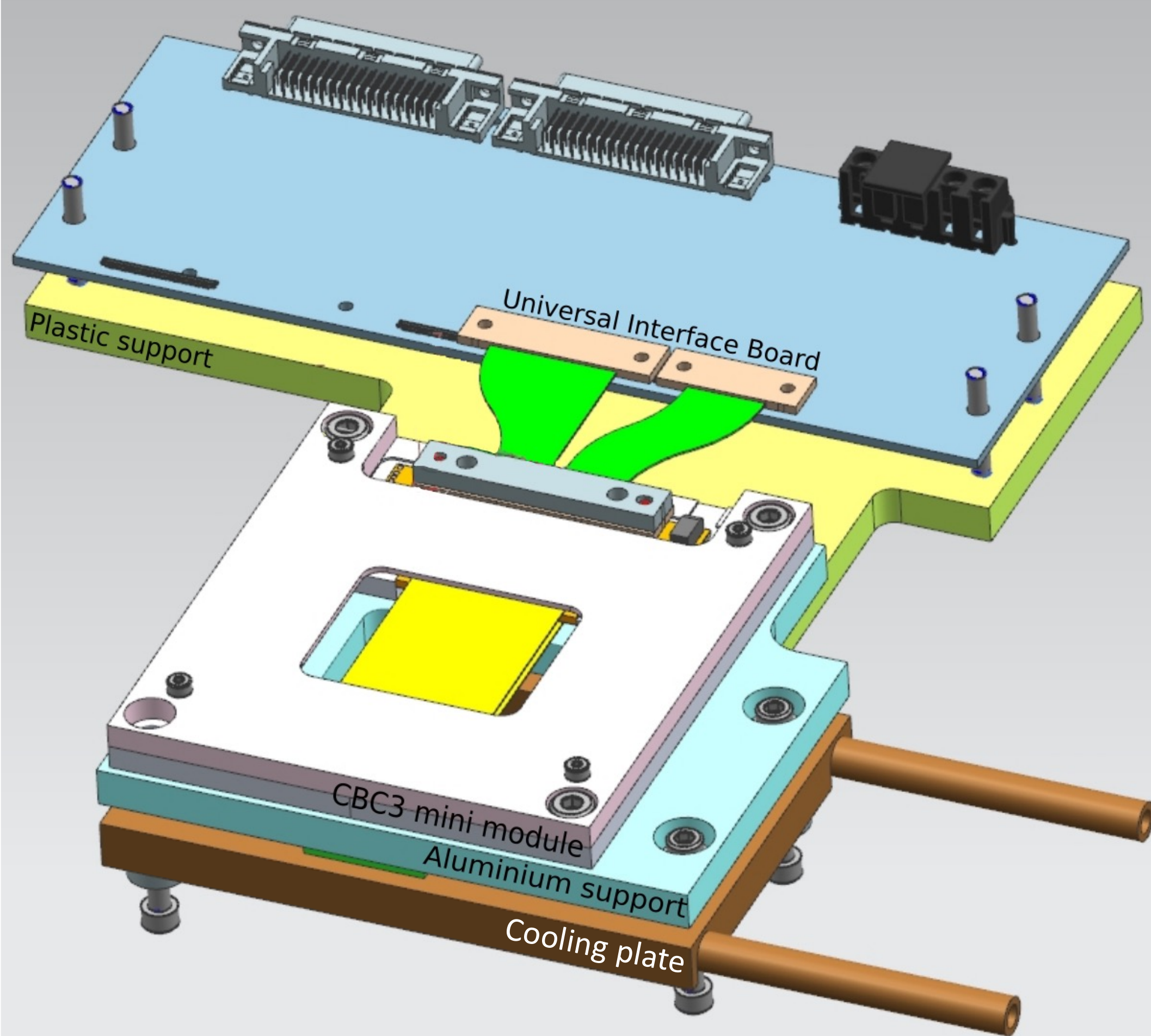}\label{fig:a}}%
	\hspace{1cm}
	\subfloat[]{\includegraphics[height=7cm]{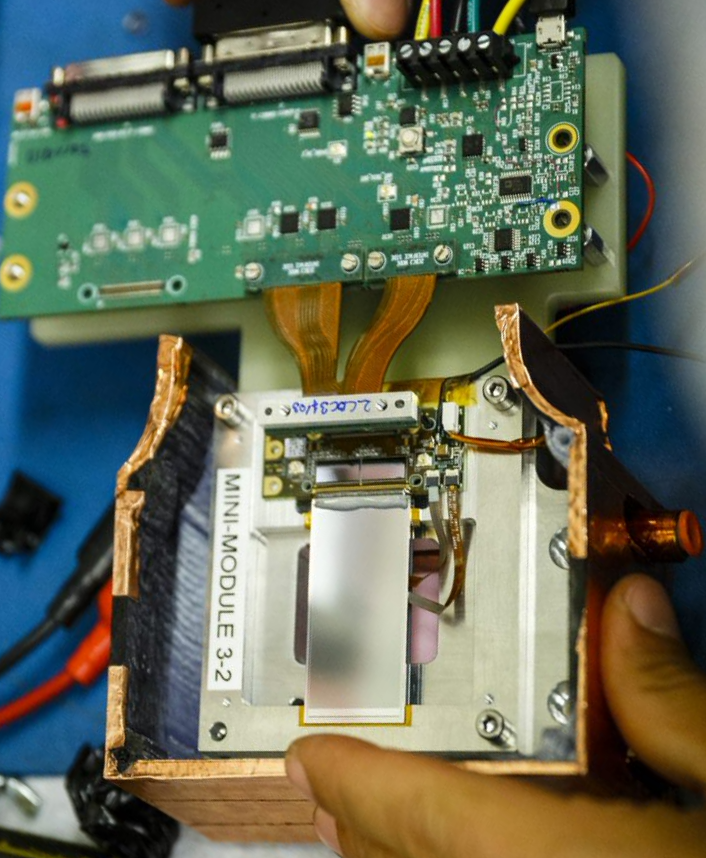}\label{fig:b}}%
	\caption{(a) Drawing of the mini-module  support structure. (b) Mini-module mounted on the support structure,
	hosted in the enclosure, and connected to the UIB.}
	\label{fig:boxdesign}
\end{figure}

The prototype \pt-module was assembled at CERN and consists of two sensors produced by Infineon~\cite{Infineon}. 
The two identical AC coupled strip sensors are made from high resistive float zone silicon, 
thinned to a physical and active thickness of 300\,$\mu$m. 
The 254 strips per sensor use polysilicon bias resistors and p-stop strip isolation and are approximately 5\,cm 
long with a pitch of 90\,$\mu$m and a width-to-pitch ratio of 0.25~\cite{InfineonSensor}. 
The strip geometry is close to the final 2S sensors and therefore also the capacitance seen by the readout chip.
The sensors are mounted on an aluminum structure to keep them separated by $d=1.8$\,mm 
(where $d$ is the distance between the mid-planes of the sensors). 
A prototype hybrid~\cite{Hybrid} with two CBC3 front-end ASICs is wire-bonded to the sensors and read out through an interface board 
(Universal Interface Board, or UIB). The CBC3 ASICs mounted on this hybrid are both version 3.0, 
while the final version of the chip that will be used in the detector is version 3.1.
The assembly is referred to as mini-module in the rest of the paper.\\
A dedicated structure was designed and manufactured to support and cool the mini-module. 
As shown in Fig.~\ref{fig:boxdesign}~(a), the cooling liquid flows through the pipes embedded in the cooling plate, and a temperature
sensor is installed on the cooling plate itself. An aluminum support hosts the mini-module and is mounted on top of Peltier elements, 
used to help regulate the temperature, and in turn is cooled down by the cooling plate.
All plates have a rectangular hole to allow the beam to pass through to minimize multiple scattering.  
The support structure and the mini-module are installed in a 3D-printed enclosure that is flushed with dry air to prevent condensation.
The enclosure, with foam used as insulation between the 
cooling plate and the plastic, is wrapped with copper tape to ensure electromagnetic shielding and to avoid penetration of light.  	
The UIB is installed outside the enclosure and connects to the back-end DAQ board, called FC7~\cite{FC7}, via a very high density cable interconnect (VHDCI). 
The final assembly is shown in Fig.~\ref{fig:boxdesign}~(b).\\
The default running configuration for the mini-module is the following, unless specified otherwise:
\begin{itemize}
	\item unirradiated: $\mathrm{V_{bias}}=-300$\,V; $\mathrm{temperature=+15}$\,$^\circ$C; \Vcth~$=560$\,DAC units;
	\item irradiated: $\mathrm{V_{bias}}=-600$\,V; $\mathrm{temperature=-20}$\,$^\circ$C; \Vcth~$=540$\,DAC units.
\end{itemize}

%% file: tex/daq.tex
\section{Data acquisition system\label{sec:daq}} 
\begin{figure}[b]
    \centering
    \includegraphics[width=1\textwidth]{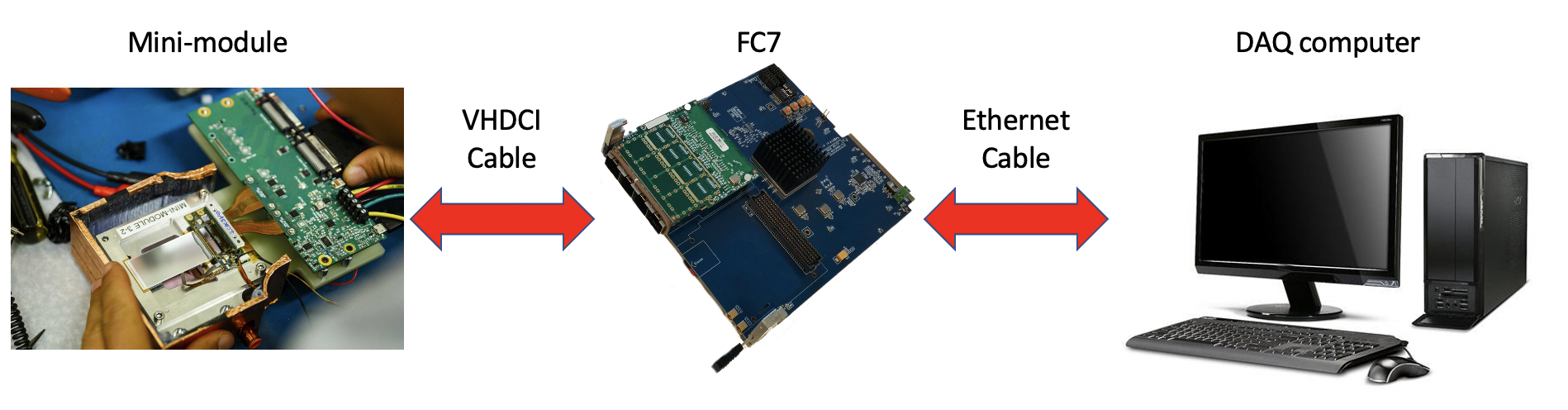}
    \caption{Data acquisition schematic view. The mini-module is connected to the FC7 back-end 
    board through a VHDCI. The FC7 is then connected to the DAQ computer with an ethernet cable.}
    \label{fig:DAQ}
\end{figure}
The data acquisition system used to read out the mini-module is based on the CMS-specific custom $\mu$TCA FC7 back-end board
which is electrically connected to the CBC3 chips on the hybrid circuit. 
A sketch of the DAQ connections is shown in Fig.~\ref{fig:DAQ}.
Both CBC3 chips are connected to a common control interface consisting of a 320\,MHz clock line, 
two lines for communication with the chips via the I2C protocol, a dedicated reset line, and a fast-command line 
on which 8-bit synchronous commands are transmitted. 
The signals that can be encoded on this line are a fast-reset, a trigger signal, a request for a test pulse, and 
a reset to zero of the on-chip trigger counter. Data from the chips are sent on six sub low voltage differential signaling (SLVDS) lines. 
Five of these lines are reserved for the stub data, which are sent bunch-synchronous at 40\,MHz to be used in the L1 trigger decision. 
The sixth line is reserved for hit data, which are only sent on reception of a trigger signal.\\ 
Since the SLVDS drivers on the CBC3 chip can only drive these signals over a very short distance, 
the UIB is used to interface the chips to the back-end electronics. 
It is connected to the hybrid via a short, flexible jumper cable and has a number of SLVDS level translators 
to drive the signals to the readout card over the 1\,m long VHDCI. 
In addition, the UIB provides the low voltage power for the hybrid circuit and has an integrated ADC 
to monitor the hybrid power consumption and temperature.\\
The CMS-specific FC7 card is the basis of the readout system 
developed for the prototyping phase of the CMS Outer Tracker. It has a Xilinx Kintex 7 FPGA and is equipped 
with two high-pin-count FPGA mezzanine card (FMC) connectors to adapt the FC7 to a variety of applications. 
The VHDCI from the UIB is connected to a custom-built mezzanine card, sitting on one of the FMC connectors, 
that routes the differential signals to the FPGA for processing. 
For this test beam, a second dedicated FMC was used to interface the FC7 to the triggering system used at the Fermilab Test Beam Facility. 
Through this second FMC board, the system clock and the trigger signal from the scintillator coincidence were connected 
to two differential inputs on the FPGA.
A differential output was used to send a busy signal to the trigger control logic to inhibit further 
triggers in case of imminent buffer overflow.\\ 
The custom firmware running on the FC7 is controlled by a common software framework written in C++ called 
Phase-2 Acquisition and Control Framework (Ph2-ACF).  This framework has been developed to cover all R\&D and testing needs
during the prototyping phase for the Phase-2 Tracker of CMS. It is based on 
the $\mu$HAL/IPBUS~\cite{IPBUS} libraries that allow for simple, yet reliable write/read 
operations of firmware registers based on the TCP/IP protocol. The core feature of the Ph2-ACF framework is 
that it introduces an abstraction layer between the $\mu$HAL calls and the higher-level user code. This allows 
full control over CBC functionalities such as configuring the chips via the I2C registers, sending fast commands, and 
reading out data, without further knowledge of the firmware structure. 
Furthermore, utilities for event data decoding and data storage are provided within the framework.\\ 
The CBC chips are reset and configured before each run.  The flow of 
triggers is enabled as soon as the system is ready. The main thread of the Ph2-ACF 
then runs a polling loop that continuously queries the state of the "data-ready" flag in the FC7 firmware.  
Once the firmware signals that data are available to be read out, the main thread performs a block-read operation of the data FIFO. 
This data block is shared with various consumer threads that store the binary raw data directly on disk without further processing.  
Consumer threads also decode the data into individual events for processing and re-formatting into the final 64-bit CMS SLINK data format.\\
A supervisor application based on the CMS XDAQ framework~\cite{XDAQ} has been developed around the Ph2-ACF libraries
to allow straightforward integration into the FTBF tracking telescope DAQ based on the 
OTSDAQ framework developed by the Fermilab Computing Division~\cite{OTSDAQ}, described in Sec.~\ref{sec:ftbf}.
This application provides all the interfaces required to communicate 
with the run control system and implements a finite state machine that ensures synchronicity of all elements of the DAQ system. 
In addition, it offers a browser-based graphical user interface that allows to easily edit configurations 
and monitor the status of the system.\\
Data quality monitoring is provided by a simple application, also based on the Ph2-ACF libraries, 
that runs asynchronously on the SLINK files once they are written to disk. 
This application provides basic histograms to monitor the recorded data within a very short turnaround 
time, which allows issues with the system to be detected immediately.

%% file: tex/ftbf.tex
\section{Fermilab Test Beam Facility\label{sec:ftbf}}

\begin{figure}[t]
	\centering
    \includegraphics[width=0.90\textwidth]{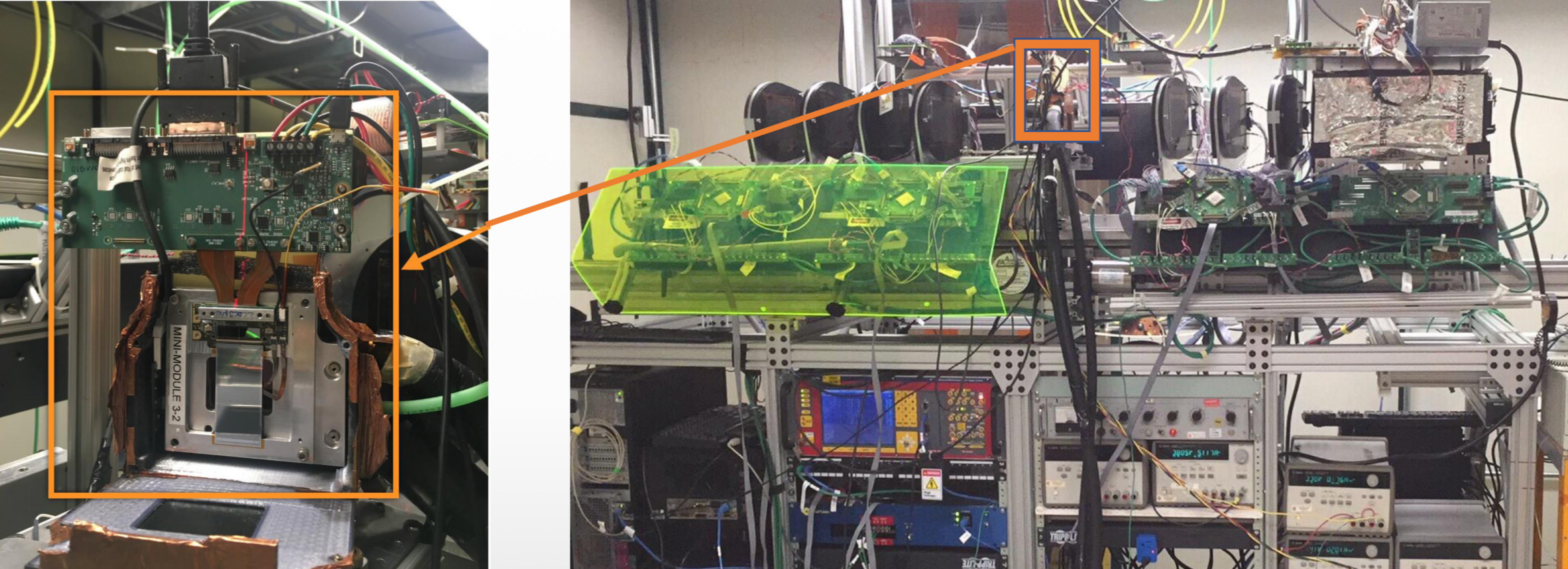}
	\caption[CBC3 Analog FrontEnd]{FTBF silicon telescope. The mini-module is installed in the middle.}
	\label{fig:FTBF}
\end{figure}

The mini-module was tested at the Fermilab Test Beam Facility~\cite{FTBF}, 
which provides a 120\,GeV bunched proton beam at rates ranging from 1 to 300\,kHz.
The beam is delivered to the facility every minute with the spill lasting 4 seconds.
The test beam campaign described in this paper used the 120\,GeV proton beam at rates between 20 and 50\,kHz.\\
Figure~\ref{fig:FTBF} shows the tracking telescope available at the FTBF and the mounting position of the mini-module.  
The mini-module was mounted on a remotely controlled rotation platform provided by the FTBF.
The facility does not have a magnet so all tests were performed in the absence of a magnetic field.
A chiller, necessary to keep the irradiated device cold to minimize the sensors' annealing, 
was already present at FTBF and used during the tests.
The telescope, which provides precision tracking information, consists of four planes equipped
with pixel modules, placed downstream, and fourteen planes equipped with strip modules~\cite{Telescope}, 
placed before and after the detector under test.
The pixel modules are identical to those used for the Phase-0 CMS Forward Pixel detector and are based on the PSI46 front-end chip~\cite{PSI46}. 
Two of the pixel layers are equipped with 2$\times$3 modules, and the other two layers with 2$\times$4 modules, where 2$\times$3 and 2$\times$4 
refer to the number of PSI46 chips present in the module.
The strip sensors were designed for the Run 2b upgrade of the D$\emptyset$ tracker~\cite{D0Tracker} and are read out by the FSSR2 front-end chip~\cite{FSSR2}. 
The telescope resolution, measured at the position where the mini-module was mounted, was approximately $7\,\mathrm{\mu m}$.\\
Two scintillator counters provided the coincidence signal from the particles in the beam. 
A Fermilab-designed FPGA-based trigger board synchronized
 the data streams of the telescope and the mini-module. This trigger board accepts
the scintillator counters' NIM signals and it also generates a 39.75\,MHz clock, 
multiplying the 53\,MHz Fermilab accelerator clock by $\frac{3}{4}$, 
thus keeping the beam and clock phases fixed relative to each other.


%% file: tex/trackingMonicelli.tex
\section{Alignment and tracking\label{sec:align}}
\begin{figure}[t]
	\centering
	\subfloat[]{\includegraphics[width=0.45\textwidth]{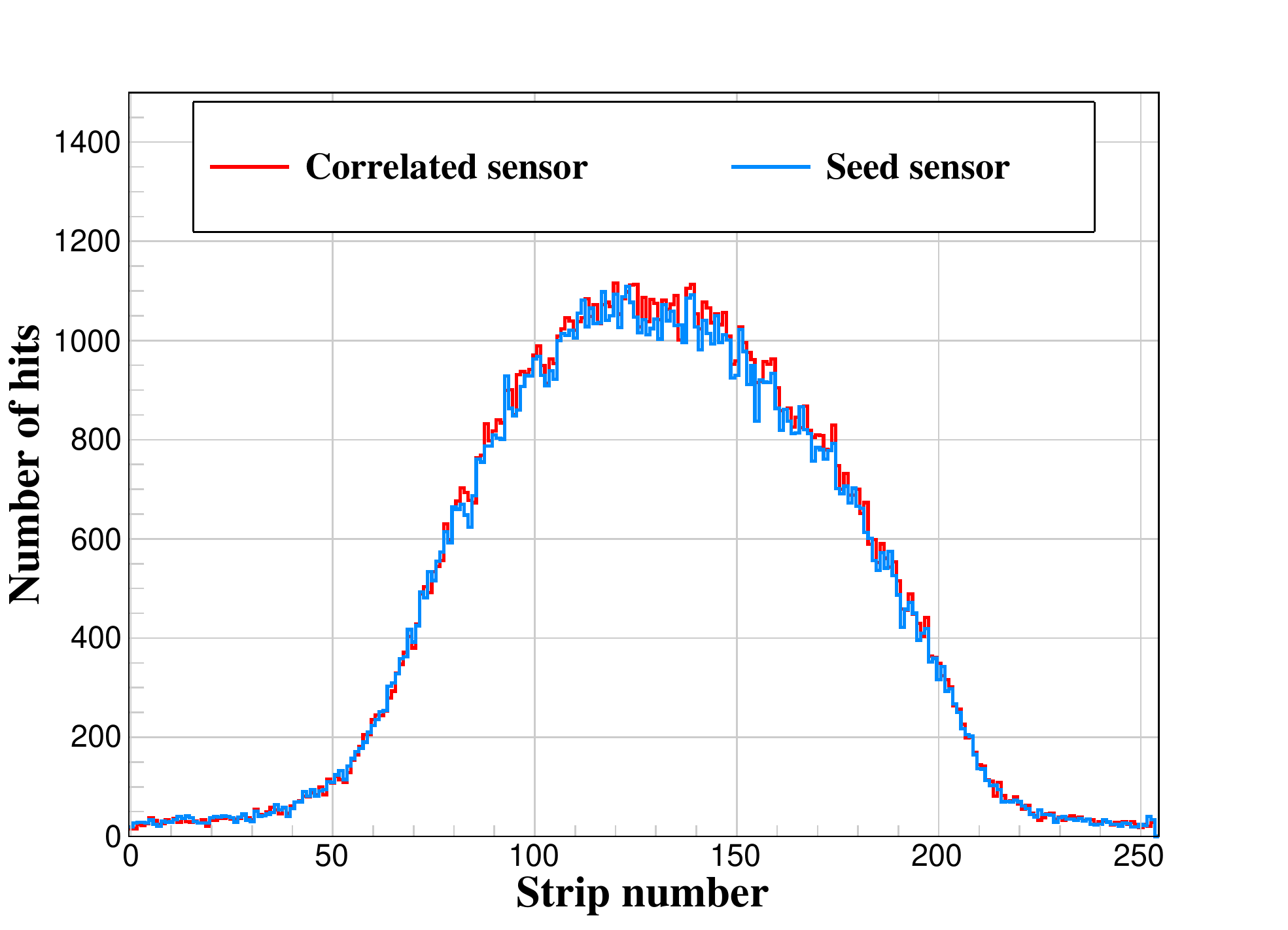}\label{fig:a}}%
	\subfloat[]{\includegraphics[width=0.45\textwidth]{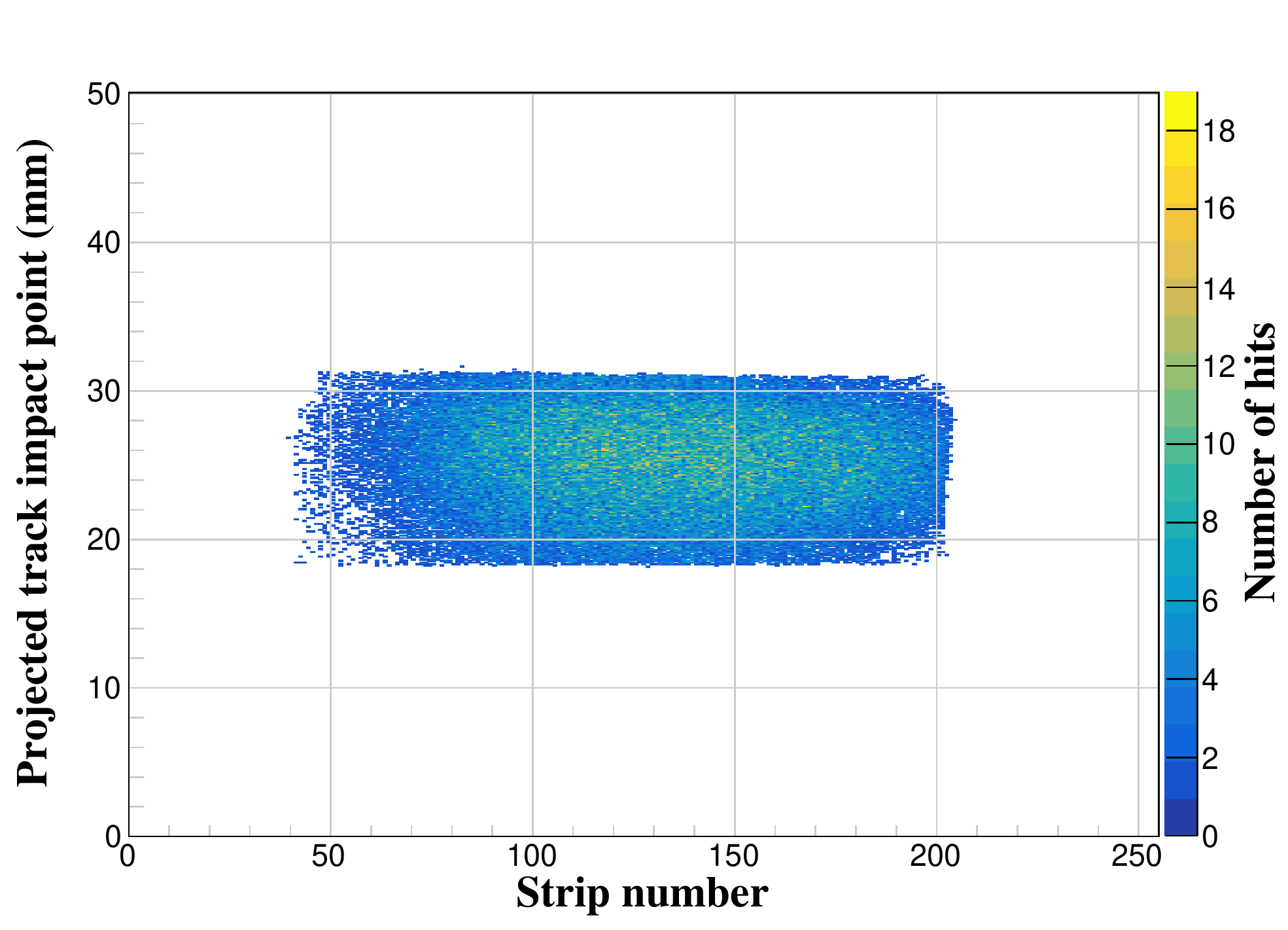}\label{fig:b}}%
	\caption{(a) Distribution of hits for the seed and correlated sensors of the unirradiated mini-module. 
	(b) 2D hit map where the $x$-axis represents the strip number and the $y$-axis the projected track impact point along the strip direction.}
	\label{fig:BeamProfile}
\end{figure}

The track reconstruction and the telescope alignment are performed by means of a dedicated package called Monicelli~\cite{Telescope}. 
This software provides the user with an iterative procedure to converge toward the optimal alignment of the telescope.
All operations can be accomplished in steps through a graphical interface that makes the software user-friendly.
The track-reconstruction code implemented in Monicelli is a generic alignment program developed for the FTBF telescope.
It performs a Kalman filter fit to the coordinates of the arrays of hits, which were preselected by pattern recognition 
based on their proximity, $\pm 500\,\mathrm{\mu m}$, to a straight line connecting a hit on the first plane and a hit on the last plane of the telescope. 
In the case of clusters of adjacent pixels or strips, the hit coordinates are calculated 
exploiting the linear relationship between the relative amount of charge shared by the two pixels or strips 
in the cluster and the proximity of the hit to the boundary between them. 
The linear relationship was measured during the first commissioning phase of the telescope. 
The errors attributed to the coordinates were also calibrated during the telescope commissioning.\\
To perform this track reconstruction process, Monicelli reads a binary file containing the telescope and mini-module merged events 
together with an XML geometry file
describing the overall configuration and geometrical details of the telescope detectors for that particular data set.
The Monicelli software output file consists of a ROOT TTree~\cite{root} containing, for each event, the reconstructed telescope tracks together with
the associated clusters of hits and the raw data, including those of the mini-module. 
This output file provides the user with all the information
needed for the analysis of the test beam data.\\
Figure~\ref{fig:BeamProfile}~(a) shows an example of a hit distribution for the seed and correlated sensors of the unirradiated 
mini-module, placed orthogonally to the incident beam. A 2D hit map, shown in Fig.~\ref{fig:BeamProfile}~(b), can be reconstructed using the strip number as one coordinate, 
with the position along the strip direction obtained by interpolating the reconstructed telescope track to the mini-module position.
The 2D map does not extend over the whole sensor due to the telescope acceptance and the beam size.

%% file: tex/datasets.tex
\newcommand{\TestBeamA}{Unirradiated~}
\newcommand{\TestBeamC}{Half fluence~}
\newcommand{\TestBeamD}{Full fluence~}

\section{Neutron Irradiations at RINSC\label{sec:irradiation}}

The Rhode Island Nuclear Science Center (RINSC) operates a 2 MW light-water cooled, 
open pool type reactor on the Narragansett Bay Campus of the University of Rhode Island, USA. 
The core consists of fuel assemblies and a combination of graphite and beryllium reflectors. 
The fuel is plate type U$_\text{{3}}$Si$_\text{{2}}$ cladded with aluminum, enriched to less than 20\% Uranium-235. 
The reactor provides several ways to irradiate samples, among others a pneumatic rabbit system that is 
suitable for small samples and beam ports that can accommodate larger samples.
The mini module was irradiated in a 6-inch beam port, which is a pipe of 6 inch inner diameter that allows 
samples to be inserted close to the reactor core. An acrylic cylinder with a diameter just under 6 inches 
and a length of 2.5 feet was used to hold the mini-module with enough padding to protect it from physical 
damage. Ultra-pure foils and silicon diodes for dosimetry and two probes to monitor the temperature of the 
module during irradiation were packaged with the mini-module. During irradiation, samples get heated up by 
the gamma radiation in the reactor. To limit the heating and the resulting annealing of the silicon sensors 
the remaining volume of the cylinder was filled with dry ice. The module was irradiated for 16 minutes two 
separate times. The reactor has to be off for the module to be loaded into the beam port. It then is ramped 
up to 100\% power and ramped down again after the desired irradiation time so that the module can be 
removed from the beam port. This procedure introduces significant edge effects that make it difficult to 
achieve a very precise irradiation time. The two irradiations resulted in fluences of 
$\mathrm{2.0\times10^{14}}$ and $\mathrm{2.4\times10^{14}\,n_{eq}/cm^{2}}$, where n$_\text{{eq}}$ stands 
for 1-MeV-equivalent neutrons. 
The maximum temperature measured in the cylinder during irradiation was $+30\,^\circ$C.

\section{Data sets\label{sec:datasets}}

The mini-module was tested during three different test beam campaigns at the FTBF.
The first run was done in November 2017 to assess the mini-module functionality and performance \mbox{after} assembly.
After this run, the mini-module was irradiated with neutrons at the RINSC reactor.
The main goal of the irradiation was to test the radiation tolerance of the sensors and 
to evaluate the overall performance of the mini-module assembly 
to make sure that the CBC3 would work as expected when connected to sensors with higher leakage current.
The CBC3 chip radiation tolerance was previously \mbox{studied} using X-rays at CERN~\cite{CBC3Irradiation}, 
since ionizing radiation is the main source of radiation damage for integrated circuits, 
and it demonstrated that the chip is capable of withstanding the \mbox{levels} of radiation expected at the HL-LHC with minimal degradation.\\
The accumulated fluence during this first irradiation was $\mathrm{2.0\times10^{14}\,n_{eq}/cm^{2}}$. 
The mini-module was then tested again in June 2018 at Fermilab.
During this second test beam campaign, due to a problem with the cooling system, the mini-module was kept at $+60\,^\circ$C for two hours,
effectively annealing part of the sensor damage due to the neutron irradiation.
This annealing time can be converted into an equivalent annealing time at $+21\,^\circ$C,
which is the temperature at which the detector is supposed to be annealed during maintenance periods in CMS,
using the current-related damage rate at different temperatures.
The conversion factor is calculated according to the parametrization given in Ref.~\cite{Moll:1999kv} and
the equivalent time at $+21\,^\circ$C is 20 days.
The future tracker is expected to spend 20 weeks at that temperature in CMS, corresponding to two weeks of maintenance every year for at least 10 years. 
After the second test beam campaign, the mini-module was irradiated a second time at RINSC and received an additional fluence of $\mathrm{2.4\times10^{14}\,n_{eq}/cm^{2}}$, 
resulting in a total accumulated fluence of $\mathrm{4.4\times10^{14}\,n_{eq}/cm^{2}}$.
This total accumulated fluence exceeds the maximum fluence of 
$\mathrm{4.0\times10^{14}\,n_{eq}/cm^{2}}$~after 4000\,fb$^{-1}$
expected for more than $95\%$ of the 2S modules in CMS~\cite{Irradiation}.
After this last irradiation, the mini-module was tested at the FTBF for the third time in December 2018.
The test beam campaigns and relative fluences are summarized in Table \ref{tab:TestBeamRuns}.
\begin{table}[h!]
    \begin{center}
      \caption{Data set summary.}
      \label{tab:TestBeamRuns}
      \begin{tabular}{|l|c|l|} 
        \hline
        \textbf{Test beam period} & \textbf{Total fluence ($\mathrm{n_{eq}/cm^{2}}$)} & \textbf{Label}\\
        \hline
        November 2017 & 0           & \TestBeamA\\
        \hline
        June 2018     & $2.0\times10^{14}$ & \TestBeamC\\
        \hline
        December 2018 & $4.4\times10^{14}$ & \TestBeamD\\
        \hline
      \end{tabular}
    \end{center}
  \end{table}

%% file: tex/calibration.tex
\section{Module calibration\label{sec:calib}}
\label{sec:ModuleCalibration}
\begin{figure}[b]
  \centering
  \includegraphics[width=7cm, height=5cm]{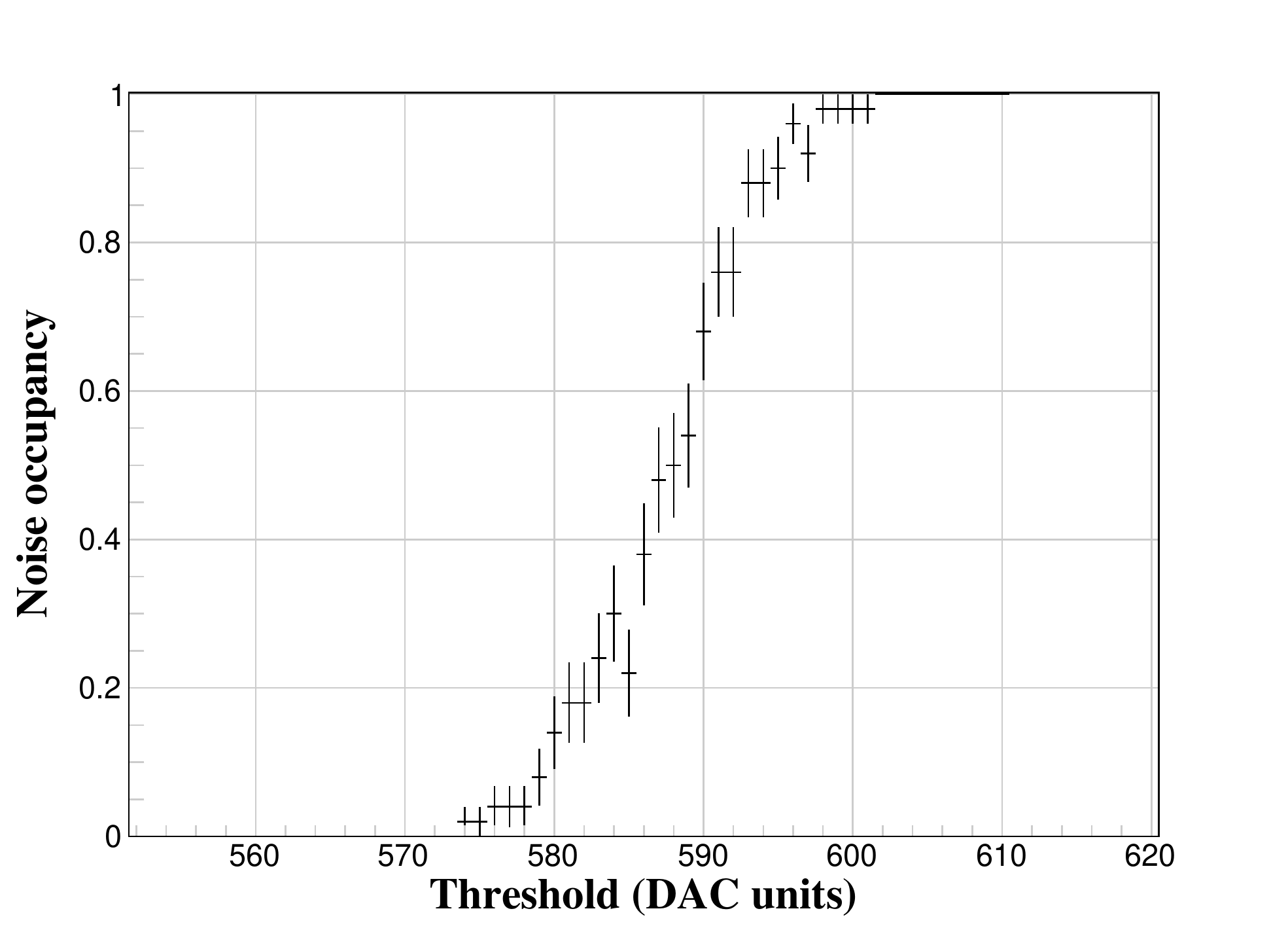}
  \caption[Example S-curve.]{Single channel S-curve for the unirradiated mini-module. 
  }
  \label{fig:CalibrationSCurve}
\end{figure}

\begin{figure}[t!]
	\centering
	\subfloat[]{\includegraphics[width=7cm, height=5cm]{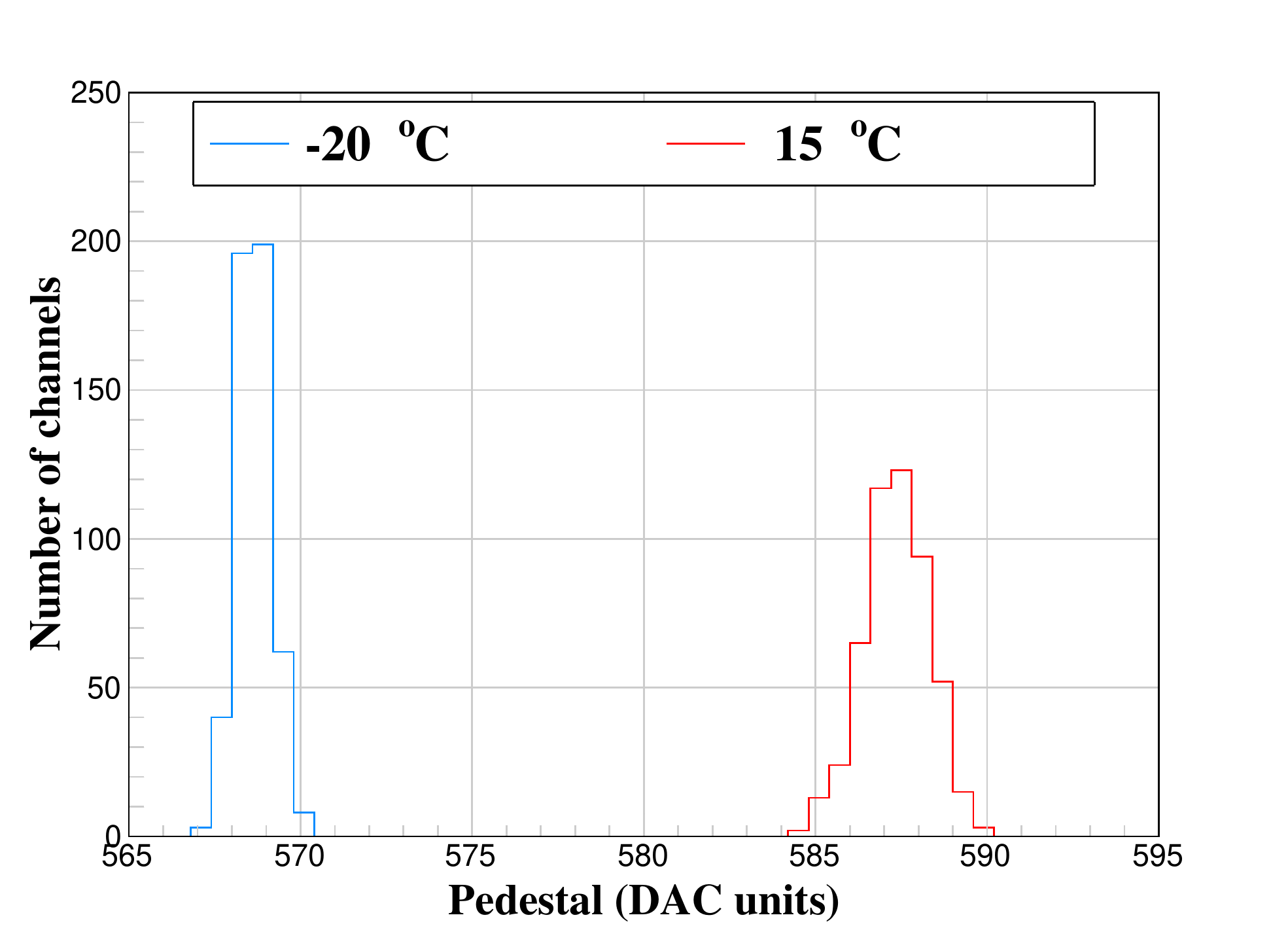}\label{fig:a}}%
	\subfloat[]{\includegraphics[width=7cm, height=5cm]{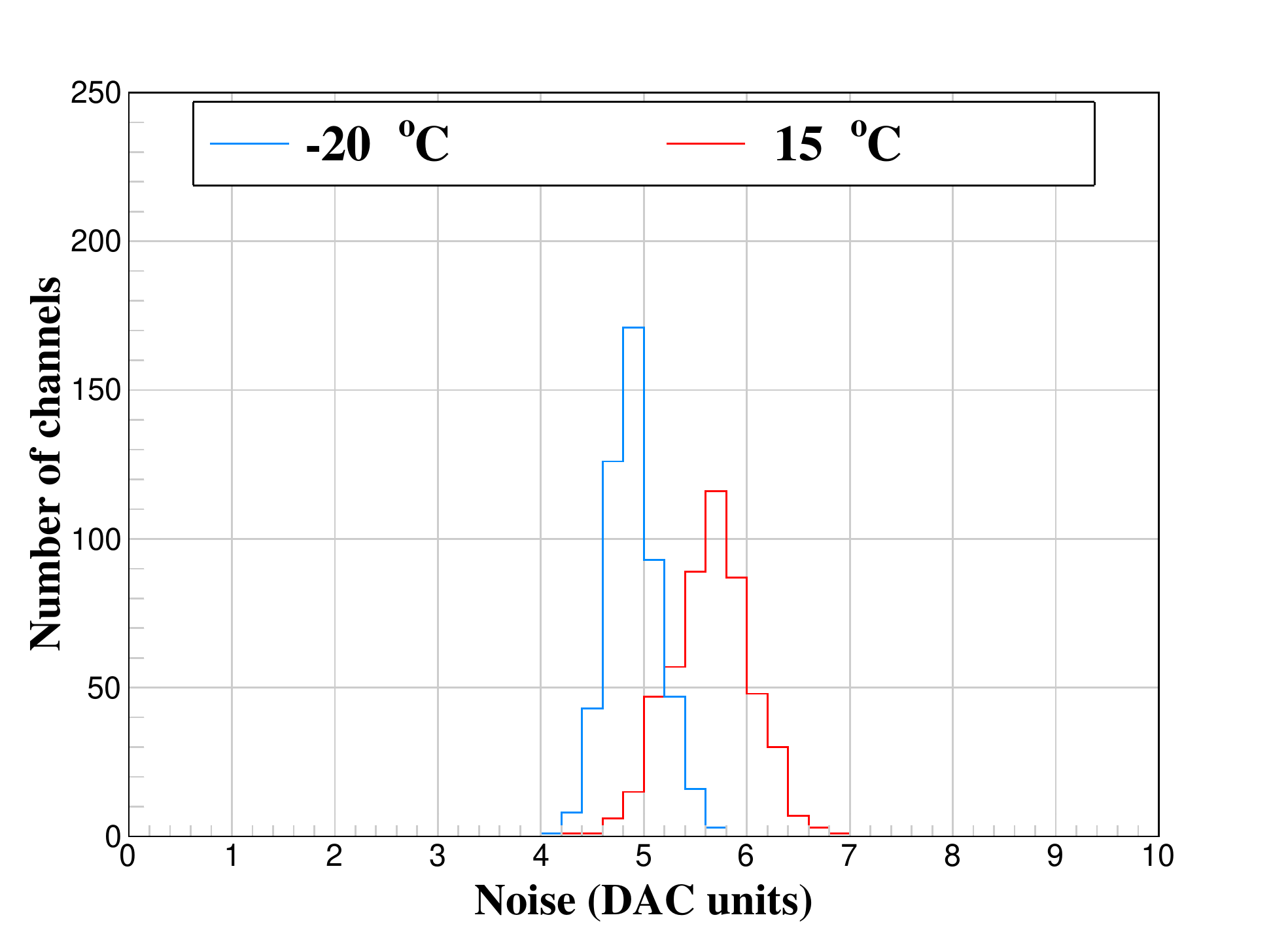}\label{fig:b}}%
	\caption{(a) Pedestal and (b) noise distributions for the unirradiated mini-module.}
	\label{fig:Calibration_Pedestal}
\end{figure}

\begin{figure}[t!]
	\centering
	\subfloat[]{\includegraphics[width=7cm, height=5cm]{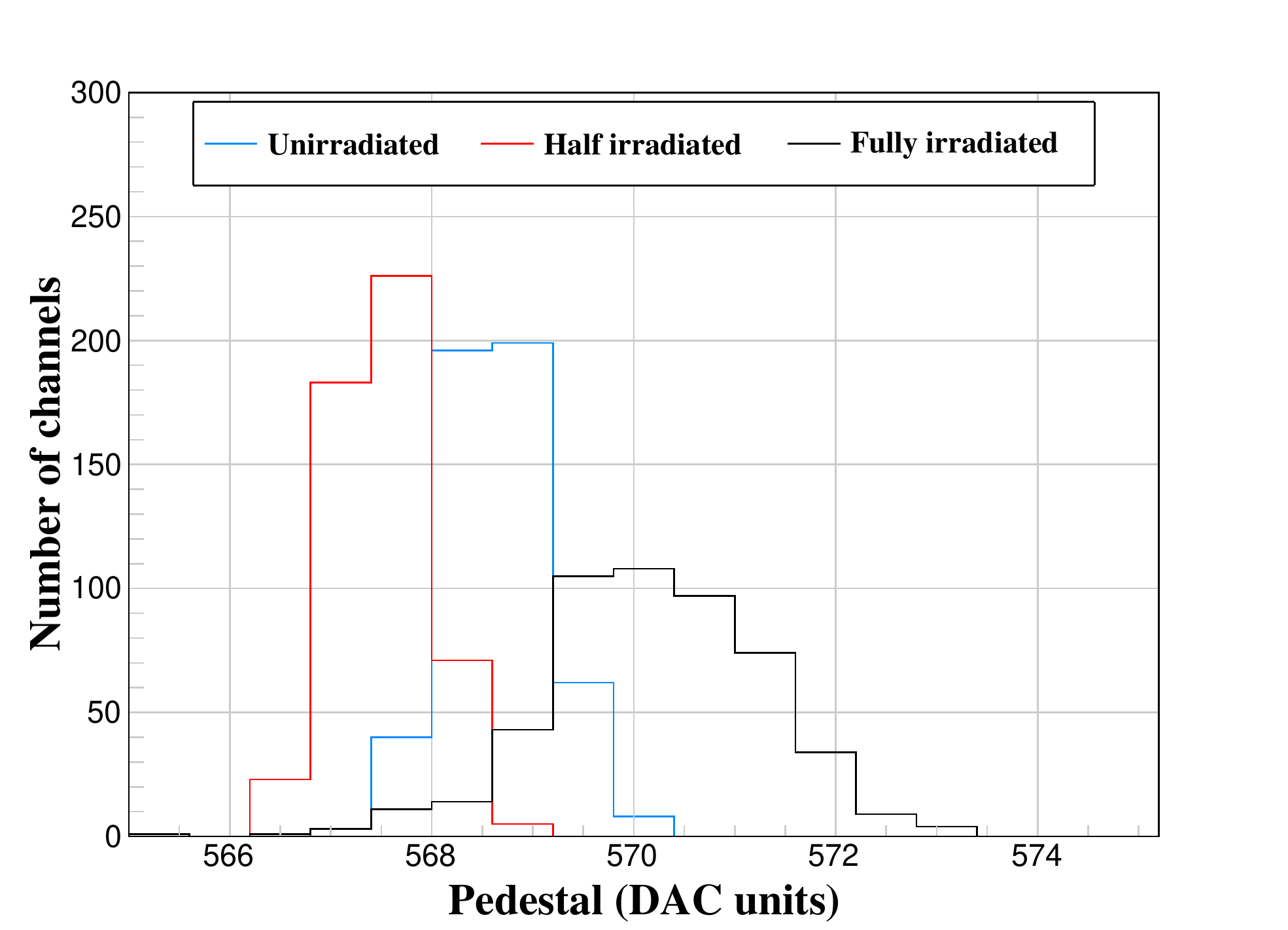}\label{fig:a}}%
	\subfloat[]{\includegraphics[width=7cm, height=5cm]{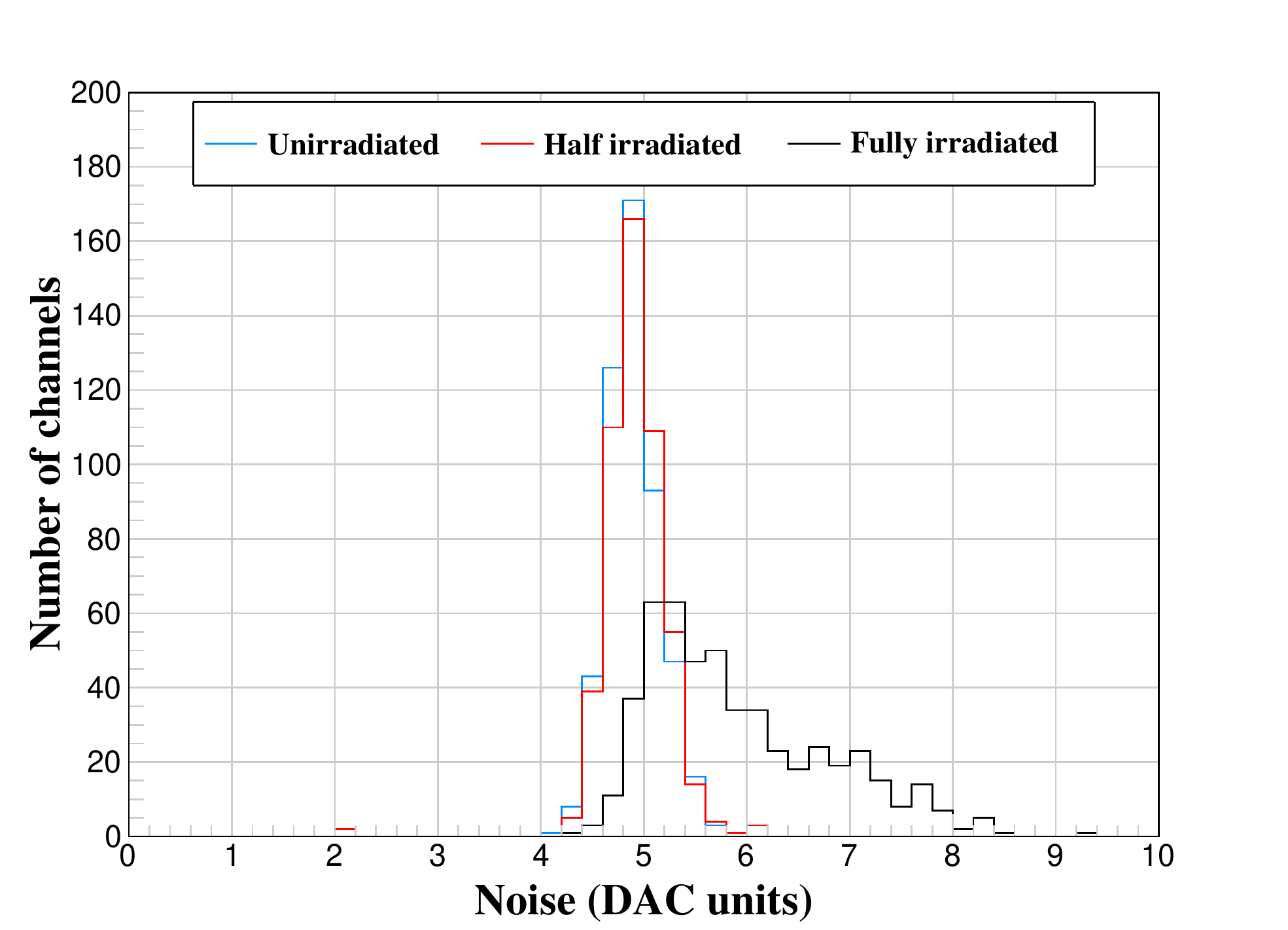}\label{fig:b}}%
	\caption{(a) Pedestal and (b) noise distributions for different irradiation fluences at $-20\,^\circ$C.}
	\label{fig:Calibration_Pedestal_Runs}
\end{figure}

As mentioned in Sec.~\ref{sec:cbc3}, the CBC3 may exhibit channel-to-channel variation of the pedestals with 
respect to the comparator threshold due to fabrication tolerances.
To ensure a uniform distribution of pedestals across one chip,
the calibration procedure (referred to as "pedestal and noise calibration" and described below) 
equalizes the distance of the pedestals from the comparator threshold, which is set globally by a register of the CBC3 (\Vcth~in Fig.~\ref{fig:CBC3AnalogueFrontEnd}).
A uniform distribution of pedestals results in the lowest possible common threshold (Sec.~\ref{sec:ThresholdOptimization}). 
The pedestal and noise calibration is performed in two steps. 
The first step is a scan of the \Vcth~register, starting from high threshold values, 
until the number of hits, or module occupancy, reaches $50\%$ of the number of triggers. 
The \Vcth~value at which the occupancy is $50\%$ is the chip average pedestal value.
After the initial determination of the pedestal, each individual channel is then calibrated changing 
its programmable offset (VPLUS in Fig.~\ref{fig:CBC3AnalogueFrontEnd}) until the channel occupancy reaches 50\%.
This individual channel pedestal shift is accomplished by adjusting the current through the 20\,k$\Omega$ resistor 
shown in Fig.~\ref{fig:CBC3AnalogueFrontEnd}.\\
The VPLUS values, calibrated during this first step of the procedure, are then saved and used for the second step 
to obtain the pedestal and noise distributions from the individual channel S-curves.
An S-curve is obtained by sweeping \Vcth~and measuring the occupancy for a fixed number of triggers.
The final pedestal value and the channel noise are extracted differentiating numerically the S-curve and 
then calculating the mean and the $\sigma$ of the obtained distribution.
The mean corresponds to the pedestal value and the $\sigma$ corresponds 
to the noise. An example of an S-curve recorded on a CBC3 is shown in Fig.~\ref{fig:CalibrationSCurve},
where higher numerical values of \Vcth~correspond to lower thresholds in the CBC3.
\\Figure~\ref{fig:Calibration_Pedestal} shows the uniformity of the front-end response and the noise 
after the pedestal and noise calibration at two temperatures, $+15\,^\circ$C and $-20\,^\circ$C. 
With the sensor biased to full depletion, which for this sensor is above 280\;V, the channel-to-channel variation of the module pedestal
is 139 and 86 electrons at~$+15\,^\circ$C and $-20\,^\circ$C, respectively. 
The conversion between DAC units and electrons is measured \mbox{using} an external X-ray source to be 156 electrons per \Vcth~DAC unit 
and it does not depend on temperature. 
The average pedestal position changes between different temperatures by 19 DAC units, from 587 at $+15\,^\circ$C to 568 at $-20\,^\circ$C. 
The noise mean value is 890 electrons when it is measured at $+15\,^\circ$C and decreases to 765 electrons at $-20\,^\circ$C.
\\The pedestal and noise were measured again at $-20\,^\circ$C after the different irradiation campaigns and the results are shown 
in Fig.~\ref{fig:Calibration_Pedestal_Runs}. 
The positions of the pedestal distributions (in DAC units) do not shift significantly with irradiation, 
but the distribution broadens after the full fluence has been delivered. 
The noise distribution does not change significantly when the detector is half irradiated, 
but broadens upwards after the full irradiation.
Using the same X-ray calibration as above it is possible to convert the \Vcth~register into electron units.
For example, for the default \Vcth~value of 560 at $+15\,^\circ$C, with the pedestal at 587, the corresponding threshold in electrons is about 4200.
A very similar value in electrons is calculated for the default \Vcth~value of 540 at $-20\,^\circ$C, since also the pedestal shifts
by 19 DAC units.

%% file: tex/detectorOptimization.tex
\subsection{Latency optimization}
\begin{figure}[t]
	\centering
	\subfloat[]{\includegraphics[width=0.45\textwidth]{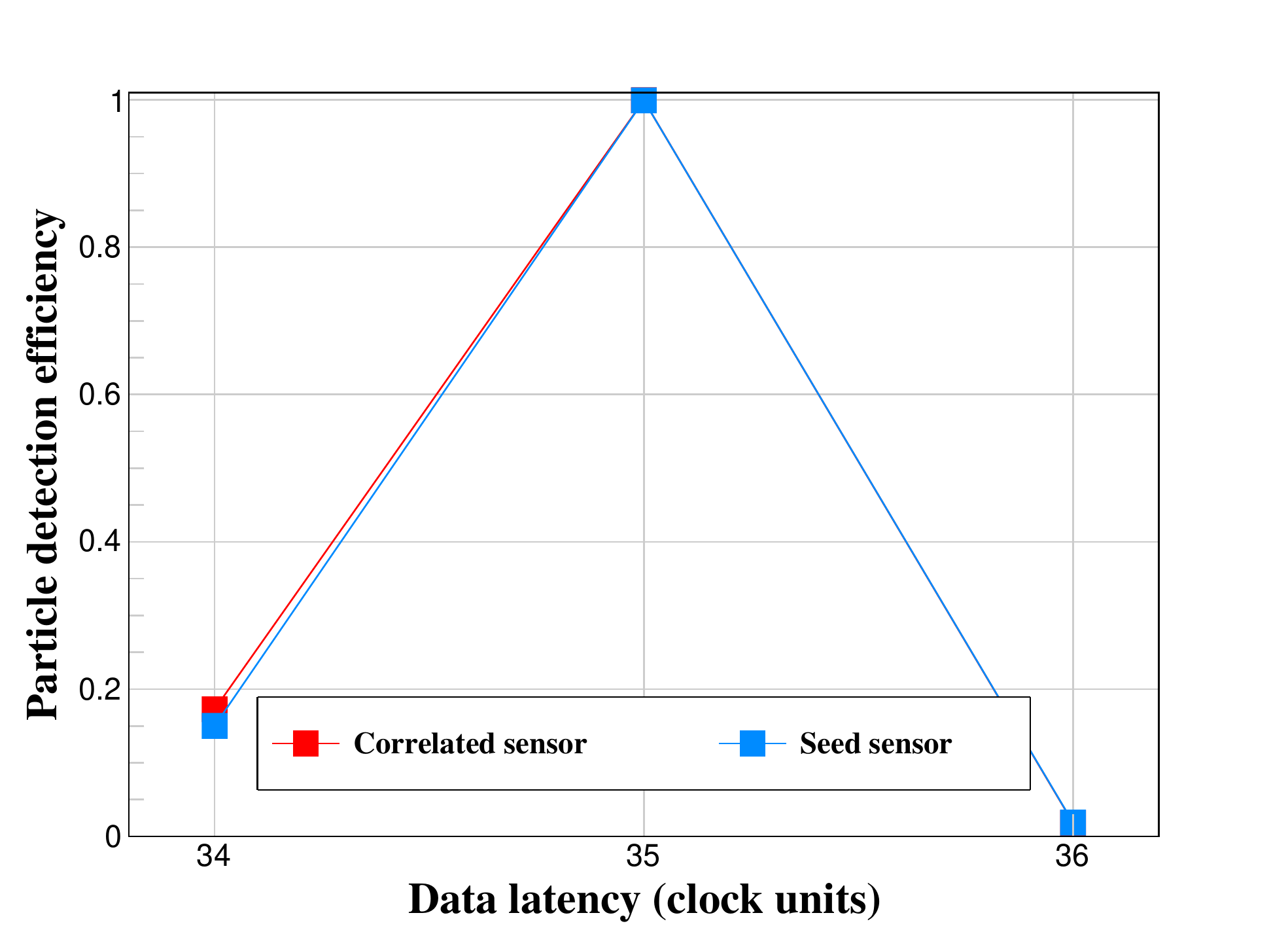}\label{fig:a}}%
	\subfloat[]{\includegraphics[width=0.45\textwidth]{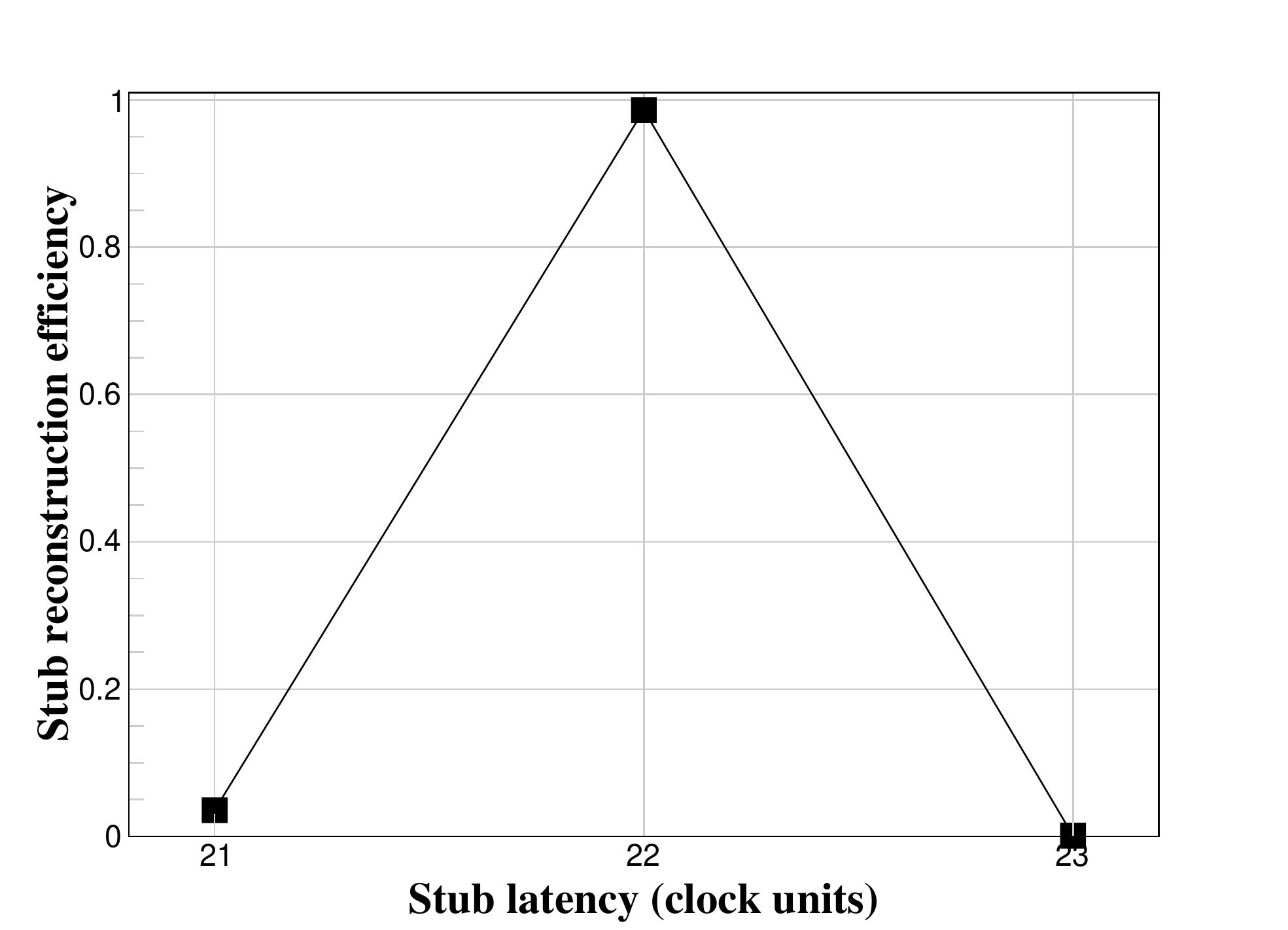}\label{fig:b}}%
	\caption{(a) Particle detection efficiency for each sensor at different values of the data latency. 
	(b) Stub reconstruction efficiency at different values of the stub latency. 
	The latencies are measured in units of 40\,MHz clock cycles. 
	Results are shown for the unirradiated module.}
	\label{fig:LatencyOptimization}
\end{figure}

For every clock cycle, each individual binary strip hit is stored into a pipeline SRAM in the CBC3.
Upon reception of a trigger, the corresponding data are sent to the DAQ back-end.
If the particle is detected at $t_0$ and the trigger, associated to that particle, is received by the CBC3 at $t_1$,
then $L_\text{{data}}$=$t_1-t_0$ represents the data latency that the CBC3 chip must take into account when it associates
a particular event stored in the SRAM at $t_0$ with the trigger received at $t_1$.
A similar latency needs to be defined for the stubs which, as soon as a particle is detected and a stub is formed, are sent to 
the DAQ back-end where they are stored in the RAM in the FPGA. 
For the stubs, $t_0$ is defined as the time when the stub data are stored in the FPGA, while $t_1$
is the time when the trigger, associated with those stubs, is received by the FPGA.
Similarly to $L_\text{{data}}$, $L_\text{{stub}}$ is defined as the time that the FPGA needs to take into account when it associates 
stubs to their corresponding trigger.
The latencies are measured in units of 40\,MHz clock cycles and while $L_\text{{data}}$ is set using an on-chip configuration register,
$L_\text{{stub}}$ is instead set using a register in the back-end FPGA.
Both latencies were specific to the trigger system used during the test beam and they both needed to be measured 
before taking any meaningful data.
The procedure to measure the two latencies is the same and relies on the fact that the beam particles, at the FTBF, are separated 
by many clock cycles. 
To determine the latency, 1000 triggers per latency setting were used.
The correct values of the data and stub latencies are the ones that maximize 
the particle detection and stub efficiencies, respectively, as shown in Fig.~\ref{fig:LatencyOptimization}.
The details regarding the determination of the particle detection and stub efficiencies are provided 
in Sec.~\ref{par:EfficiencyAnalysis}.
\begin{figure}[b]
	\centering
	\subfloat[]{\includegraphics[width=0.45\textwidth]{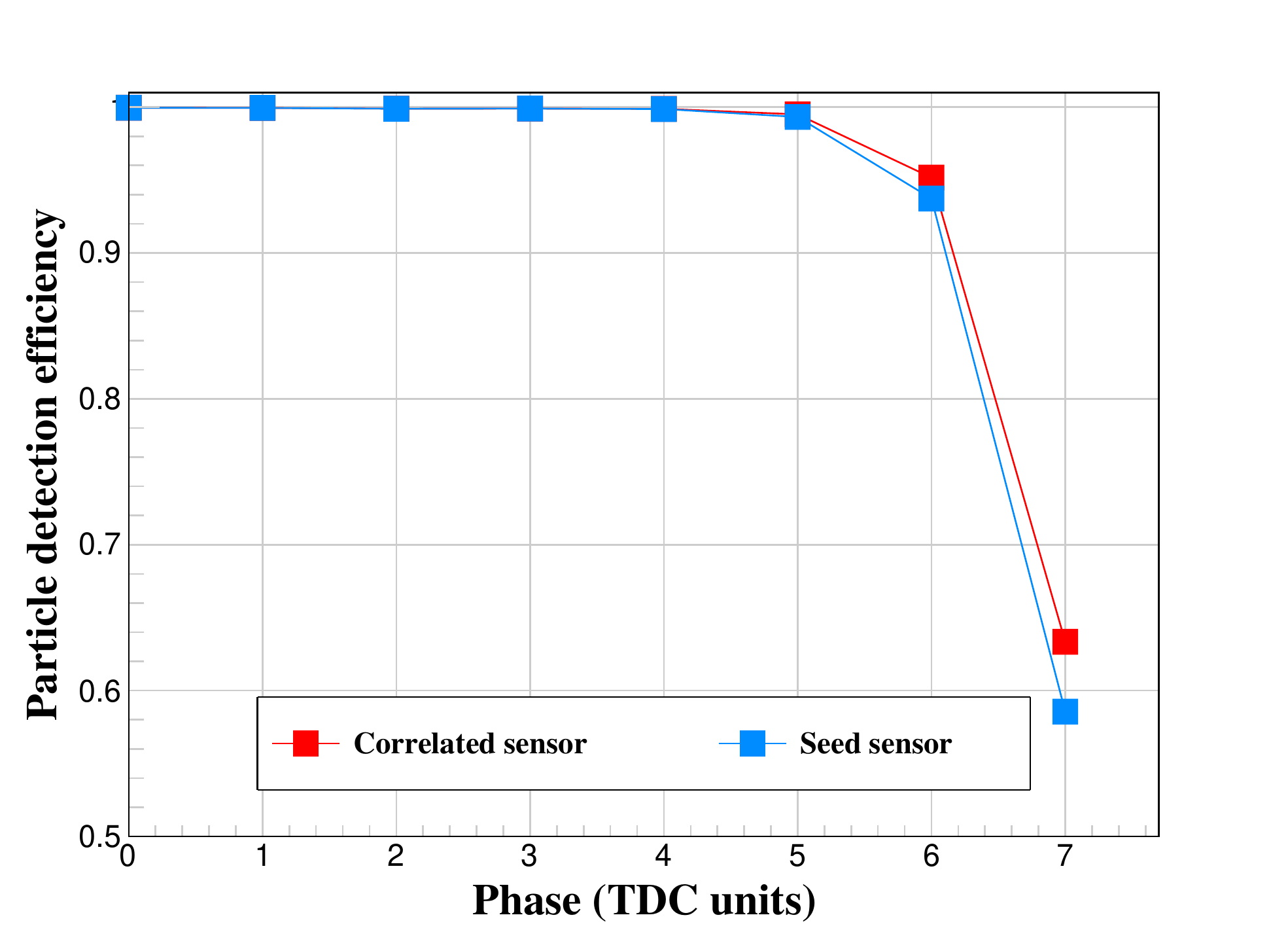}\label{fig:a}}%
	\subfloat[]{\includegraphics[width=0.45\textwidth]{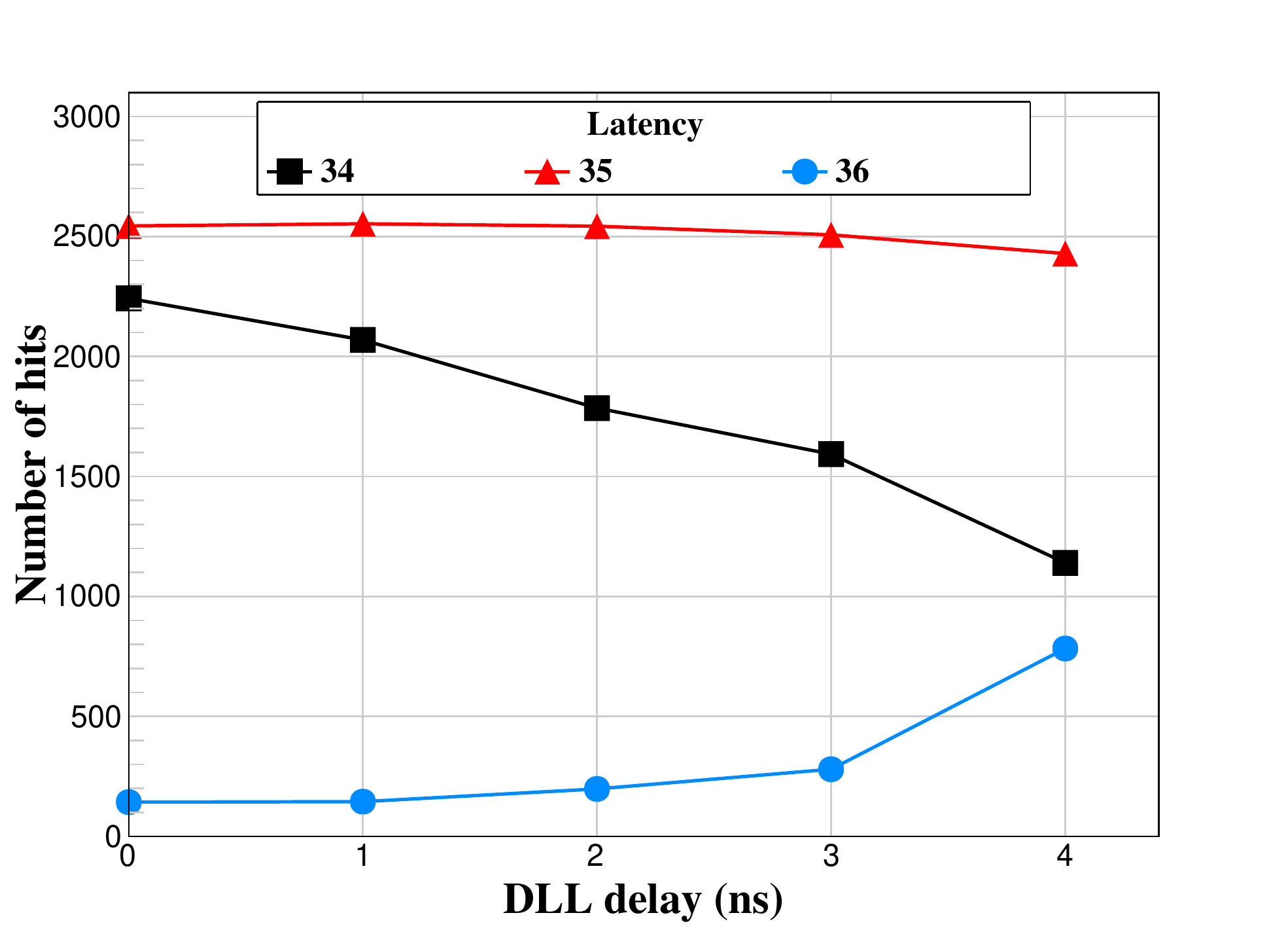}\label{fig:b}}%
	\caption[TDC Scan.]
	{(a) Particle detection efficiency as a function of the TDC phase, in units of 3.125 ns 
	for the seed and correlated sensors of the unirradiated mini-module. 
	(b) Number of hits as a function of the DLL delay for three different latencies.}
	\label{fig:TDCScan}
\end{figure}
Two more steps are necessary to tune the relative phase of the CBC3 clock and the beam. 
The beam synchronization with the CBC3 clock is done using the trigger board as described in Sec.~\ref{sec:ftbf}.
This optimization is done using a high resolution time-to-digital converter\,(TDC) implemented in the trigger board FPGA.
The TDC is a 3 bit counter\,(0-7) operating at 320\,MHz, and can thus shift the clock delivered 
to the back-end board and used by the mini-module by 3.125\,ns.
The procedure to optimize the TDC phase is done changing the counter \mbox{every} 1000 triggers and measuring, 
for each value of the counter, the particle detection efficiency. 
The TDC scan for the unirradiated mini-module is shown in Fig.~\ref{fig:TDCScan}~(a).
The value of the TDC phase used during data taking is the one which maximizes the particle detection efficiency 
and it is the farthest away in time from the phases where the efficiency is dropping.
From Fig.~\ref{fig:TDCScan} the TDC operating value chosen was 0.
The relative phase is further optimized scanning the CBC3 internal DLL register, which has a time resolution of 1\,ns 
(Fig.~\ref{fig:TDCScan}~(b)).
The optimization of the DLL is done changing the register every 1000 triggers and counting, 
for each latency and for every register value, the number of hits recorded on the mini-module.
The optimal DLL setting was determined to be the one just before the number of hits were increasing 
for latency 36 and decreasing for \mbox{latency 35}. Figure~\ref{fig:TDCScan}~(b) shows that this condition was met between 
2 and 3\,ns. The DLL operating value chosen was 1. 

\subsection{Threshold optimization\label{sec:ThresholdOptimization}}
\begin{figure}[b]
	\centering
	\includegraphics[width=0.45\textwidth]{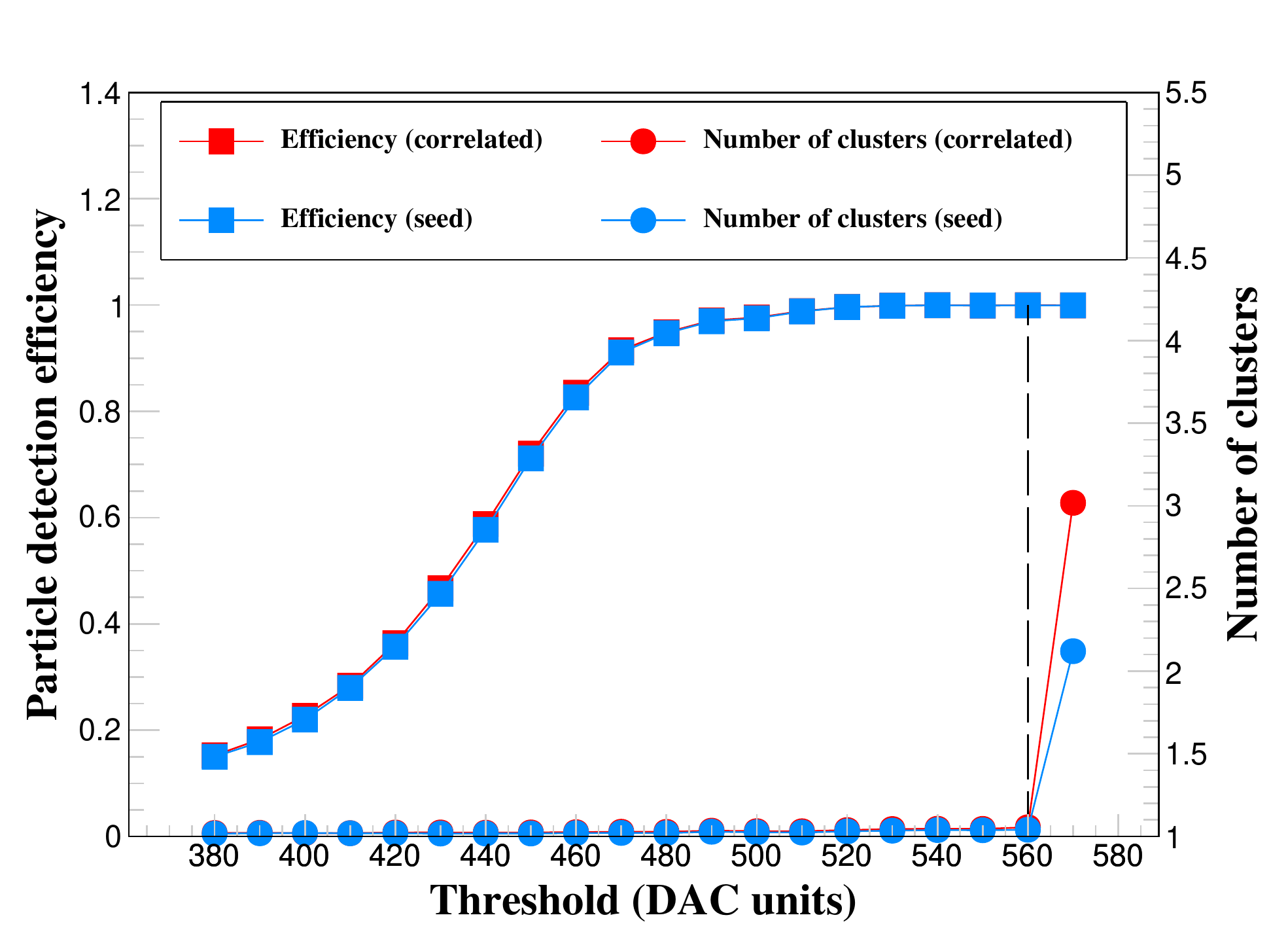}
	\caption{Particle detection efficiency and number of clusters detected per event as a function of \Vcth, for the unirradiated mini-module. 
	The dashed line corresponds to the optimal threshold.}
	\label{fig:ThresholdOptimizationScan}
\end{figure}

\begin{figure}[t]%
	\centering
	\subfloat[]{\includegraphics[width=0.37\textwidth]{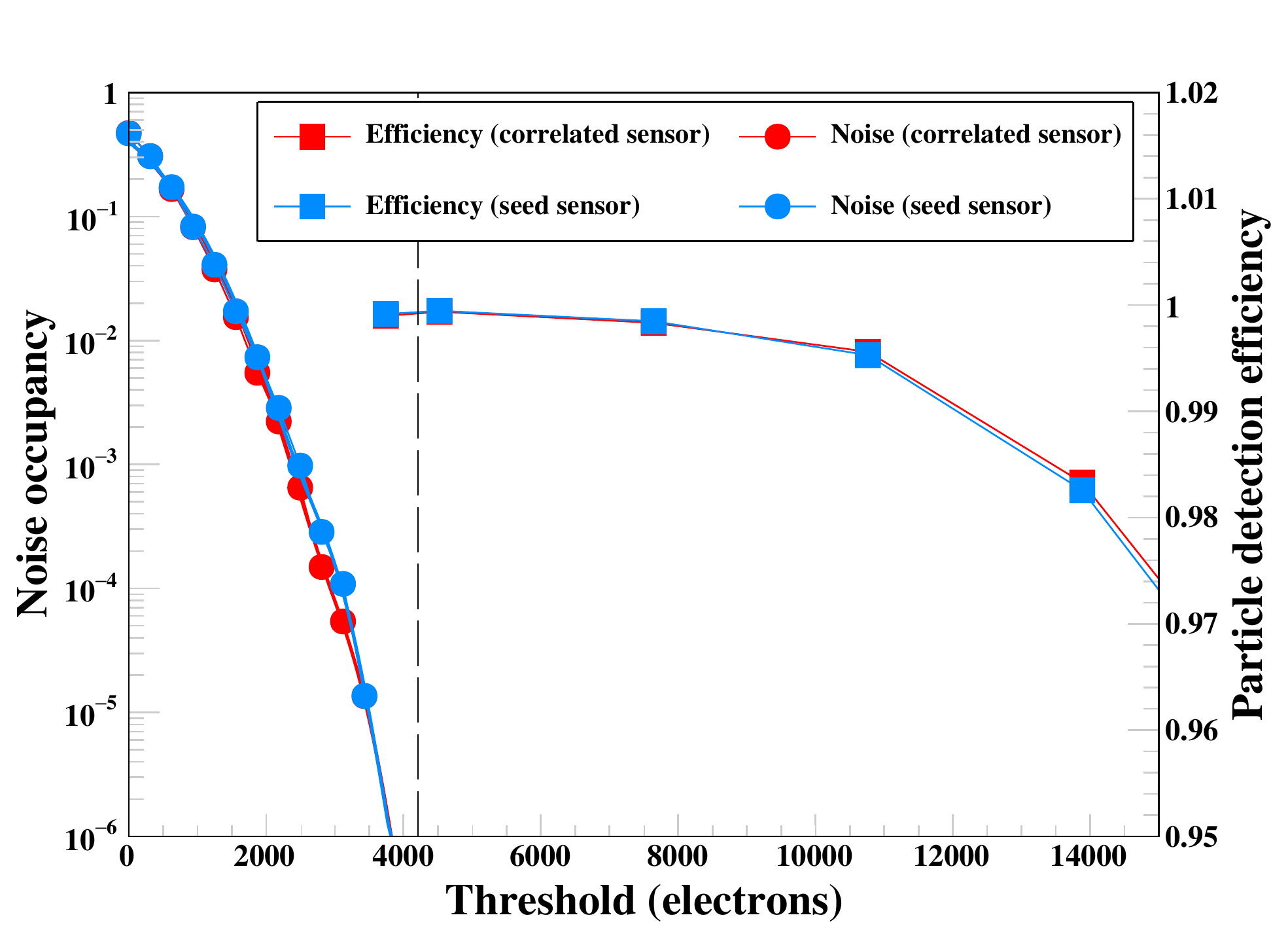}\label{fig:a}}%
	\subfloat[]{\includegraphics[width=0.37\textwidth]{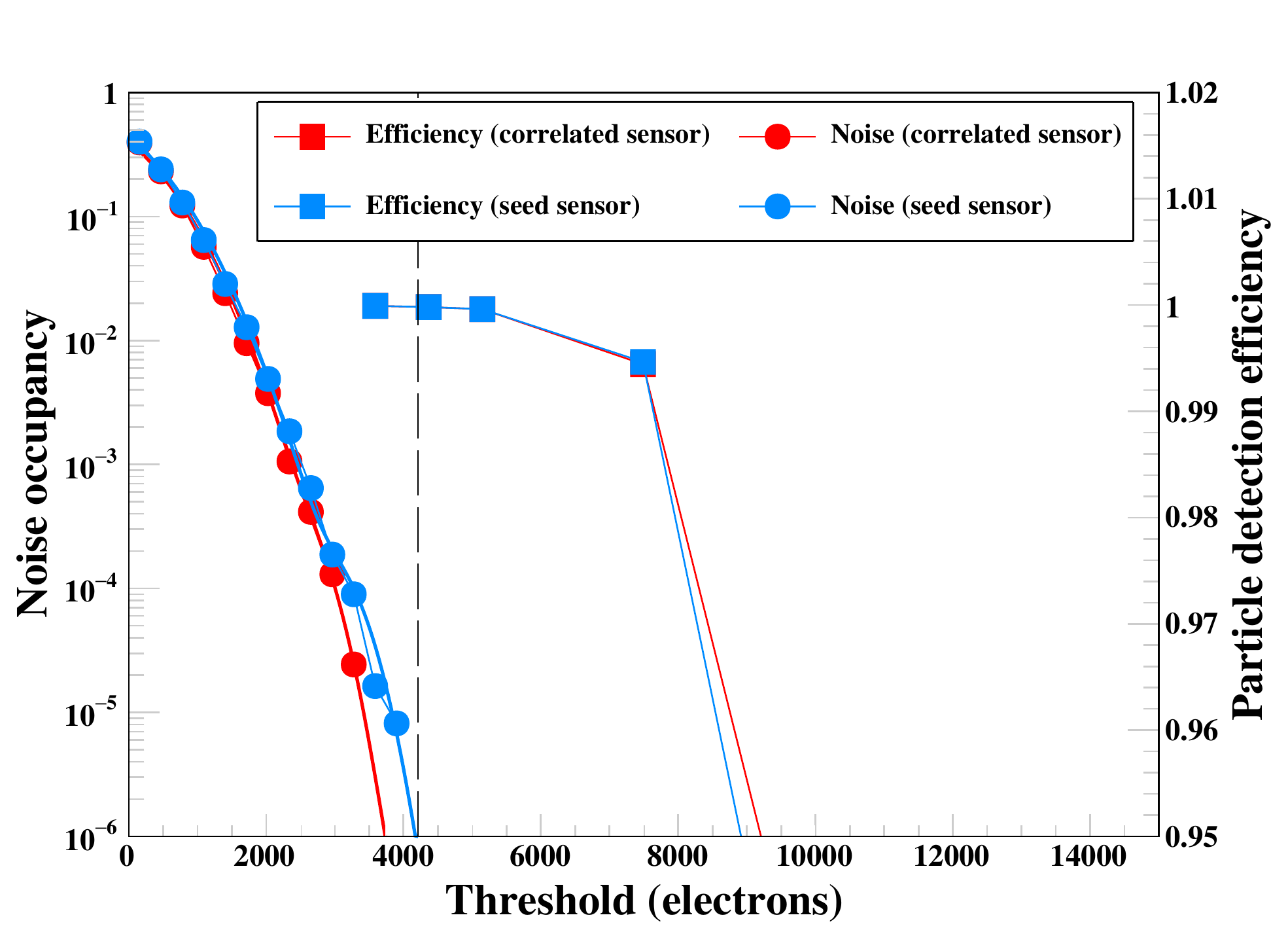}\label{fig:b}}\\
	\subfloat[]{\includegraphics[width=0.37\textwidth]{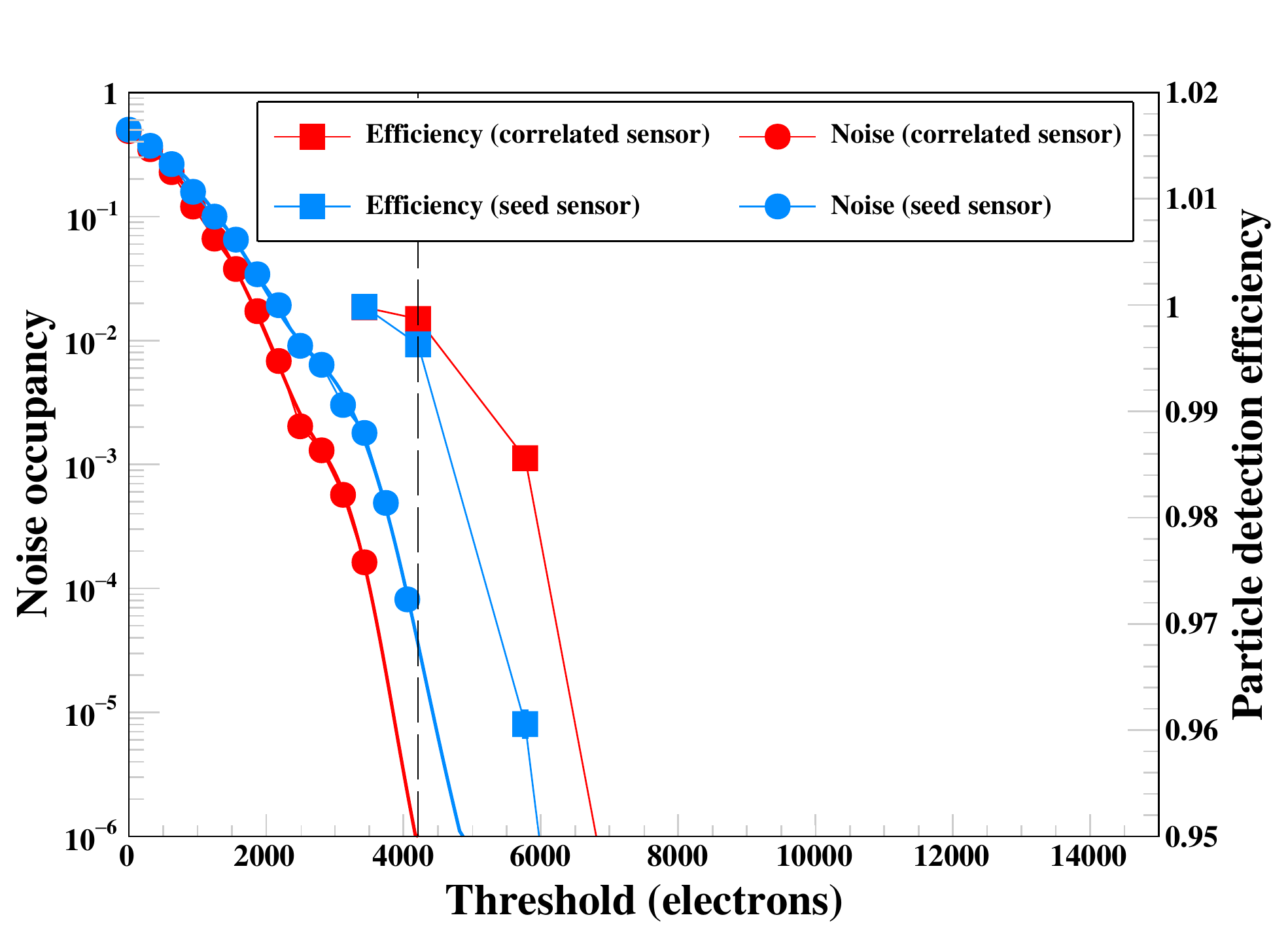}\label{fig:c}}%
	\caption{Particle detection efficiency and noise occupancy as a function of \Vcth~for the unirradiated (a), 
	half irradiated (b), and fully irradiated (c) mini-module. The dashed lines correspond to the optimal threshold.}%
	\label{fig:ThresholdAndNoiseScan}%
\end{figure}

The optimal comparator threshold is defined as the threshold that maximizes
the particle detection efficiency and maintains the noise hit occupancy, 
defined as the probability of recording a noise hit in a channel per bunch crossing,
below 10$^{-4}$. The optimal \Vcth~value was 
determined to be 560 DAC units by measuring the particle detection efficiency and the number of clusters as a function of the threshold. 
The number of clusters increases substantially, due to noise, only when the threshold is close to the pedestal, 
as shown in Fig.~\ref{fig:ThresholdOptimizationScan} for the \mbox{unirradiated} mini-module. 
Figure~\ref{fig:ThresholdAndNoiseScan} shows the efficiency and the noise hit occupancy as a 
function of \Vcth~for the unirradiated, half irradiated, and fully irradiated mini-module
and indicates that a threshold of 4200 electrons is optimal in all cases.
Figure~\ref{fig:ThresholdAndNoiseScan} also shows that in all runs the threshold could be lowered 
to values that kept the particle detection efficiency well above $99\%$.
In \mbox{particular}, for the fully irradiated mini-module, where the efficiencies start 
dropping earlier due to the reduced amount of charge collected, 
it is possible to achieve high efficiencies with a noise level below 10$^{-4}$.
%


%% file: tex/alignmentBeamParameters.tex
\subsection{Calibration of the rotation angle}
\begin{figure}[h]
	\centering
		\includegraphics[ width=0.45\textwidth]{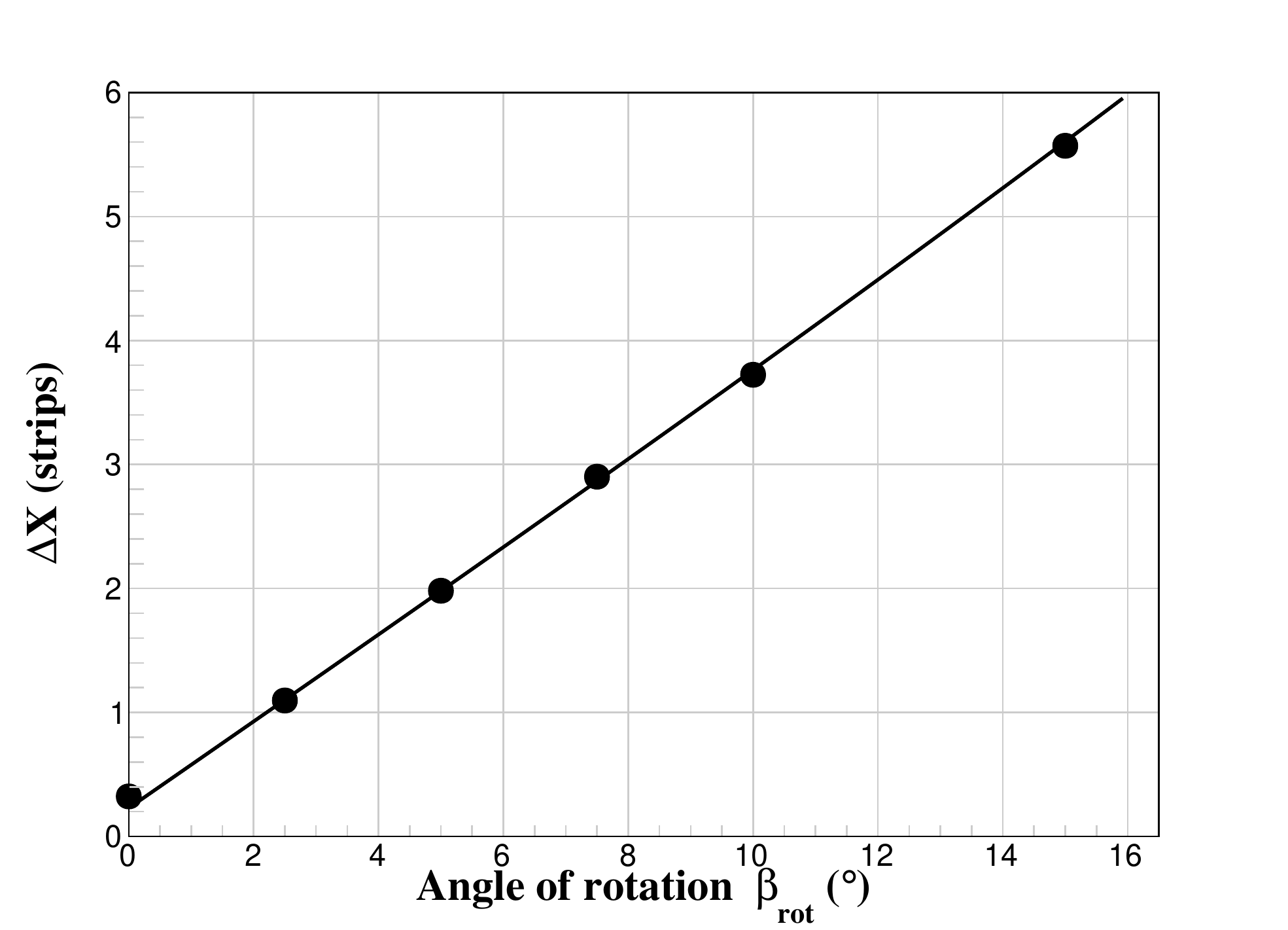}
	\caption{ \label{fig:alignment}
	Mean stub displacement measured for different angles of the mini-module.
	}	
\end{figure}

The last step in the setup procedure was to determine the offset in the rotation angle between the mini-module and
the beam. The mini-module was mounted on a rotation stage with the strips pointing in the vertical direction. 
The module was rotated around the vertical axis.
The angle of the mini-module could be changed in steps of half a degree using the rotation stage.
Information about the alignment is extracted from the measurement of the impact point difference between the two mini-module sensors, 
which is shown in Fig.~\ref{fig:alignment}.
The stub displacement $\Delta X$ can be written as
\begin{equation}
	\Delta X = \Delta X_{0} + \frac{d}{p} \tan(\beta_\text{rot} + \beta_\text{0}),
	\label{eq:alignment}
\end{equation}
where $\beta_\text{rot}$ is the angle as read off from the rotation stage, $\beta_\text{0}$ the angular offset between the beam
and the normal to the mini-module at $\beta_\text{rot} = 0$, $d = 1.8$\,mm the distance between the two sensors, $p = 90\,\mu$m
the strip pitch, and $\Delta X_{0}$ the translational misalignment of the two sensors.
Using eq.~\ref{eq:alignment}, it is possible to fit the data, as shown in Fig.~\ref{fig:alignment} and measure 
$\beta_\text{0}$ or, equivalently, $\Delta X_{0}$ at $\beta_\text{rot} = 0$. 
Fixing $\Delta X_{0} = 0$, the fit returns the angular misalignment $\beta_\text{0} = 0.385^{\circ} \pm 0.004^{\circ}$,
that is needed to correct the beam incident angle. 
A similar analysis has been done for each data set and the measured misalignment 
values have been used to correct the beam incident angle for each run. 
All plots and results presented in this paper have these corrections applied.

%% file: tex/efficiencyAnalysis.tex
\subsection{Particle detection efficiency}
{\label{par:EfficiencyAnalysis}}
To measure the particle detection efficiency, tracks were selected with the following criteria:
\begin{description}
\item [$\bullet$] a track must traverse all telescope planes or miss at most one;
\item [$\bullet$] there cannot be more than two clusters on any telescope plane;
\item [$\bullet$] there can only be one reconstructed track;
\item [$\bullet$] the track must have a $\chi^{2}$/ndof $< 5$.
\end{description}
These stringent criteria lead to a high quality data set.
Inefficiencies due to edge effects were avoided
by excluding cases where the predicted track pointed to a masked region or outside the active
area of the sensor under study. 
The above criteria defined the sample of events for the measurement (with size $N_\text{tot}$). 
The particle detection efficiency 
is then measured as the number of events where the projected track is matched to a cluster within a window
of $\pm 135\,\mu$m divided by $N_\text{tot}$.
In what follows, the efficiency is presented as a function of \Vcth, the strip number, 
and the relative position of the track with respect to the center of the strip.

\subsubsection{Dependency on \Vcth}
\begin{figure}[b]
	\centering
	\subfloat[]{\includegraphics[width=0.45\textwidth]{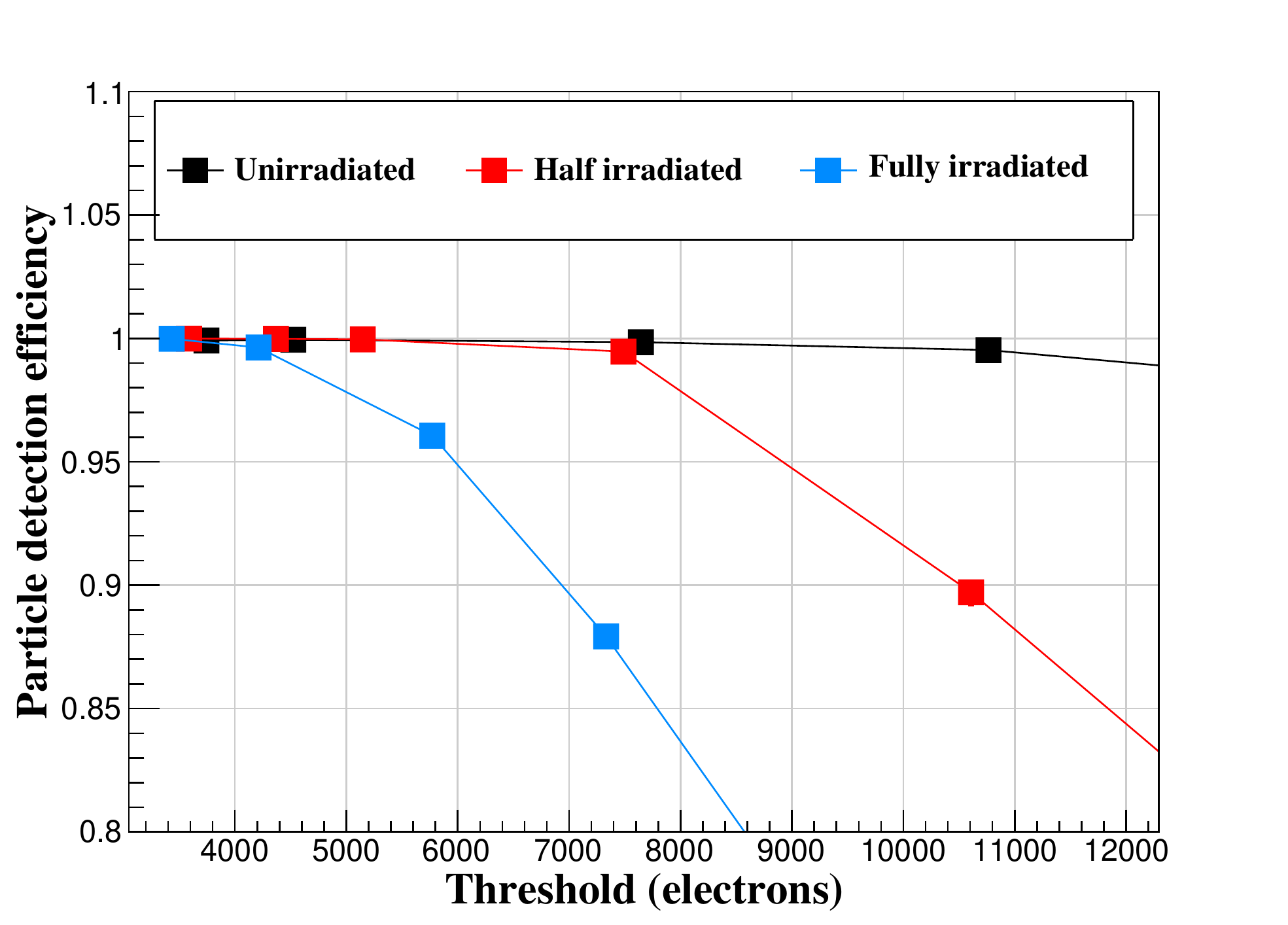}\label{fig:a}}%
	\subfloat[]{\includegraphics[width=0.45\textwidth]{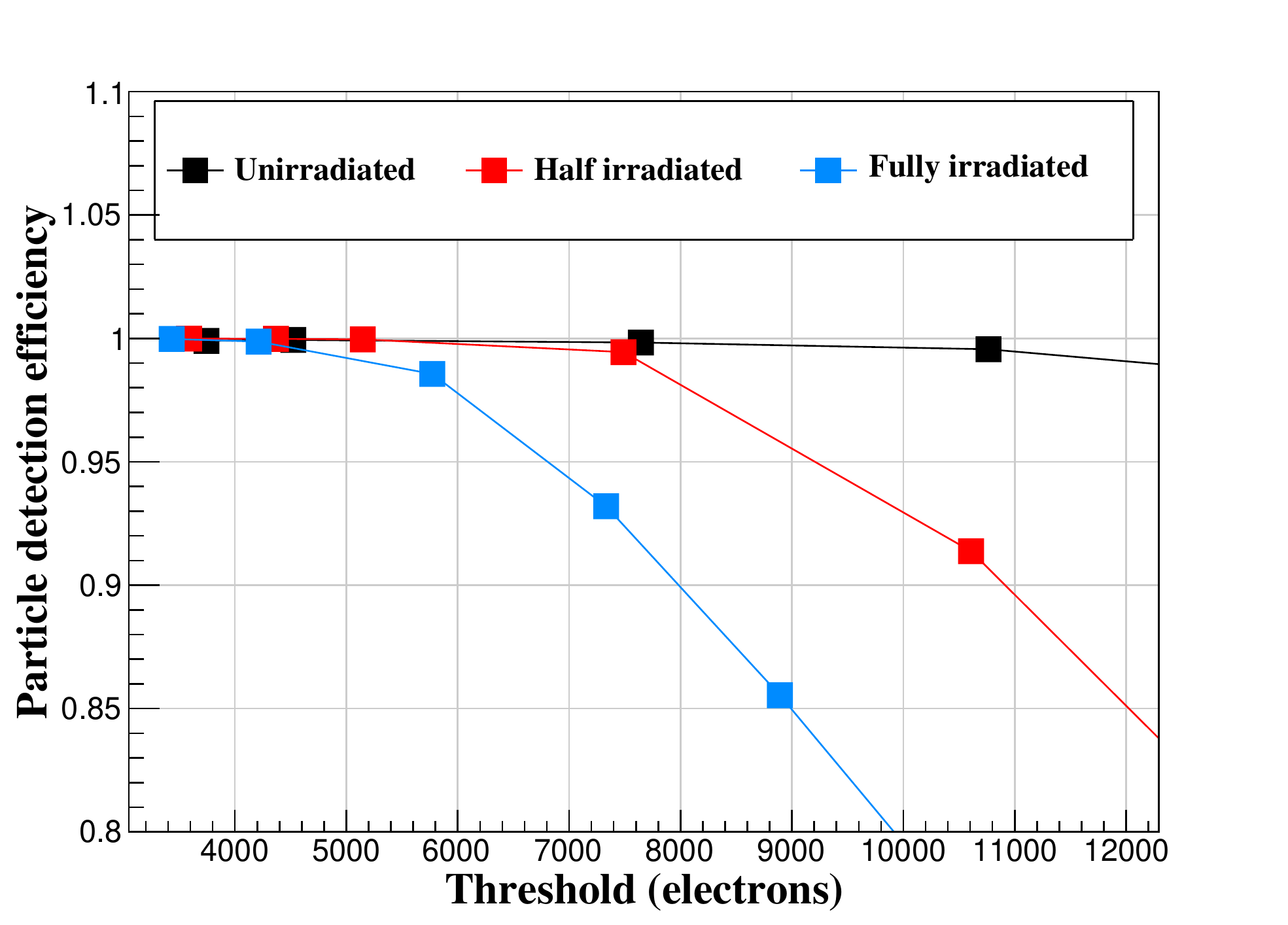}\label{fig:b}}%
	\caption{Particle detection efficiency vs. \Vcth~for the different irradiation fluences. (a) Seed sensor, (b) correlated sensor.}
	\label{fig:EfficiencySummaries}
\end{figure}

Efficiency scans were performed to measure the impact of irradiation on the mini-module's performance.
Figure~\ref{fig:EfficiencySummaries} compares the efficiency measured in three threshold scans corresponding to the 
different irradiation conditions. 
The mini-module is more than $99.5\%$ efficient for thresholds as high as 7000 electrons when unirradiated 
and half irradiated, far beyond the standard working threshold of 4200 electrons. 
After the full irradiation, the mini-module is still more than $99\%$ efficient at a threshold of 4200 electrons, 
but the efficiency starts dropping rapidly for values above 5000 electrons.
These scans thus show that the module works efficiently at thresholds lower than 5000 electrons
even when the charge collected is reduced by irradiation, see Sec.~\ref{par:ChargeCollectionMeasurement}.


\subsubsection{Dependency on the strip number}
\begin{figure}[t]
	\centering
	\subfloat[]{\includegraphics[width=0.45\textwidth]{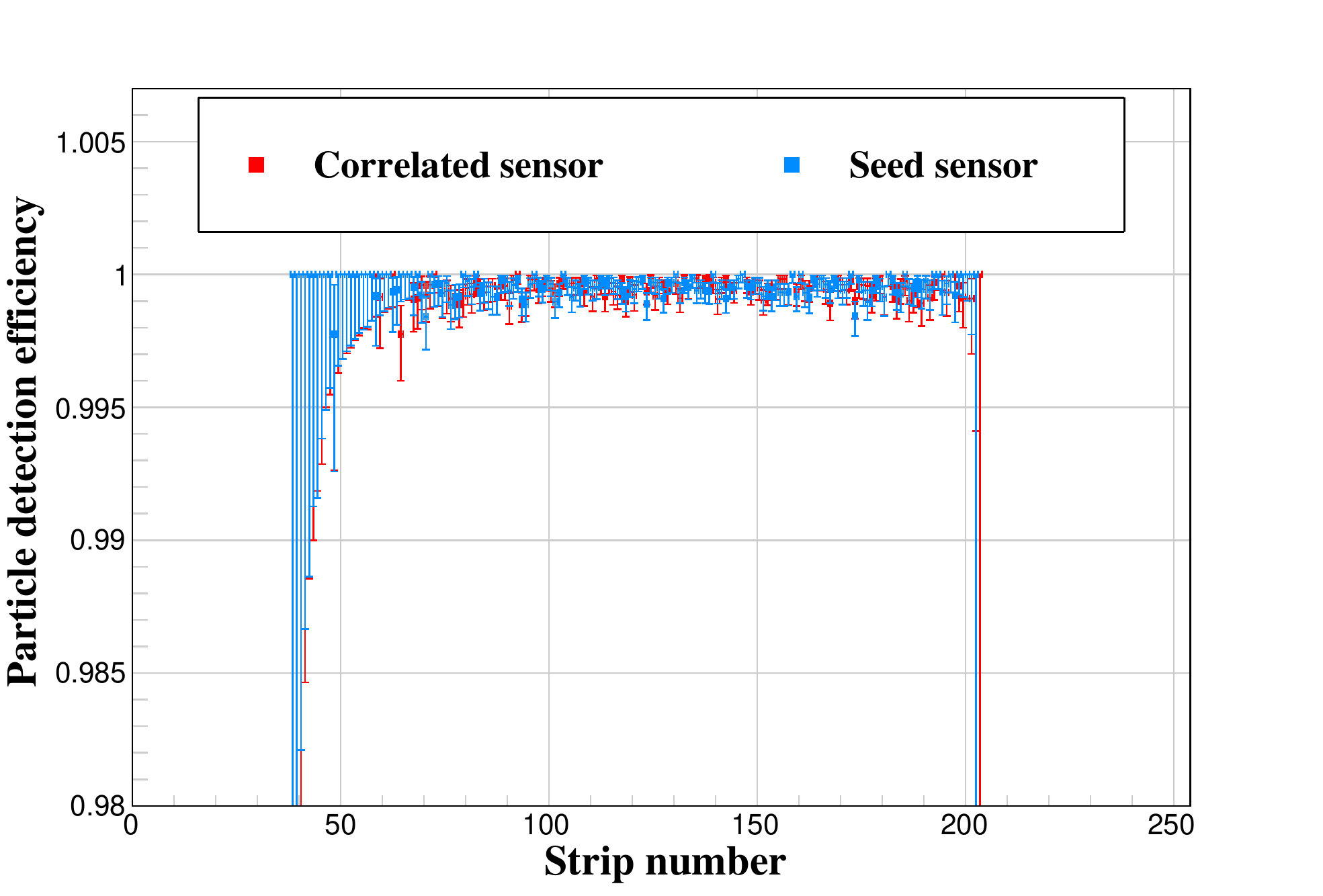}\label{fig:a}}%
	\subfloat[]{\includegraphics[width=0.45\textwidth]{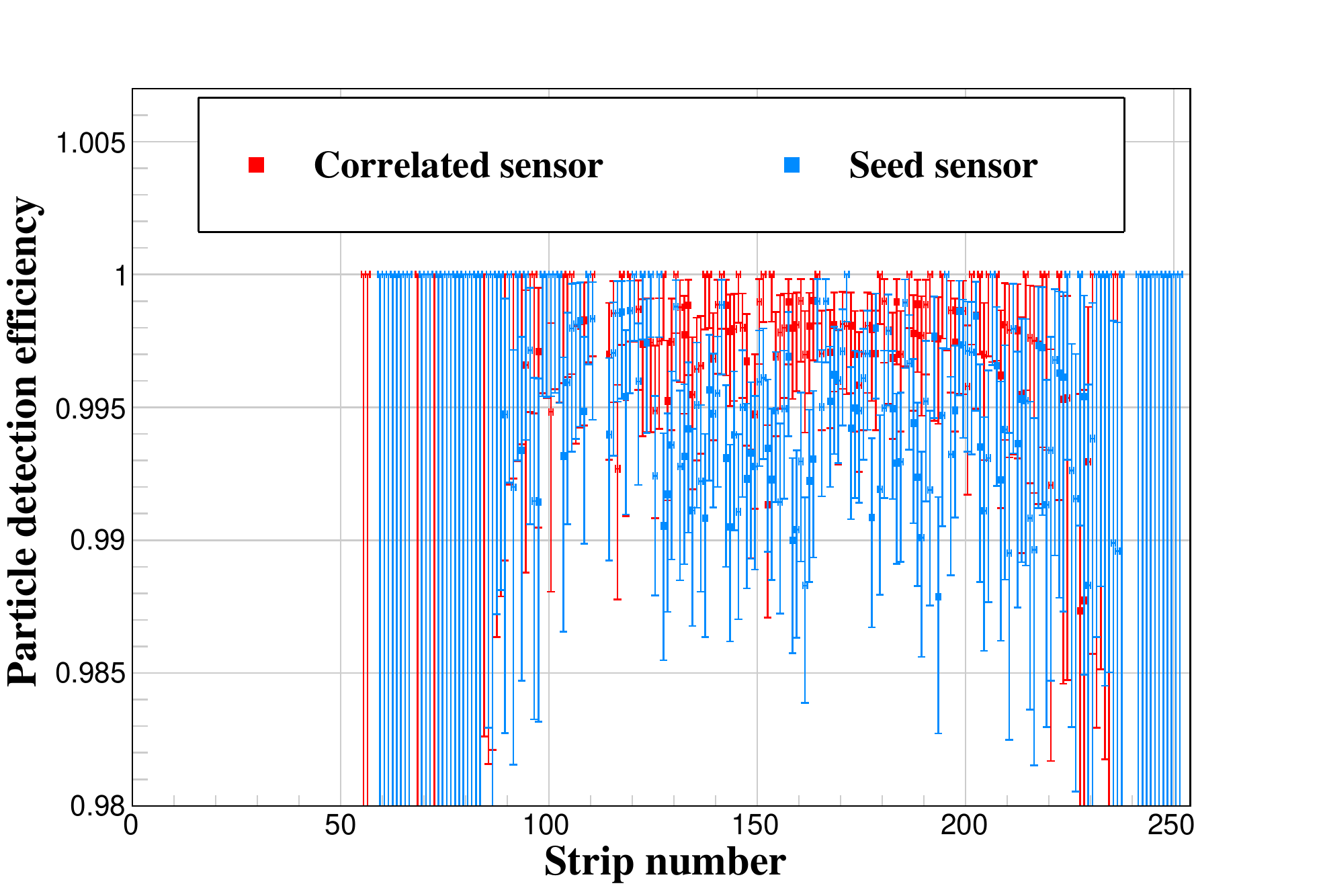}\label{fig:b}}%
	\caption{Particle detection efficiency along the sensor for the unirradiated (a) and the fully irradiated (b) mini-module.}
	\label{fig:ClusterSensorEfficiencyAll}
\end{figure}

Figure~\ref{fig:ClusterSensorEfficiencyAll} shows that the efficiency as a function of the strip number for the 
unirradiated mini-module is close to $100\%$, while for the fully irradiated mini-module the efficiency is slightly reduced 
but it still exceeds $98.5\%$ everywhere. In both cases the threshold was set to the standard value of 4200 electrons. 
The particle detection efficiency has a very large statistical uncertainty 
at the edges of the narrow beam which is mostly hitting the center of the sensors.


\subsubsection{Dependency on relative position with respect to the center between adjacent strips}
\begin{figure}[h!]
	\centering
	\subfloat[]{\includegraphics[width=0.45\textwidth]{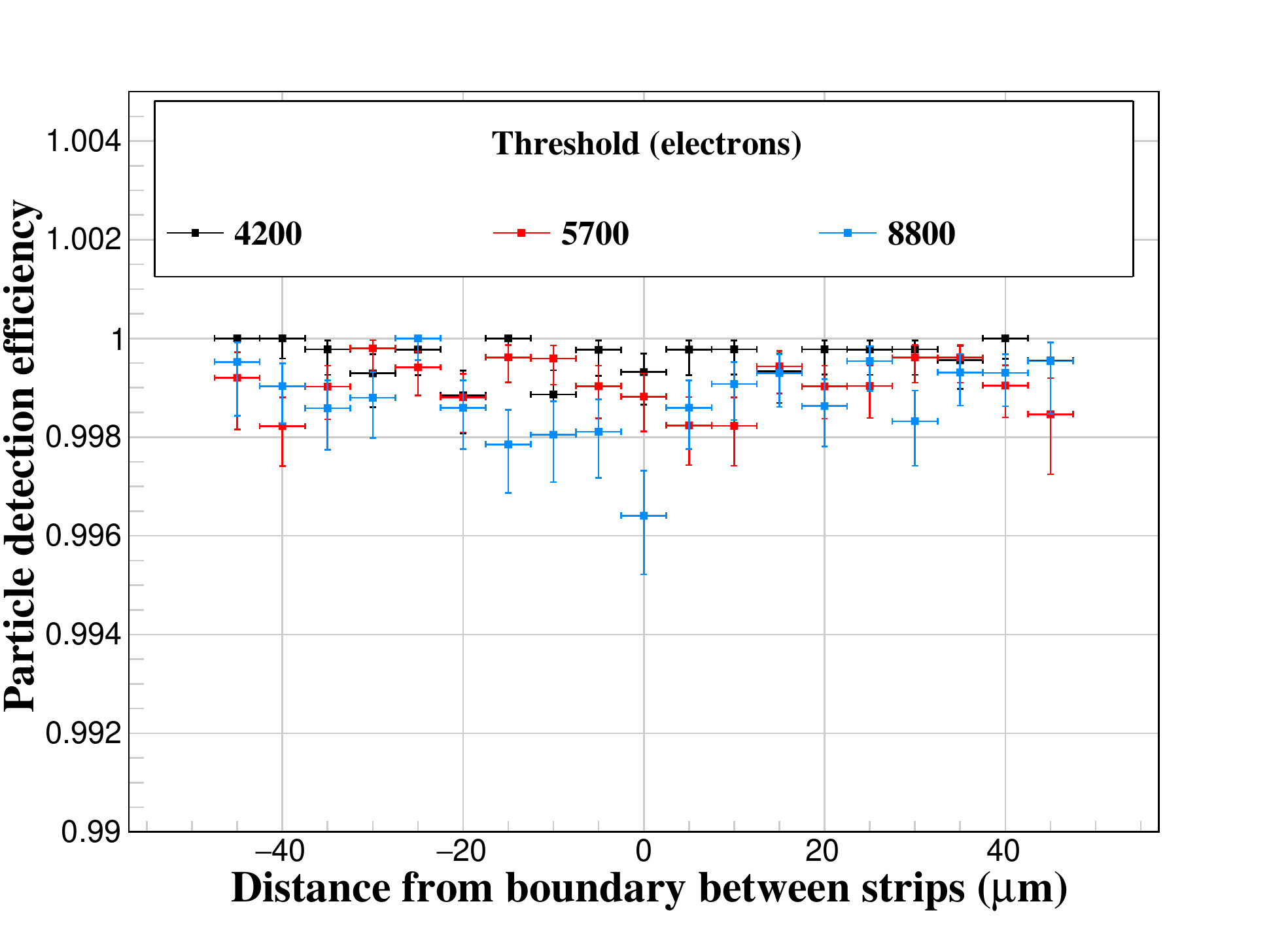}\label{fig:a}}%
	\subfloat[]{\includegraphics[width=0.45\textwidth]{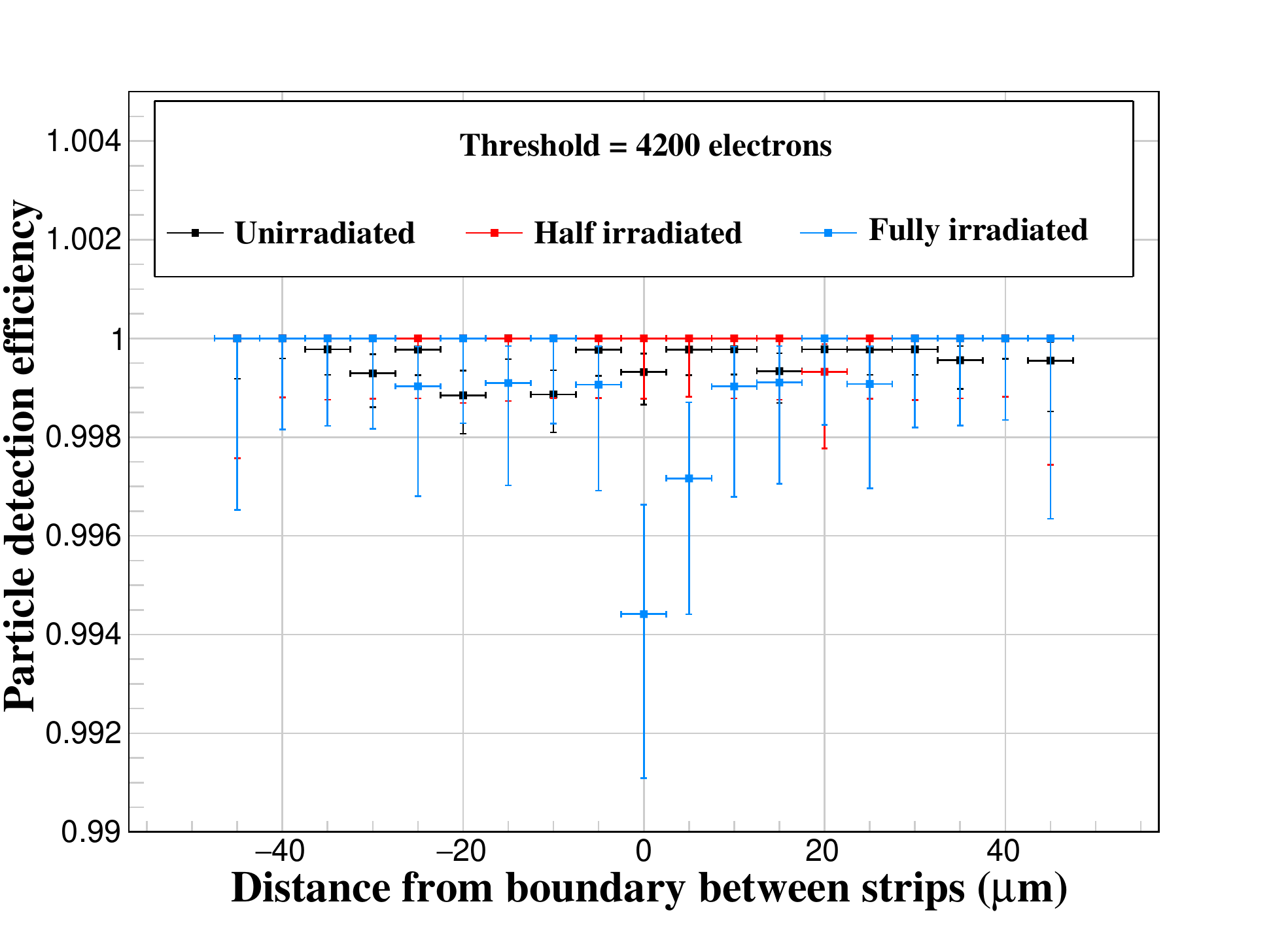}\label{fig:b}}%
	\caption{Particle detection efficiency as a function of the track location with respect to the center between adjacent strips 
	for different thresholds of the unirradiated mini-module (a) and for different irradiation fluences (b).}
	\label{fig:FirstSecondHitEfficiency}
\end{figure}

The particle detection efficiency was also measured as a function of the relative position of 
the track to the center between adjacent strips to check for possible inefficiencies due
to the reduced amount of charge collected by individual strips when the charge is shared.  
The study was performed for different thresholds, Fig.~\ref{fig:FirstSecondHitEfficiency}~(a), 
and for the different irradiation configurations, Fig.~\ref{fig:FirstSecondHitEfficiency}~(b). 
The results indicate that, due to charge sharing, 
there is a small inefficiency only at very high thresholds or when the mini-module is fully irradiated.

%% file: tex/timeWalkAnalysis.tex
\subsection{Pulse shape}
\begin{figure}[t]
	\centering
	\includegraphics[width=0.42\textwidth]{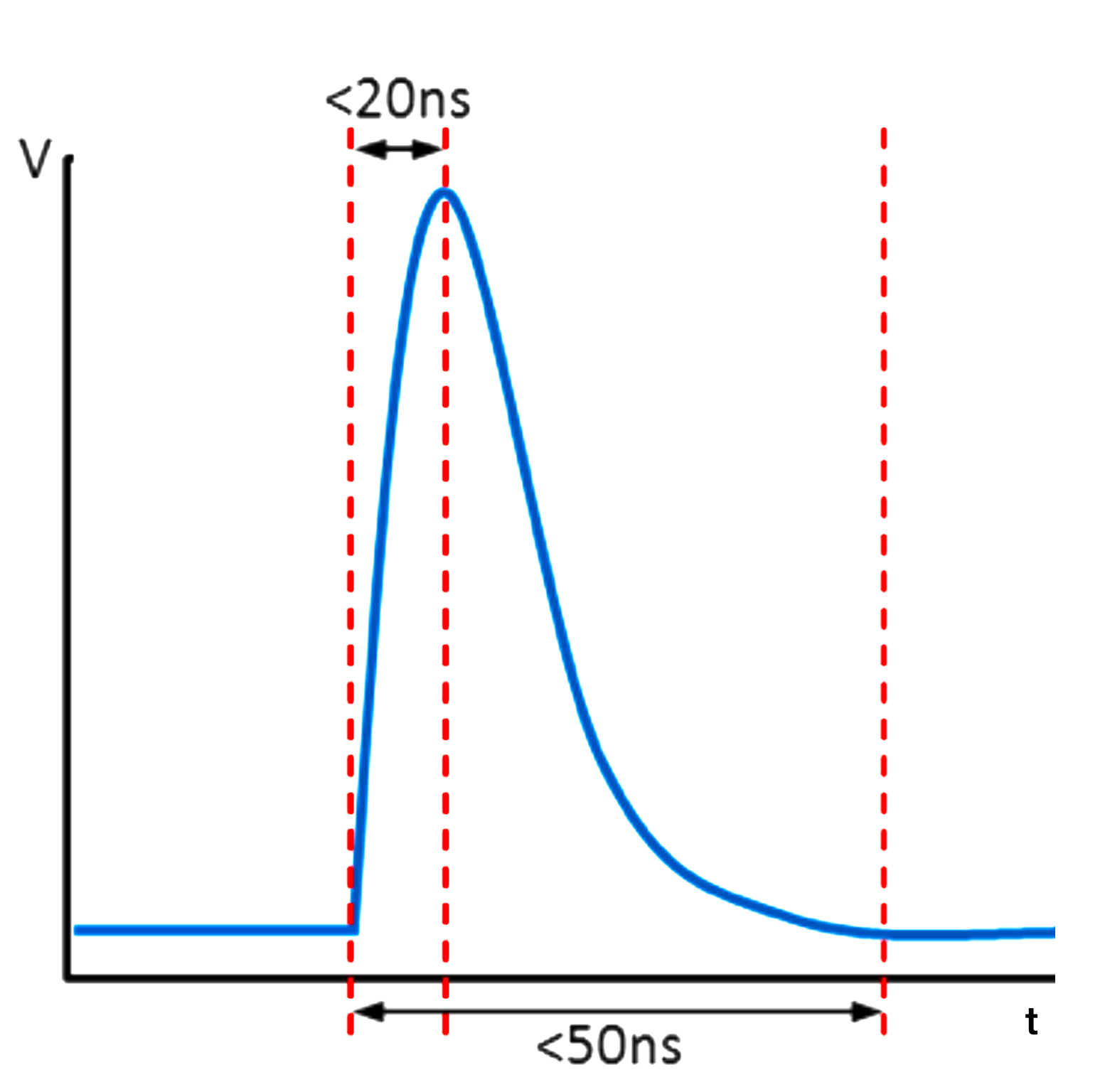}
		\caption{ \label{fig:CBC3PulseShaping}
		Sketch of the CBC3 pulse shape.}	
\end{figure}

\begin{figure}[b]
	\centering
	\subfloat[]{\includegraphics[width=0.42\textwidth]{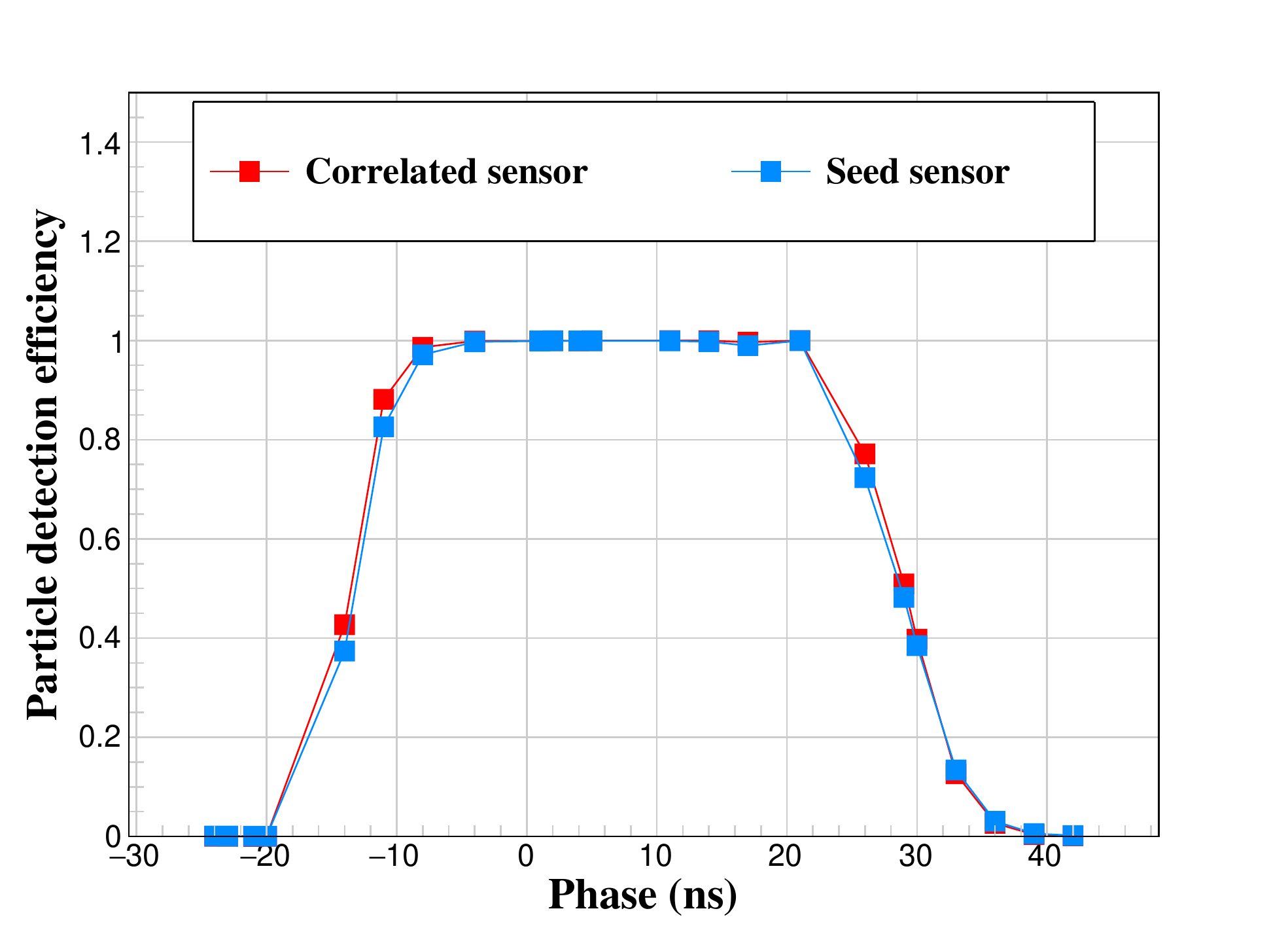}\label{fig:a}}%
	\subfloat[]{\includegraphics[width=0.42\textwidth]{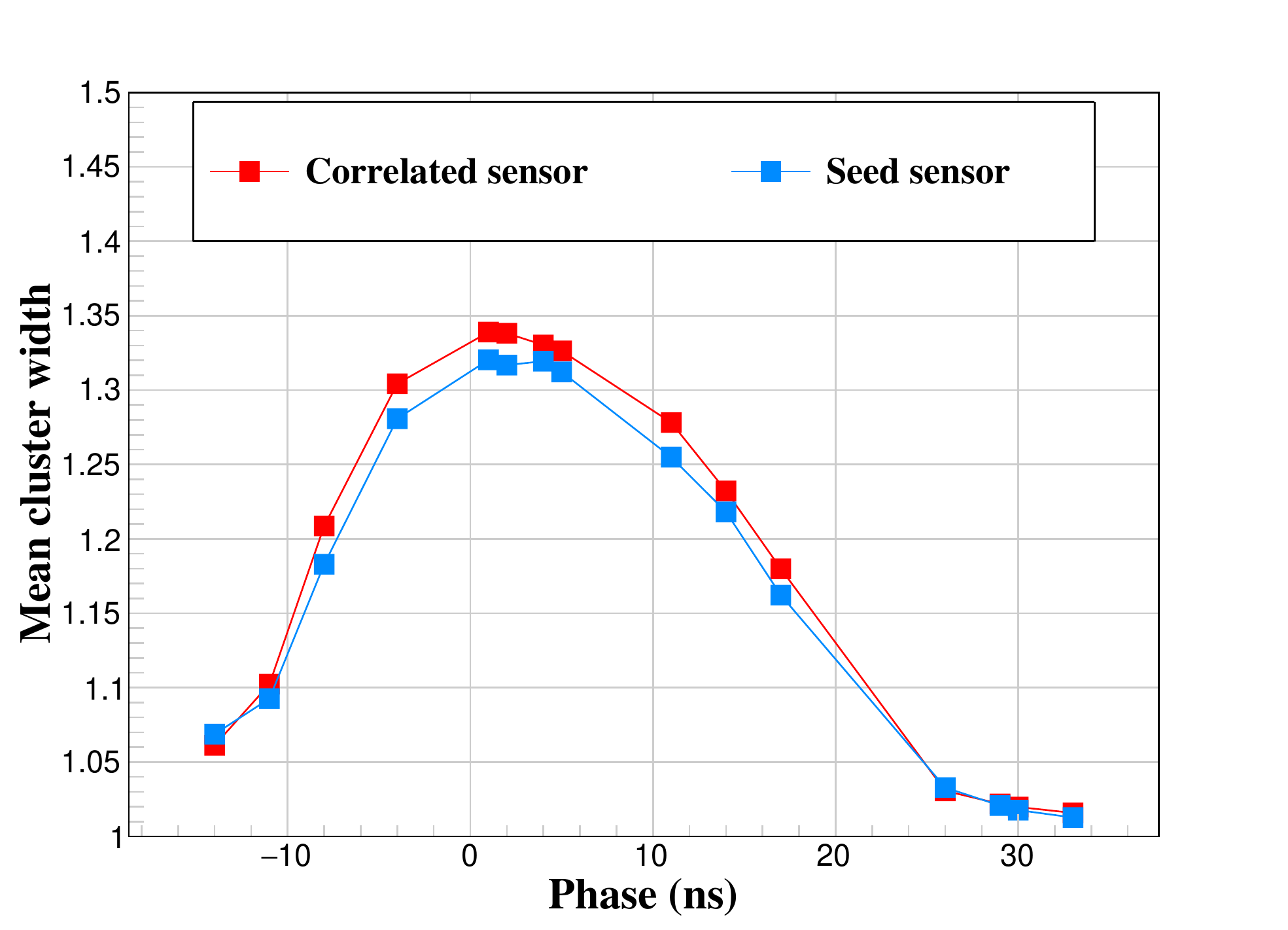}\label{fig:b}}%
	\caption{(a) Particle detection efficiency as a function of the clock phase and 
	(b) mean cluster width as a function of the clock phase, for the fully irradiated mini-module. }
	\label{fig:timeWalk}
\end{figure}

A measurement of the CBC3 amplifier pulse shape and peaking time has been performed 
during the final test beam campaign when the mini-module was fully irradiated to verify that 
the chip continues to meet the requirement of a peaking time around 20\,ns and a return to baseline within 50\,ns.
This measurement was done scanning the DLL settings of the CBC3 which allows to shift the sampling clock by units of 1\,ns.
The register controlling the DLL was varied over 3 latency values because
the amplifier pulse shape takes about 50\,ns to return to the baseline, as shown in Fig.~\ref{fig:CBC3PulseShaping}.
\\By measuring the particle detection efficiency at each setting, the amplifier pulse shape can be reconstructed.
Figure~\ref{fig:timeWalk}~(a) indicates that the amplifier pulse shape is above threshold for less than 50\,ns 
having a full width at half maximum of $\approx42$\,ns as measured from the plot.
The peaking time, which is independent of the amount of charge collected, can be determined by analyzing 
the average cluster width which is maximized when the sampling clock is aligned with the charge peak.
Figure~\ref{fig:timeWalk}~(b) shows that the mean cluster width peaks when the phase is at $+2$\,ns, 
corresponding to a CBC3 chip clock phase shifted by 20\,ns with respect to the \mbox{rising} edge of the signal, 
which, as shown in Figure~\ref{fig:timeWalk}~(a), starts to rise at around $-18$\,ns.
This measurement is in agreement with the CBC3 specifications.


%% file: tex/clusterWidthAnalysis.tex
\subsection{Cluster width}
\begin{figure}[t]
	\centering
	\includegraphics[width=0.45\textwidth]{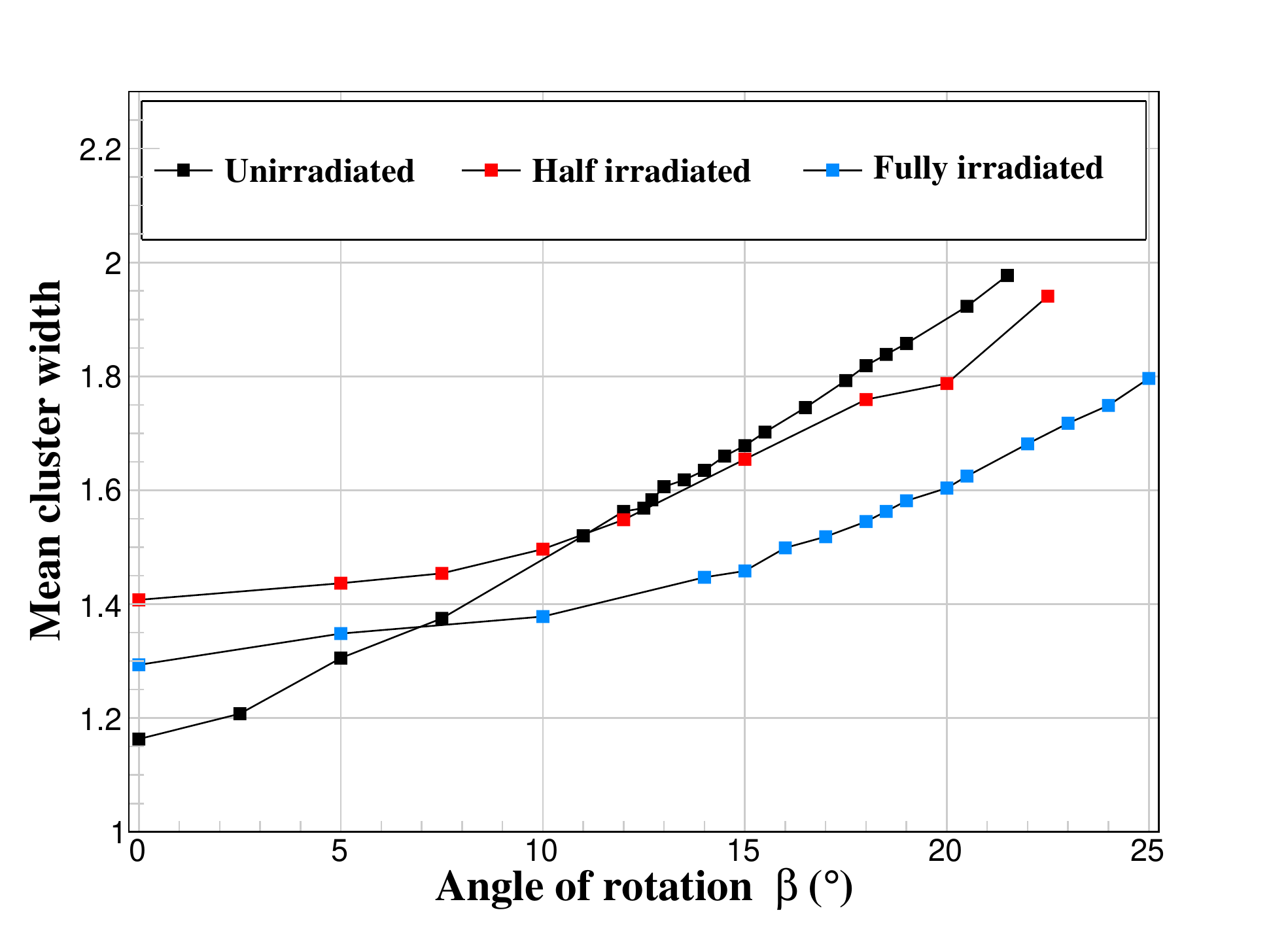}
	\caption{ \label{fig:MeanClusterWidthDifferentAngles}
	Mean cluster width as a function of incident angle.
	Results are shown for the correlated sensor of the unirradiated, 
	half irradiated and fully irradiated mini-module.}	
\end{figure}

{\label{par:ClusterWidthAnalysis}}
\begin{figure}[b!]
	\centering
	\subfloat[]{\includegraphics[width=0.45\textwidth]{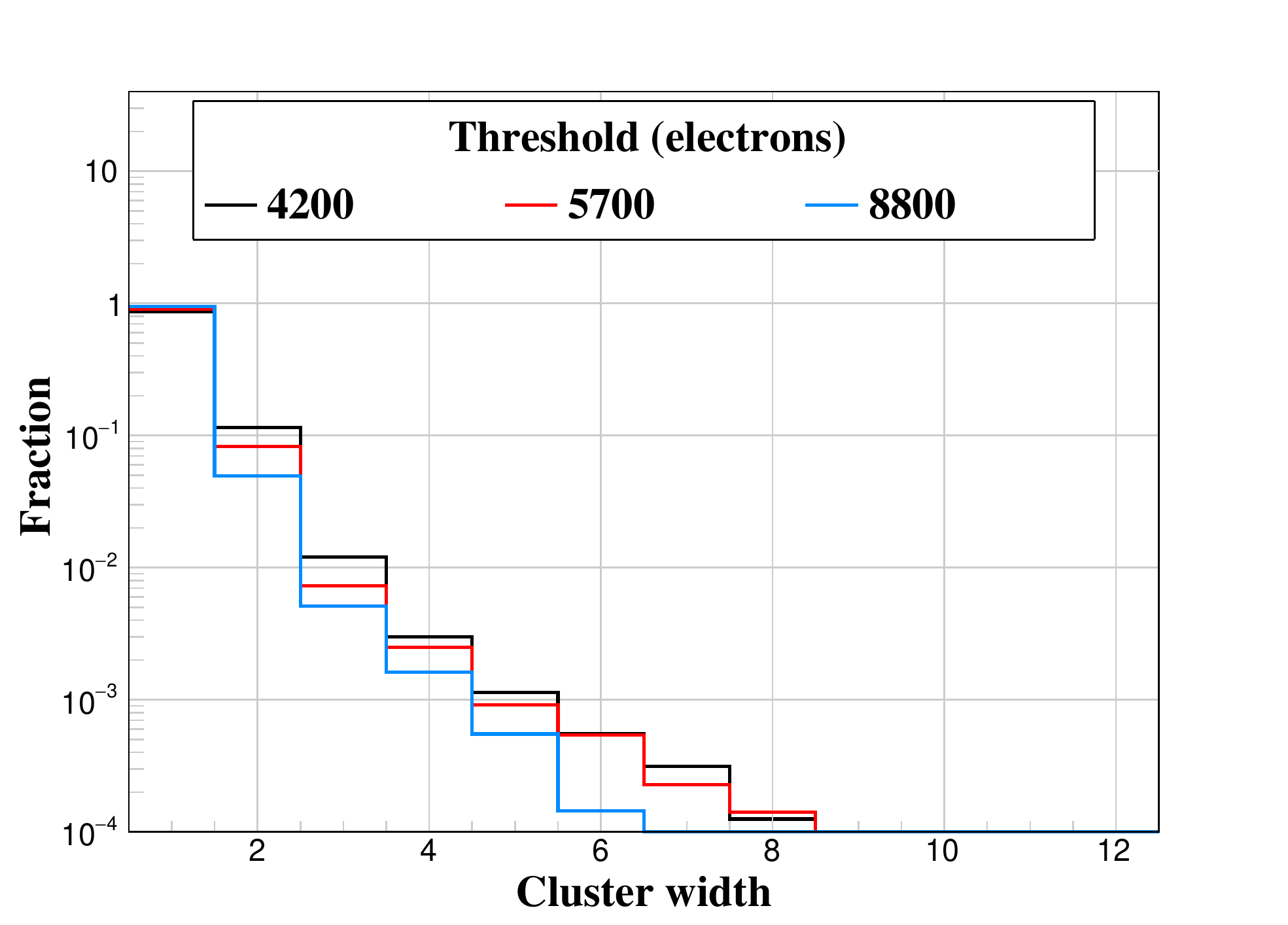}\label{fig:a}}%
	\subfloat[]{\includegraphics[width=0.45\textwidth]{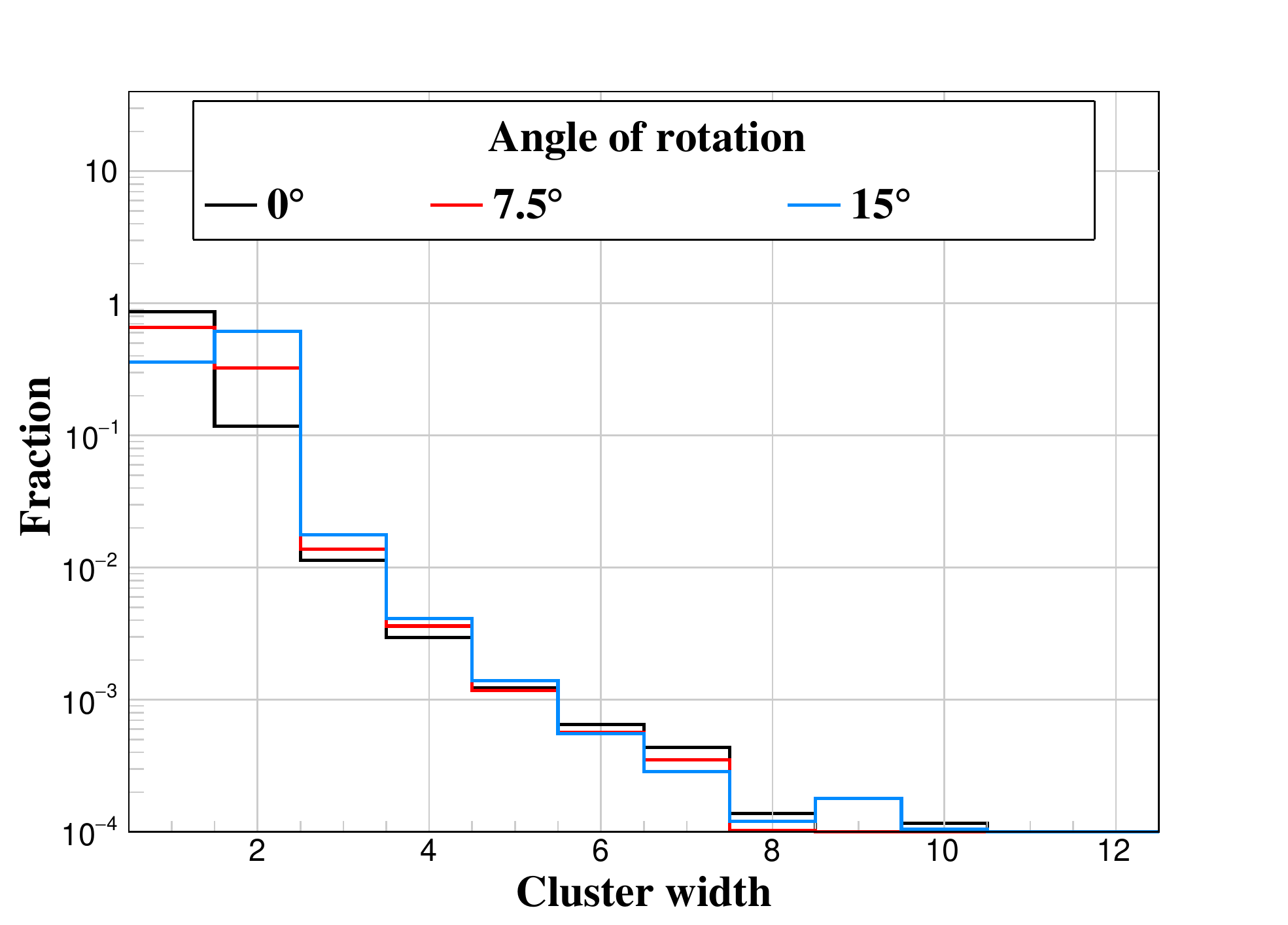}\label{fig:b}}%
	\caption{Cluster width for selected values of \Vcth~(a) and incident angle (b) for the unirradiated mini-module.}
	\label{fig:ClusterWidthDifferentAngles}
\end{figure}
\begin{figure}[b!]
	\centering
	\subfloat[]{\includegraphics[width=0.45\textwidth]{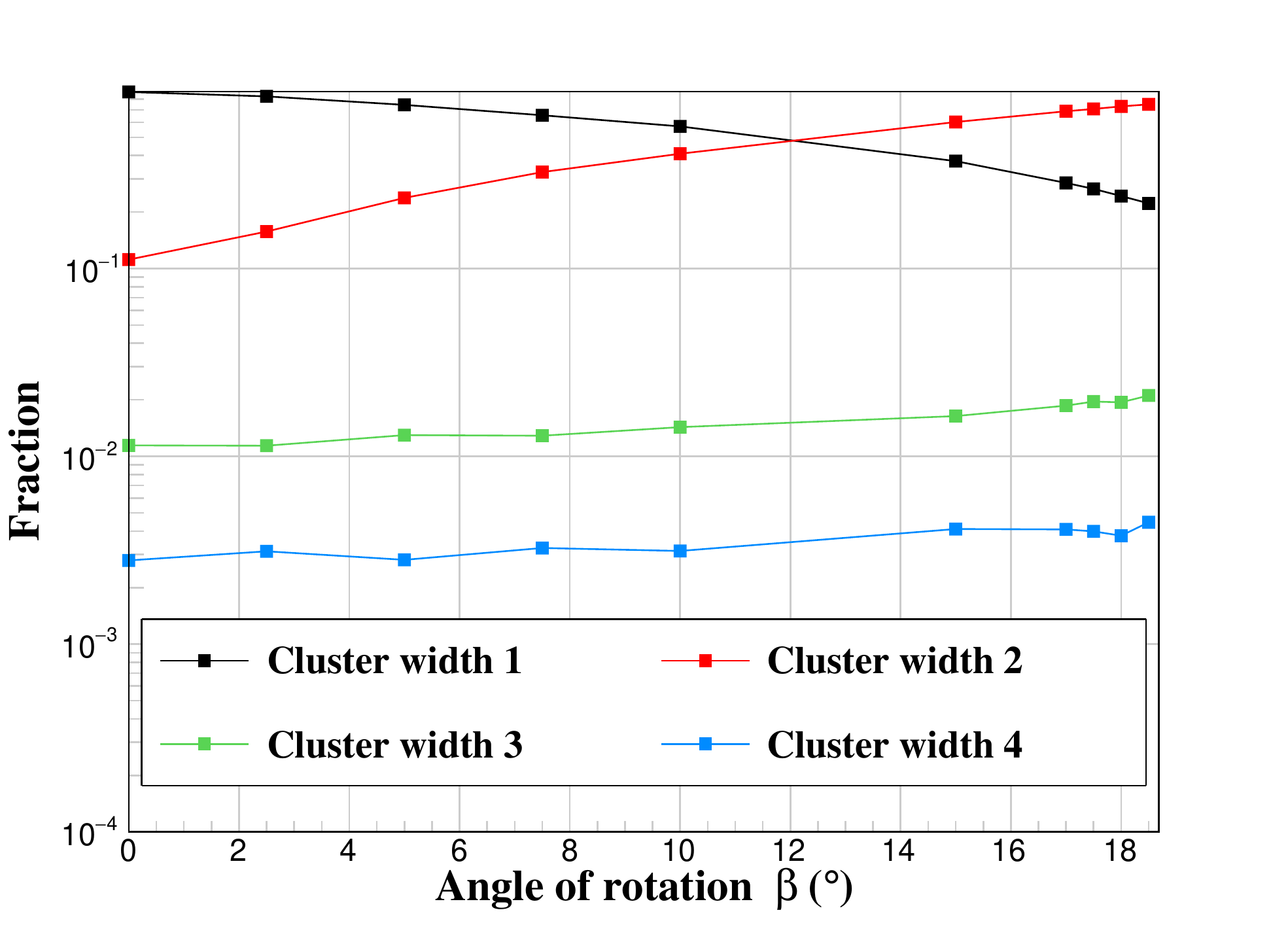}\label{fig:a}}%
	\subfloat[]{\includegraphics[width=0.45\textwidth]{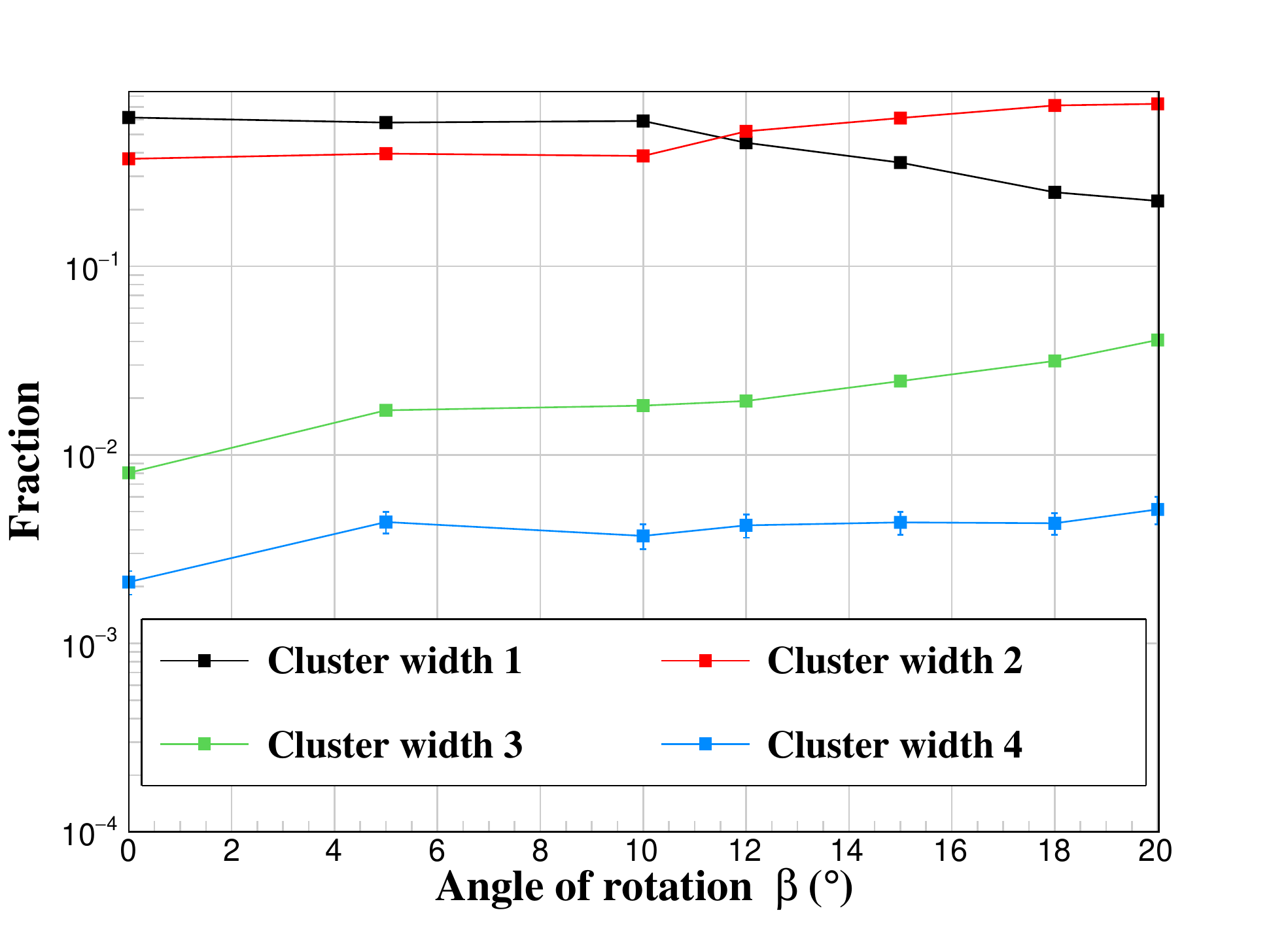}\label{fig:b}}\\
	\subfloat[]{\includegraphics[width=0.45\textwidth]{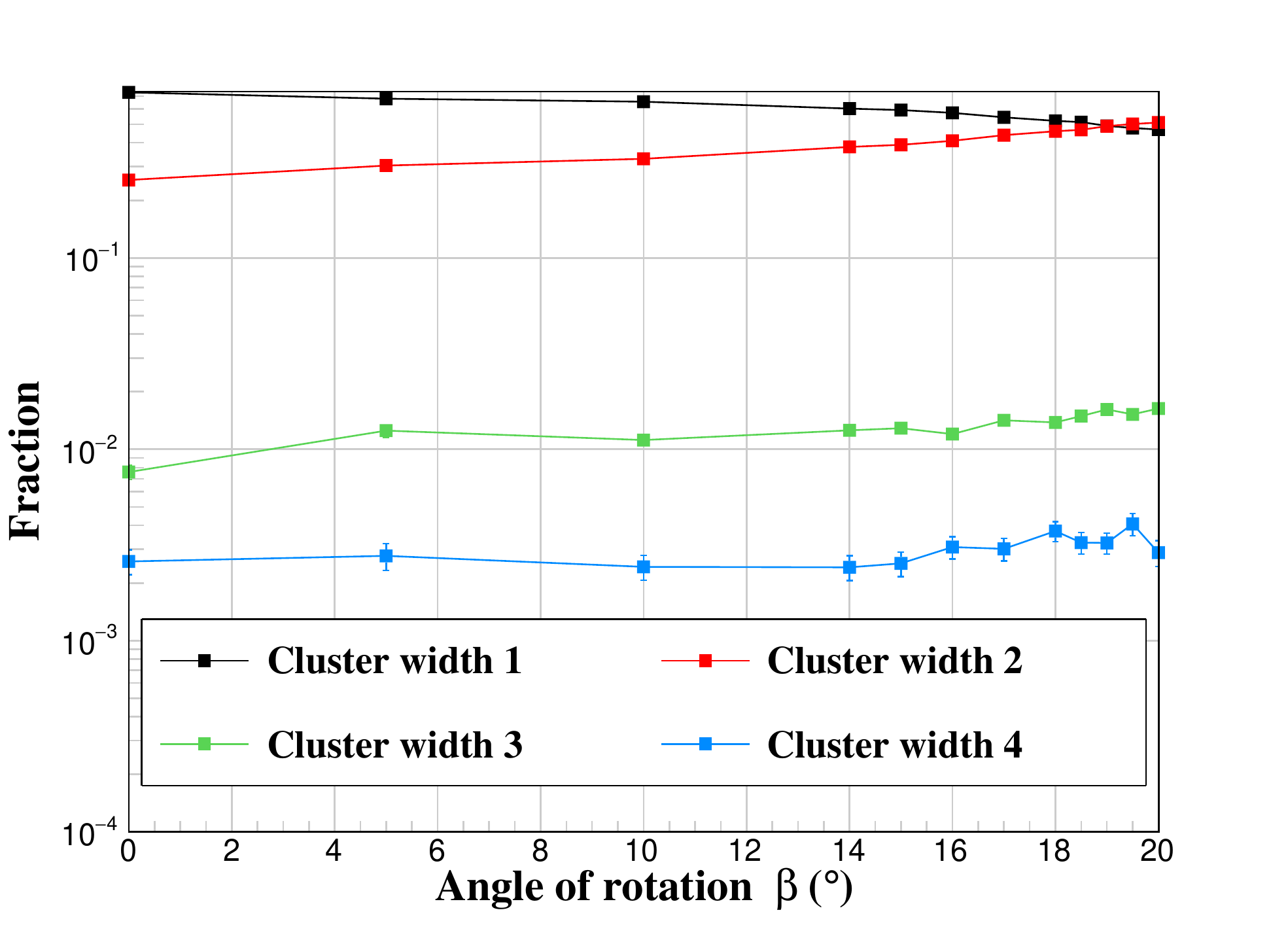}\label{fig:c}}%
	\caption{\label{fig:FractionClusters}
	Fraction of clusters with different widths shown as a function of the incident angle. 
	Results are shown for the correlated sensor of the (a) unirradiated , 
	(b) half irradiated and (c) fully irradiated mini-module.}	
\end{figure}
While the resolution is the key parameter to assess the module performance, the resolution itself depends on the cluster width.
Figure~\ref{fig:ClusterWidthDifferentAngles} shows distributions of cluster width in the seed sensor 
for different thresholds (a) and incident angles (b).
The two plots show the expected broadening of the clusters for lower thresholds and bigger incident angles ($\beta$). 
In both cases, a long tail of large cluster widths, likely due to delta rays, is visible.
\\The fraction of clusters of different widths is shown in Fig.~\ref{fig:FractionClusters} as a function of incident angle. 
For the unirradiated mini-module, at $\beta = 0^{\circ}$, close to $90\%$ of clusters were single-strip, about $10\%$ were two-strip, and
approximately $1\%$ were multi-strip clusters. The fraction of single-strip clusters decreases with increasing incident
angle. At around $\beta = 12^{\circ}$, the fractions of one and two-strip clusters is approximately equal. Two-strip clusters 
dominate at higher angles, reaching about $75\%$ at $18^{\circ}$. 
A similar behavior is also observed for the half irradiated mini-module where the crossing point between single-strip 
and two-strip clusters happens at around $\beta = 12^{\circ}$, while for the fully irradiated mini-module single-strip 
and double-strip clusters are still almost equally split at $\beta = 20^{\circ}$.
These plots also show that, when the mini-module is irradiated, the fraction of two-strip clusters increases significantly at $\beta = 0^{\circ}$ 
with respect to the unirradiated detector.
It should be noted that, in all runs, the fraction of 
clusters with width three and four, which are dominated by delta rays, do not show a strong dependence on the incident angle.
All observations are in agreement with prior measurements in the literature~\cite{CBC2Testbeam}.
Figure~\ref{fig:MeanClusterWidthDifferentAngles} compares the cluster width for the unirradiated, 
half irradiated, and fully irradiated mini-module.

%% file: tex/resolutionAnalysis.tex
\subsection{Resolution}
\begin{figure}[t]
	\centering
	\subfloat[]{\includegraphics[width=0.45\textwidth]{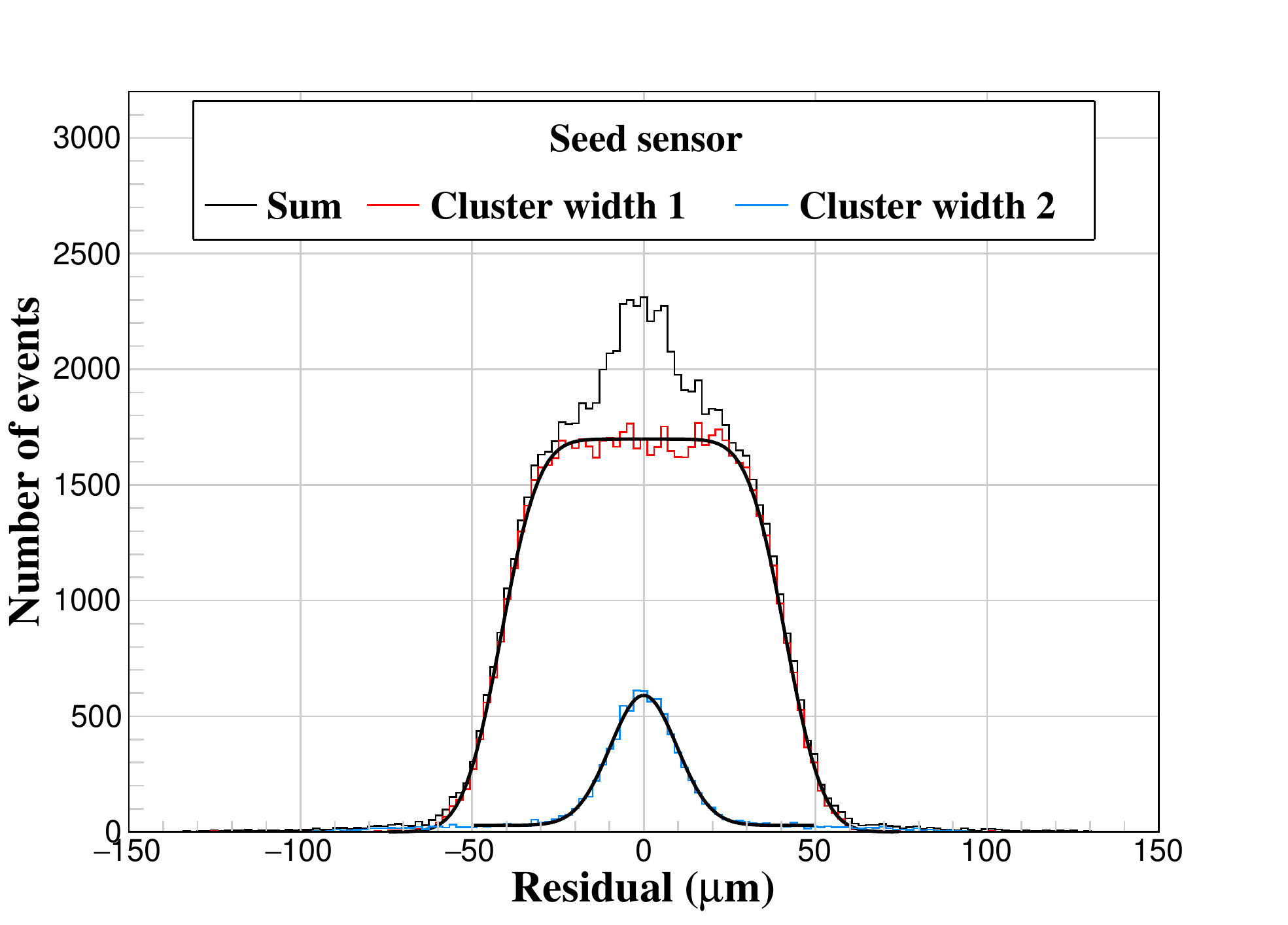}\label{fig:a}}%
	\subfloat[]{\includegraphics[width=0.45\textwidth]{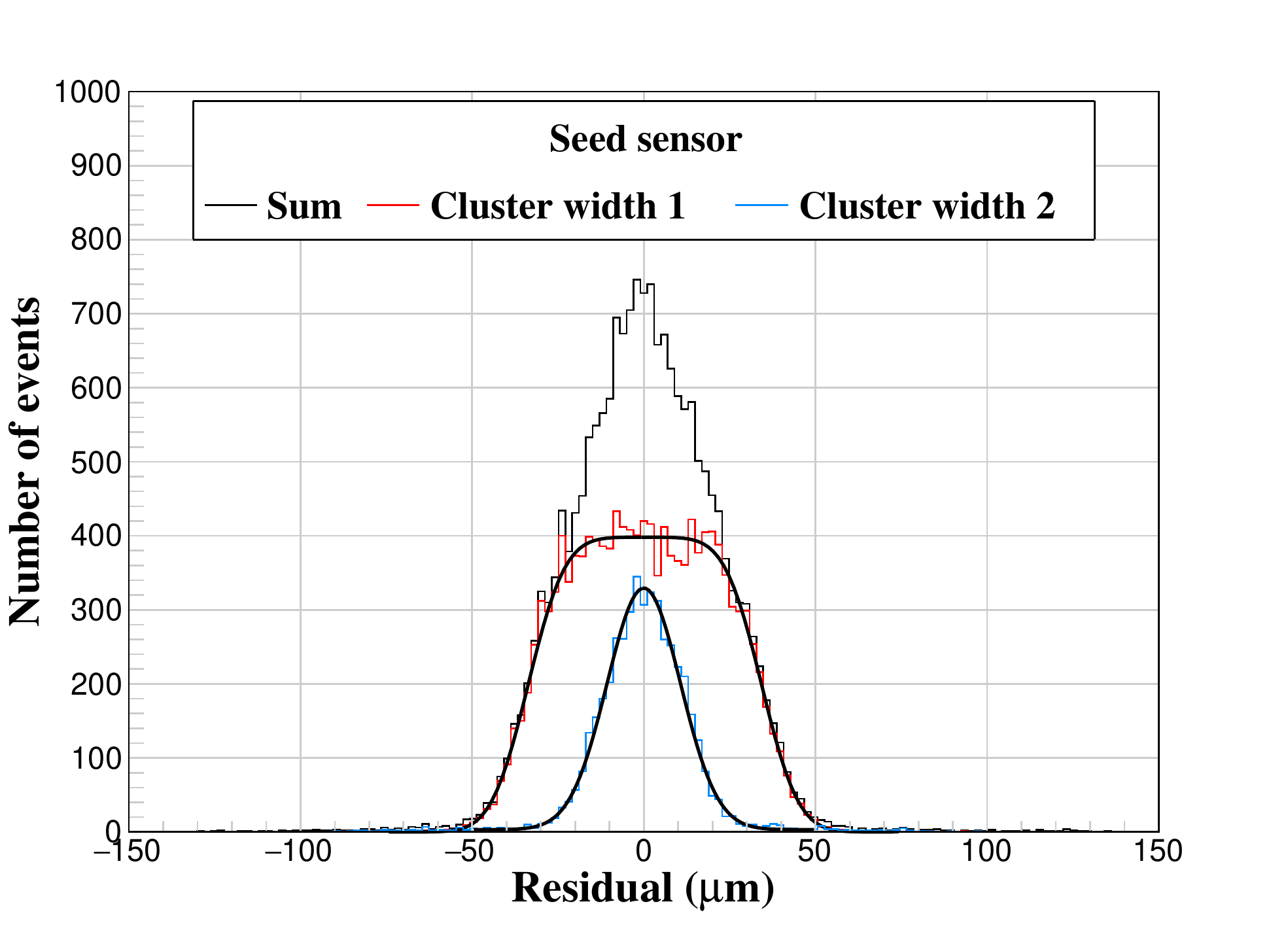}\label{fig:b}}\\
	\subfloat[]{\includegraphics[width=0.45\textwidth]{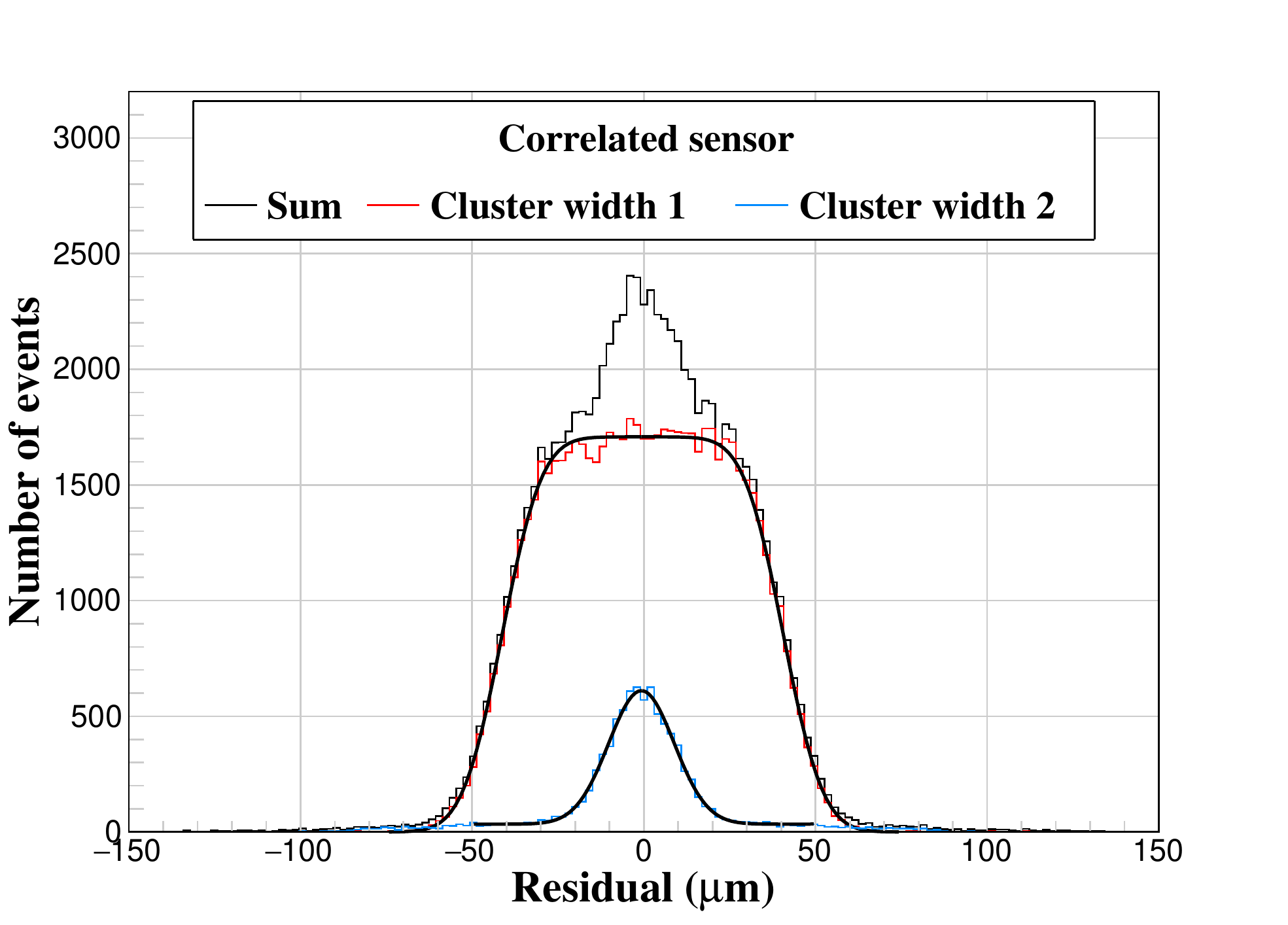}\label{fig:c}}%
	\subfloat[]{\includegraphics[width=0.45\textwidth]{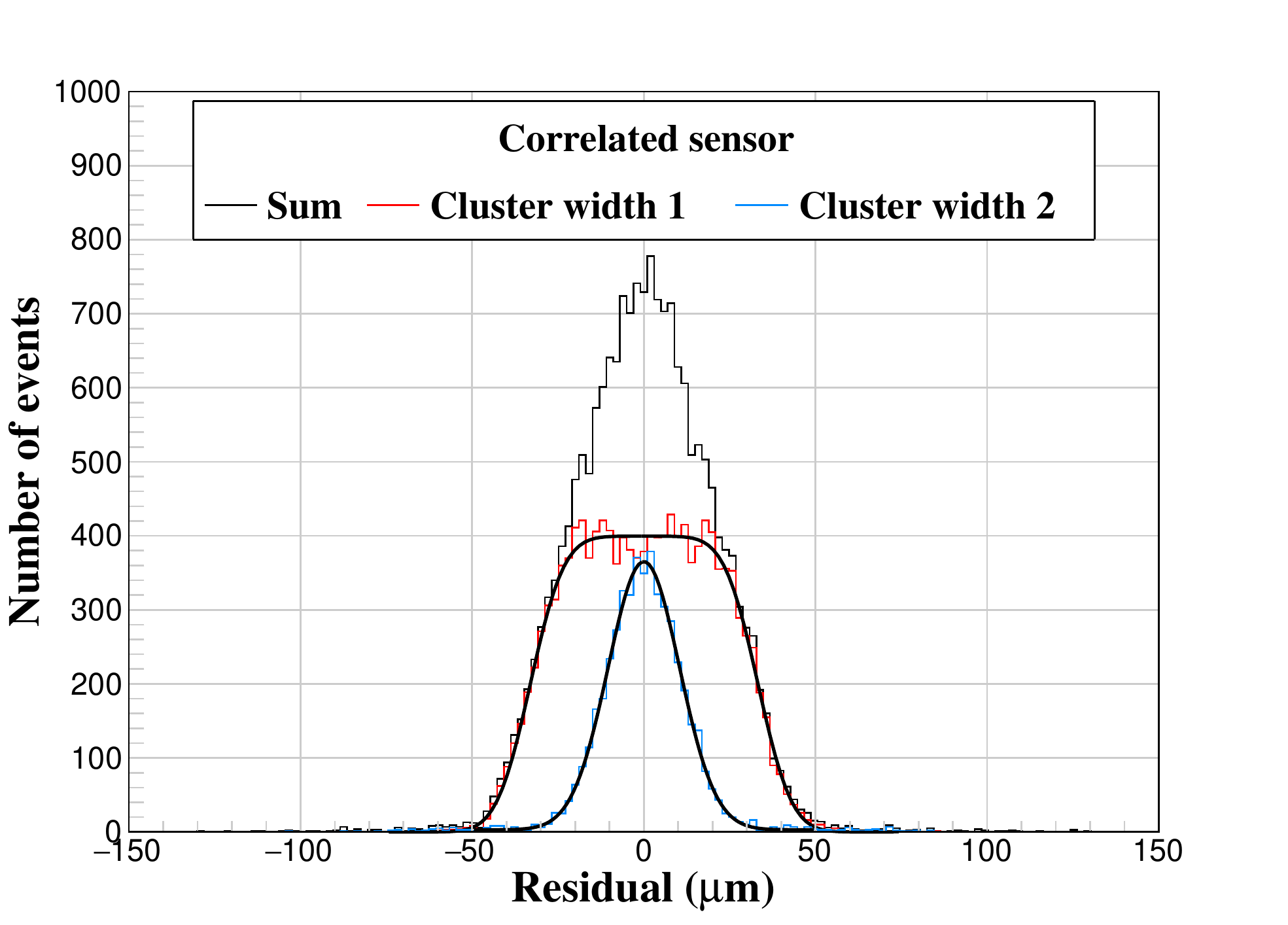}\label{fig:d}}
	\caption{\label{fig:residualsSeedCorrelated}
		Residual distributions for clusters of width one and two for the unirradiated (a), fully irradiated (b) seed
		and unirradiated (c), fully irradiated (d) correlated mini-module sensors. 
		The cluster of width one distribution was fitted using a constant linear function convolved with a Gaussian function. 
		The cluster of width two distribution was fitted using a Gaussian function with an offset.
	}
\end{figure}

\begin{figure}[t]
	\centering
		\subfloat[]{\includegraphics[width=0.45\textwidth]{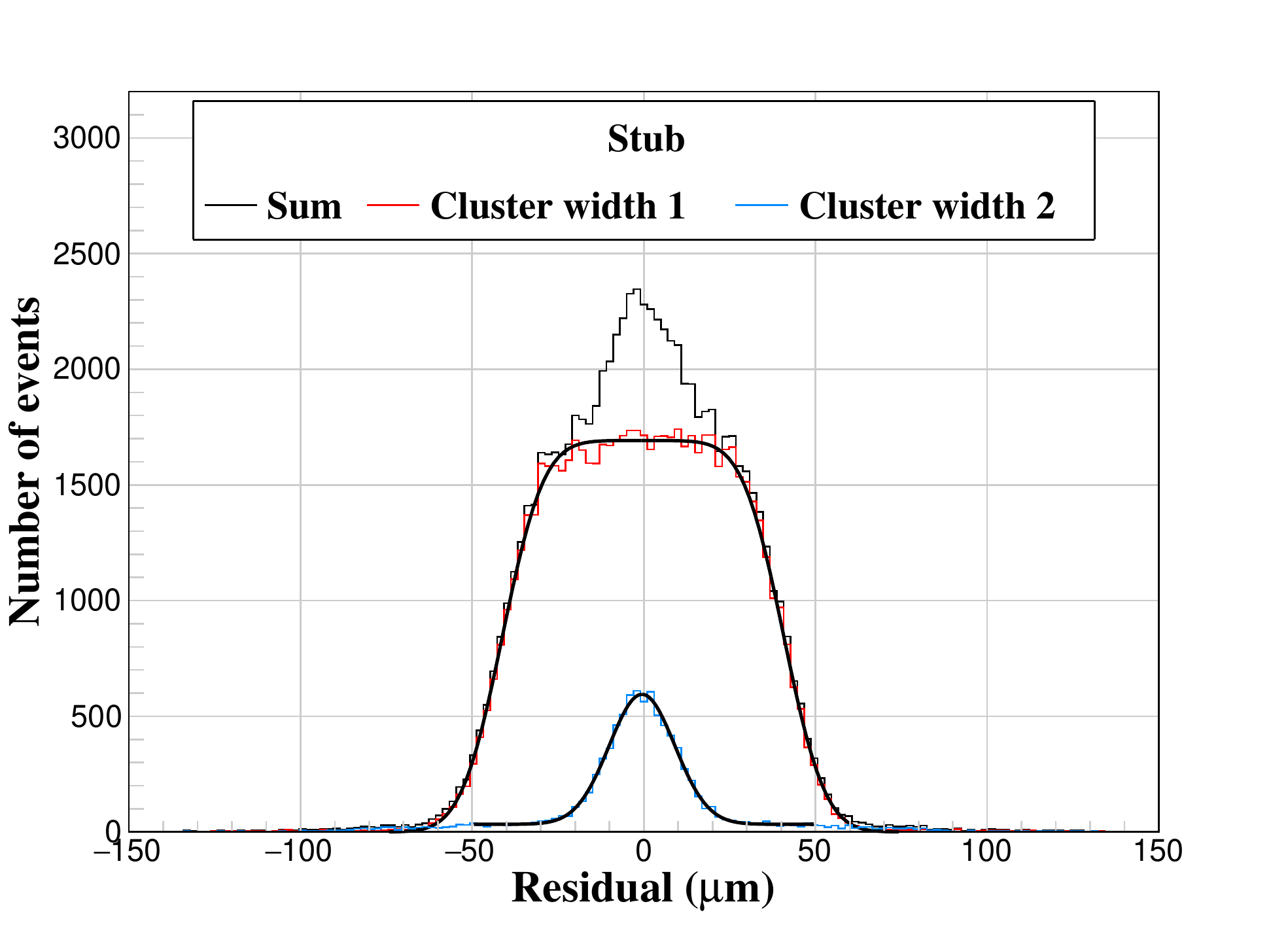}\label{fig:a}}%
		\subfloat[]{\includegraphics[width=0.45\textwidth]{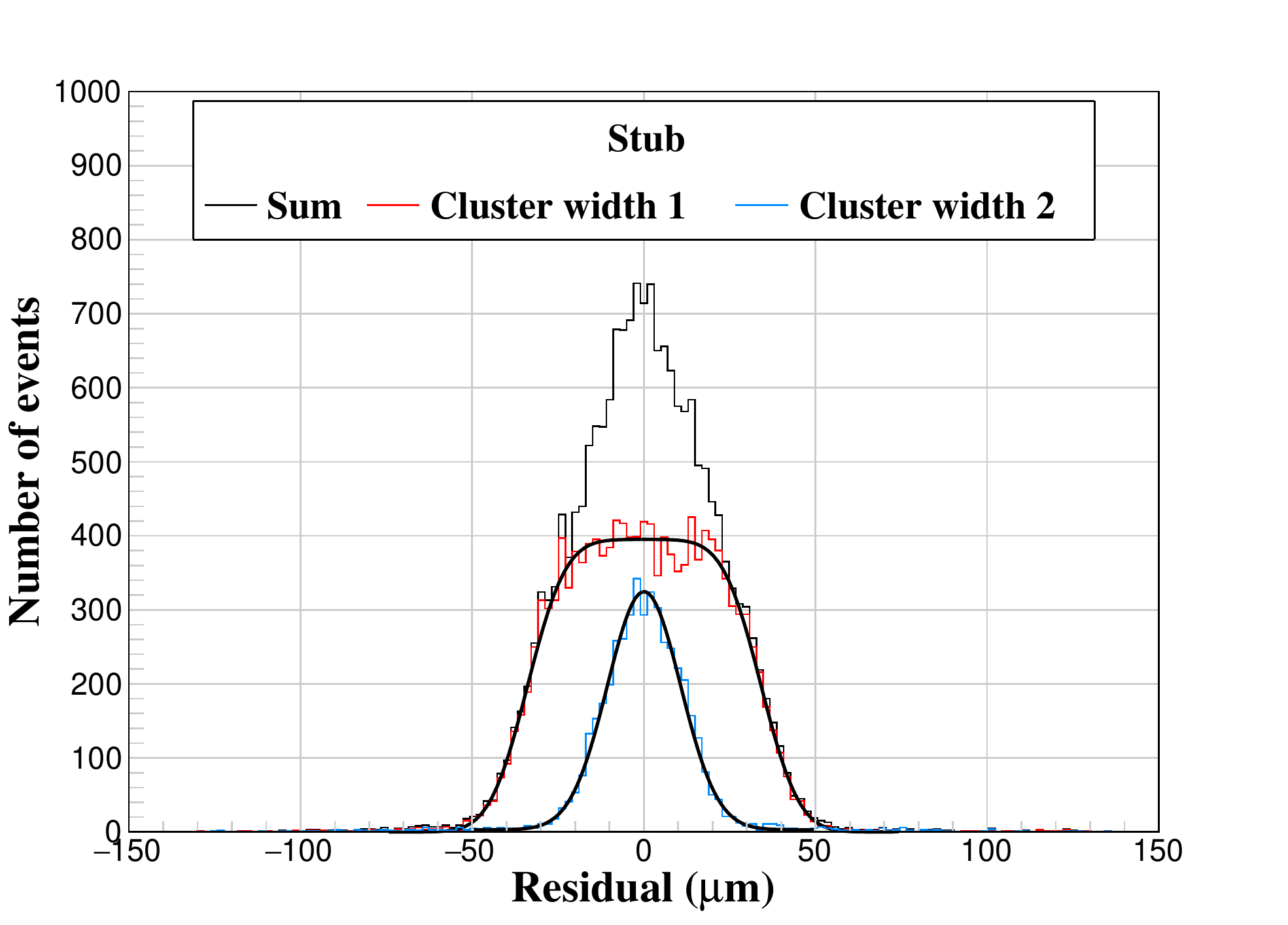}\label{fig:b}}\\
	\caption{ \label{fig:residualsStub}
	Stub residual distributions for the unirradiated (a) 
	and fully irradiated (b) mini-module. The cluster of width one distribution was fitted using a constant linear function convolved with a Gaussian function. 
	The cluster of width two distribution was fitted using a Gaussian function with an offset.
	}	
\end{figure}

\begin{figure}[t]
	\centering
	\subfloat[]{\includegraphics[width=0.45\textwidth]{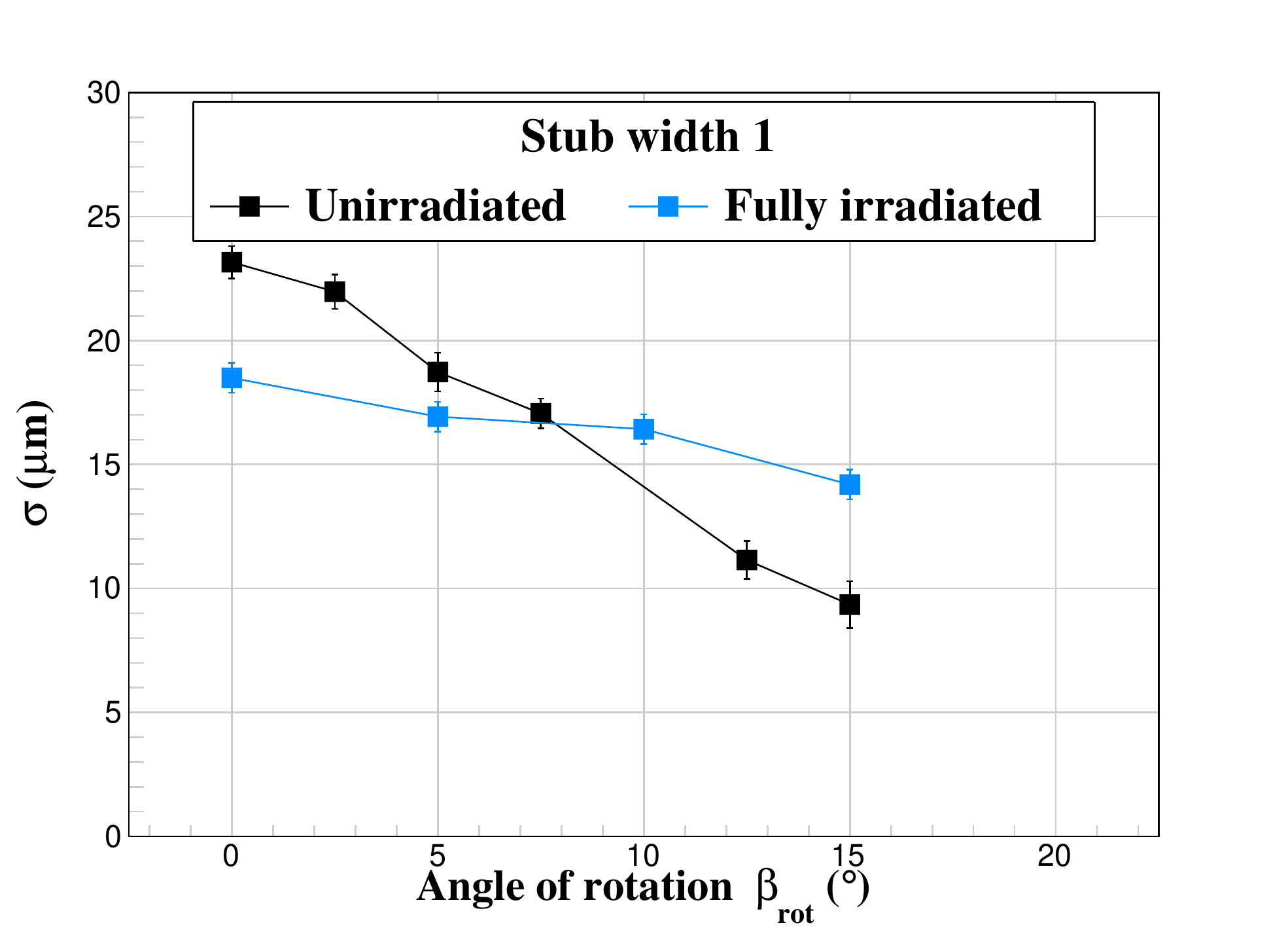}\label{fig:a}}%
	\subfloat[]{\includegraphics[width=0.45\textwidth]{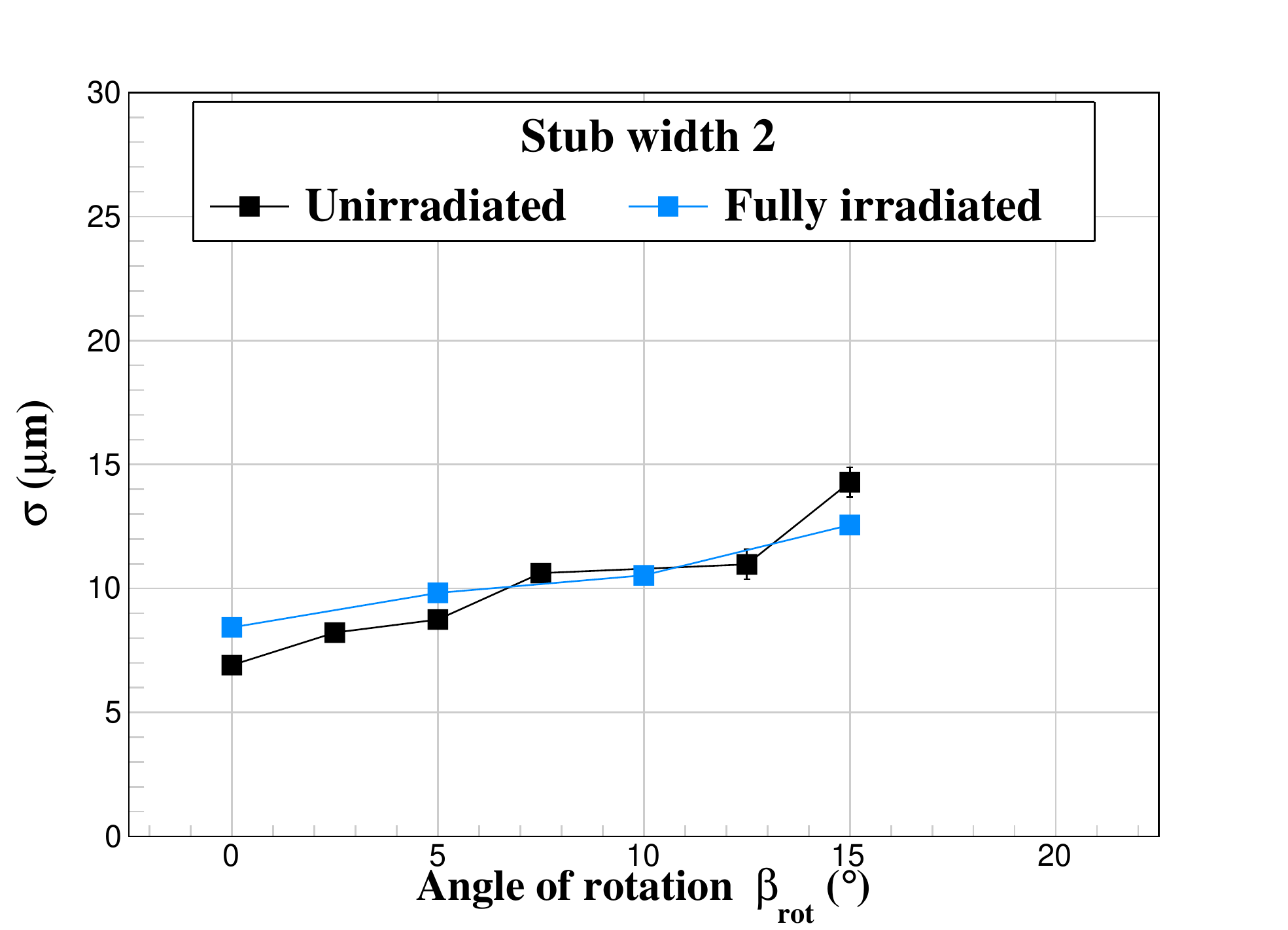}\label{fig:b}}\\
	\caption{ \label{fig:resolutionVsAnglesStub}
	Stub resolution as a function of incident angle for clusters of width one (a) and width two (b). 
	Results are shown for the stubs of the unirradiated and fully irradiated mini-module.
	}	
\end{figure}

In order to assess the expected reconstruction performance throughout the detector lifetime, the detector 
resolution has been studied for different irradiation fluences.
The resolution for clusters of width one is extracted from the fit to their residual distribution. 
The function used for the fit is a constant linear function convolved with a Gaussian function and 
the resolution is the sigma of the constant linear function. 
The function used to fit the residual distribution for clusters of width two is a Gaussian function with an offset.
The resolution for clusters of width two is the sigma of the Gaussian component of the fit.
The residuals are defined, both for clusters and stubs, as the difference between the projected track 
and the reconstructed position on the sensor.
All measurements presented in this paragraph have been done using the standard threshold of 4200 electrons.
Figure~\ref{fig:residualsSeedCorrelated} shows the seed and correlated sensor residual distributions 
for the unirradiated (a), (c) and the fully irradiated (b), (d) mini-module. 
The distributions of clusters of width one and width two are shown separately, with their sum overlaid.
Wider clusters, which are likely due to delta rays, are not included in the resolution study.\\
For the unirradiated sensors, Fig.~\ref{fig:residualsSeedCorrelated}~(a) and (c), 
the residual distributions for clusters of width one show the box like shape that is expected for a $90\,\mathrm{\mu m}$ pitch strip sensor.
Smearing at the edges is caused by the telescope resolution and due to tracks near the edges creating two-hit clusters. 
For the irradiated case the clusters of width one are concentrated in a narrower band around the strip center, 
which indicates that the charge sharing increases when the sensor is irradiated.
As a consequence, only about $10\%$ of the clusters have width two in the unirradiated sensor, 
as shown previously in Fig.~\ref{fig:FractionClusters}, while the percentage increases when the detector is fully irradiated, 
as can be also seen in Figure~\ref{fig:residualsSeedCorrelated}~(b) and Fig.~\ref{fig:residualsSeedCorrelated}~(d).
\\The resolution of the stubs has been measured for both the unirradidated and the fully irradiated mini-module, with results   
presented in Fig.~\ref{fig:residualsStub}, and is compared to the values for the clusters in Tab.~\ref{tab:ResolutionTable}.
All values are extracted from the fits and calculated subtracting in quadrature the estimated telescope resolution of $7\,\mathrm{\mu m}$.
The stub residual distributions show that the CBC3 reconstructs clusters of width one and two very efficiently 
giving a stub distribution which is very similar to the individual sensor distributions. 
\begin{table}[b!]
    \begin{center}
      \caption{Resolution summary for normal incident tracks. Width 1 and 2 refer to the cluster width.}
      \label{tab:ResolutionTable}
	  \resizebox{\textwidth}{!}{\begin{tabular}{|l|c|c|c|c|c|c|c|}
		\hline
		\multirow{2}{*}{Data set} & \multicolumn{2}{c|}{Seed} & \multicolumn{2}{c|}{Correlated} & \multicolumn{2}{c|}{Stub} \\ \cline{2-7} 
	 	& Width 1 ($\mu m$) & Width 2 ($\mu m$) & Width 1 ($\mu m$) & Width 2 ($\mu m$) & Width 1 ($\mu m$) & Width 2 ($\mu m$) \\ \cline{1-7} 
		 Unirradiated & 22.7 & 7.0 & 23.0 & 7.2 & 22.7 & 7.0\\ \hline
		 Fully irradiated & 18.0 & 8.1 & 18.5 & 8.4 & 18.5 & 8.4\\ \hline
		  \end{tabular}}
	\end{center}
\end{table}
\begin{figure}[b!]
	\centering
	\subfloat[]{\includegraphics[width=0.45\textwidth]{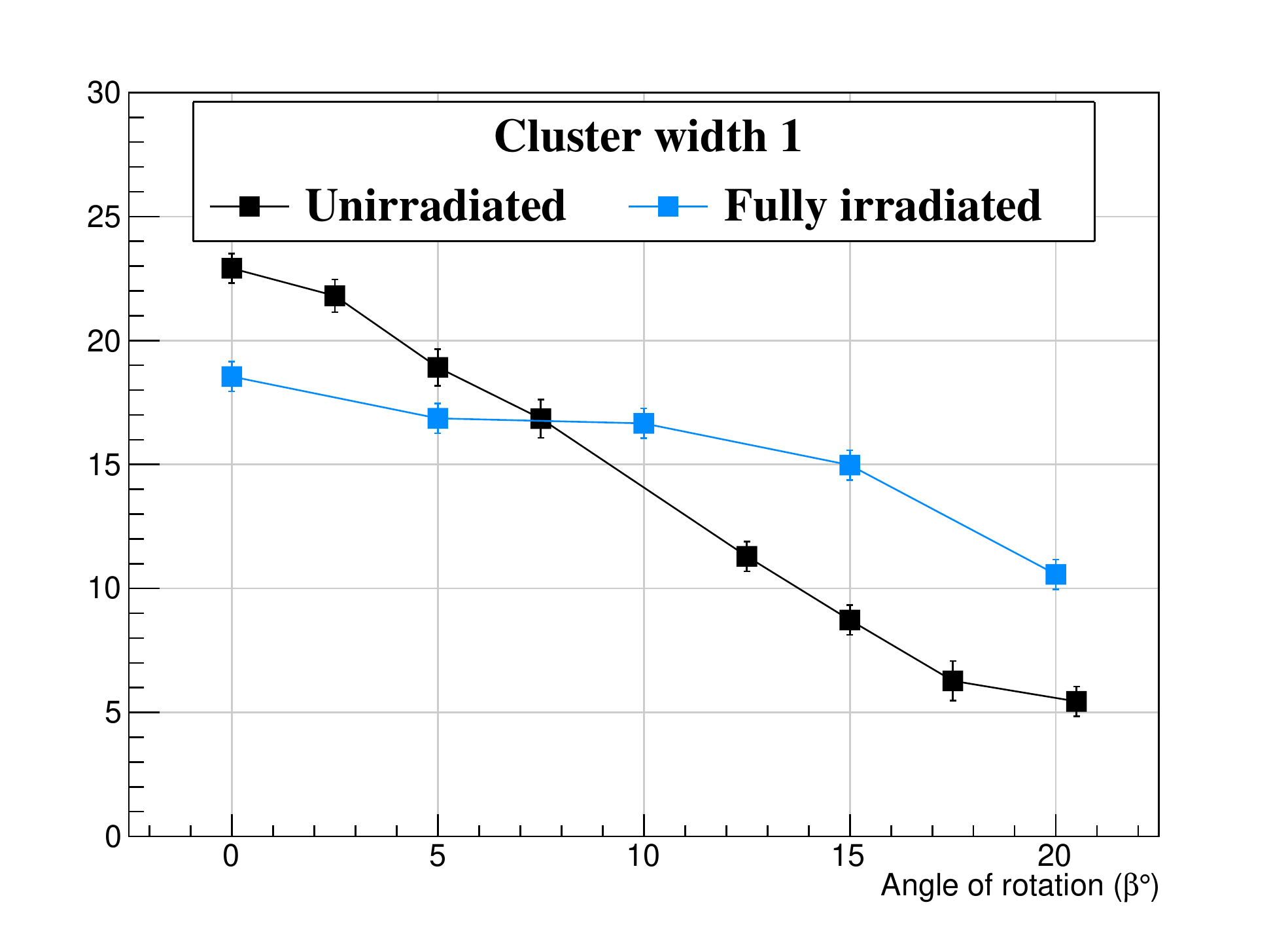}\label{fig:a}}%
	\subfloat[]{\includegraphics[width=0.45\textwidth]{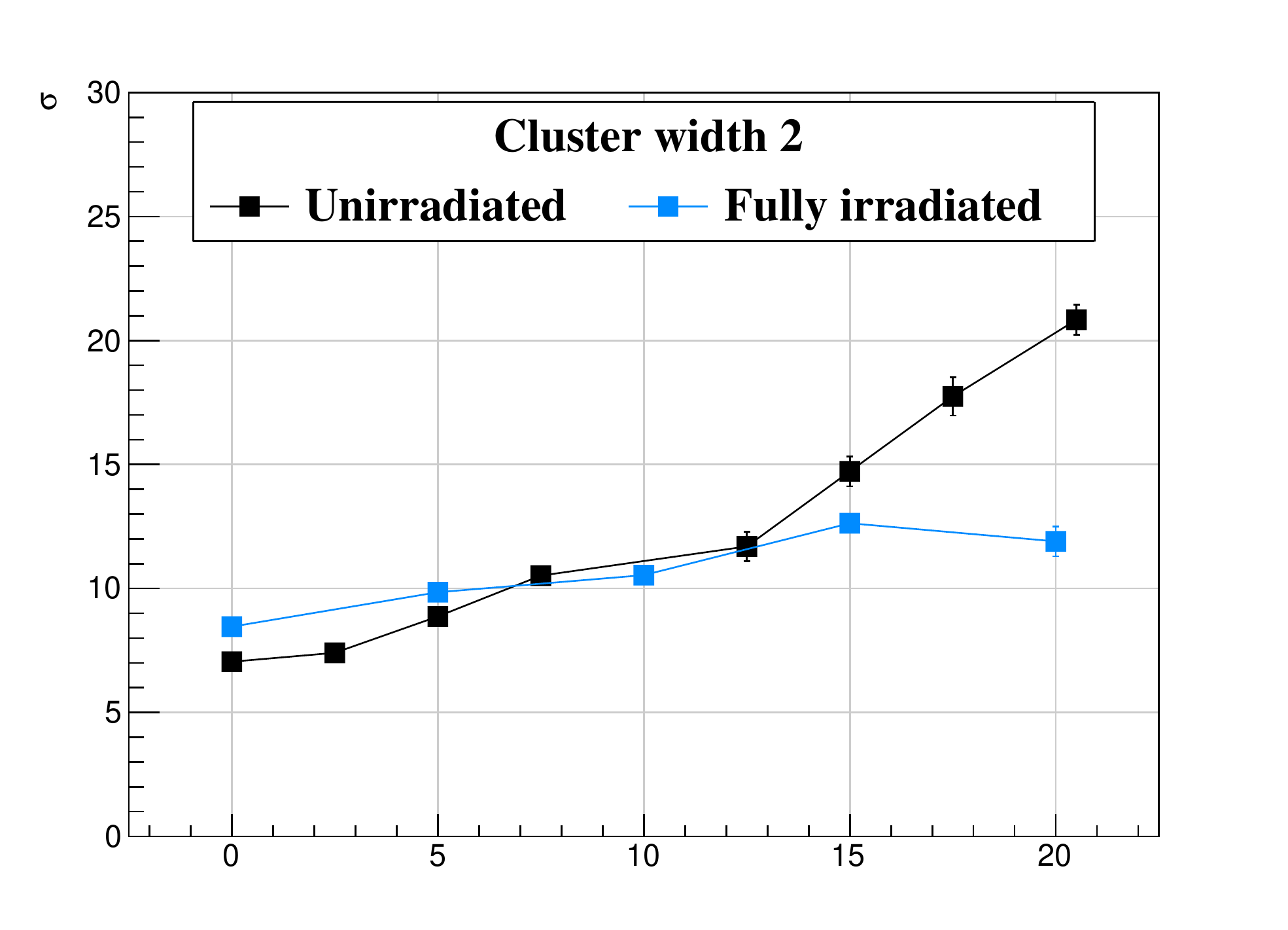}\label{fig:b}}\\
	\caption{ \label{fig:resolutionVsAngles}
	Resolution as a function of incident angle for clusters of width one (a) and width two (b). 
	Results are shown for the correlated sensor of the unirradiated and fully irradiated mini-module.
	}	
\end{figure}	
The detector and stub resolution for clusters of width one and two 
have also been studied at different incident angles. This is relevant for different parts of the detector along the beamline.
Figure~\ref{fig:resolutionVsAngles} and Fig.~\ref{fig:resolutionVsAnglesStub} show how the detector and stub resolution, respectively, 
varies as a function of incident beam angle for clusters of width one and two.
All values are extracted from the fits to the residual distributions and calculated subtracting in quadrature the estimated telescope 
resolution of $7\,\mathrm{\mu m}$.
As expected, with increasing incident beam angle, the resolution for clusters of width one improves while for clusters of width two it deteriorates. 
At any angle the resolution is always between $7\,\mathrm{\mu m}$ and $23\,\mathrm{\mu m}$ and it does not change significantly between the unirradiated and the 
irradiated detector. 
These results confirm that the clusters and stubs can be reconstructed with the expected resolution to allow precise tracking 
also for trigger purposes even at the end of the detector's lifetime.

%% file: tex/chargeAnalysis.tex
\subsection{Charge collection}
\label{par:ChargeCollectionMeasurement}
\begin{figure}[b]
	\centering
	\subfloat[]{\includegraphics[width=0.45\textwidth]{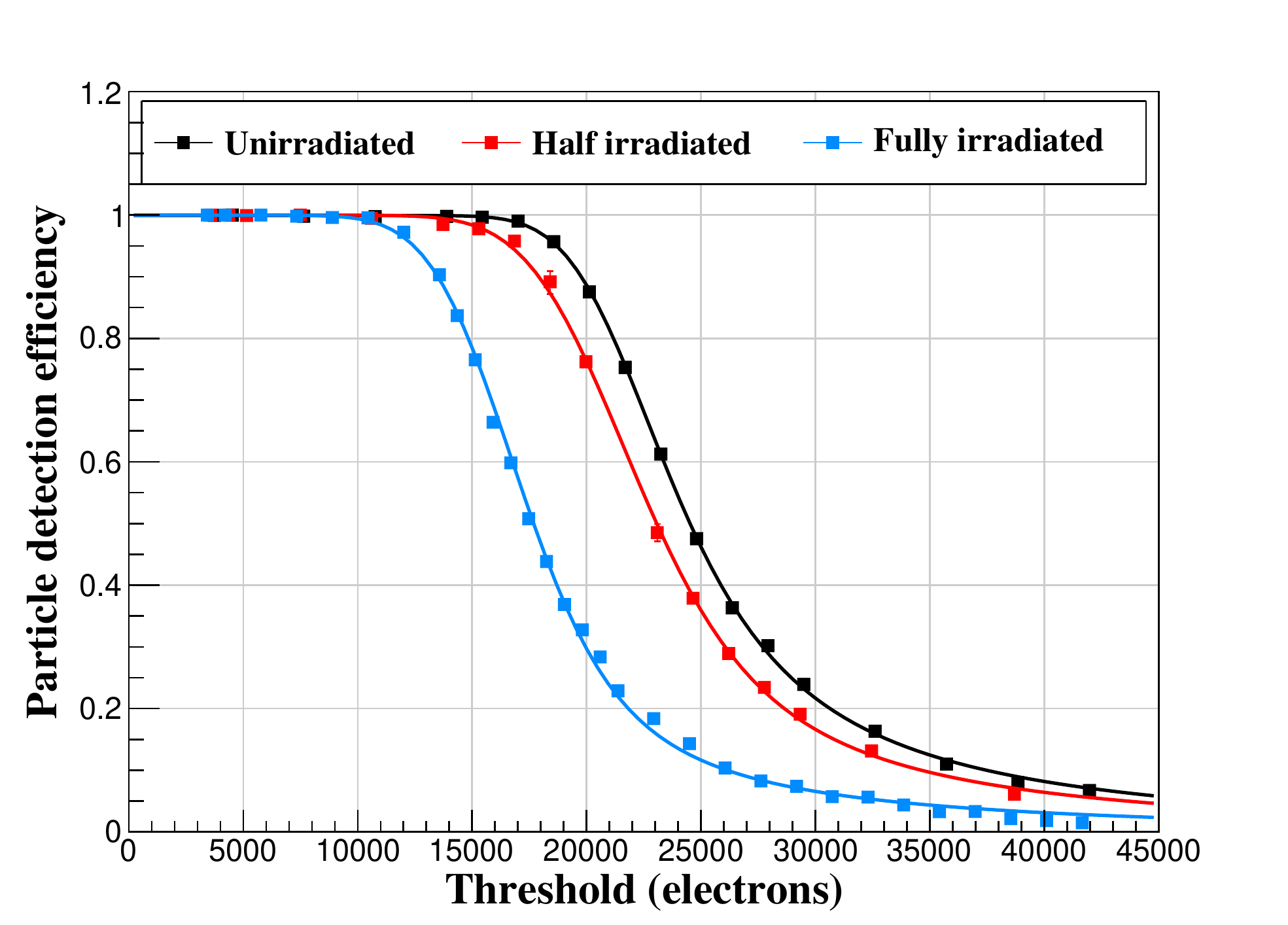}\label{fig:a}}%
	\subfloat[]{\includegraphics[width=0.45\textwidth]{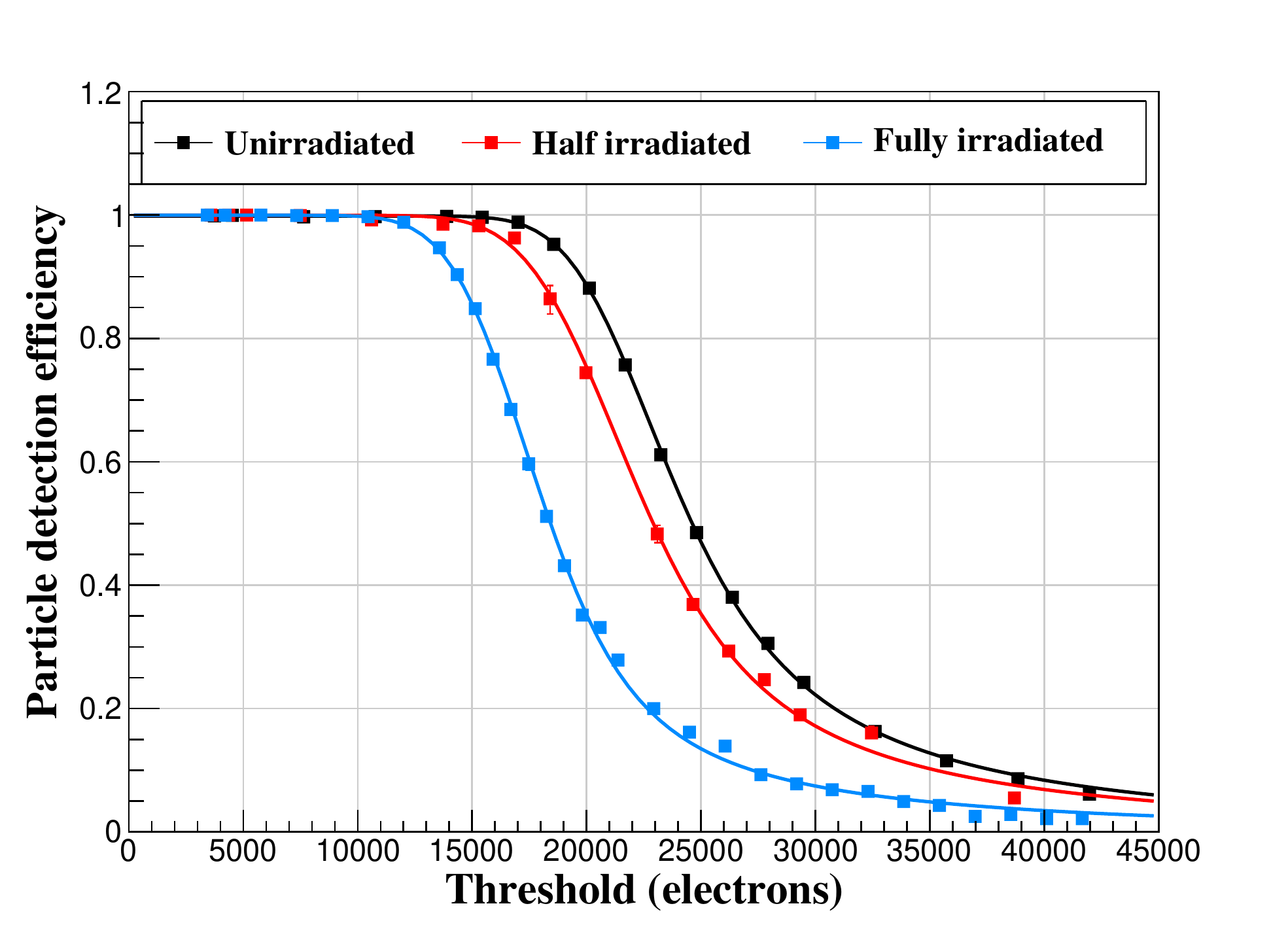}\label{fig:b}}%
	\caption{\label{fig:ThresholdSummaryCenter}Particle detection efficiency as a function of the threshold, using selected tracks 
	pointing within $\pm5\,\mu$m around the strip center, for the (a) seed sensor and (b) correlated sensor. 
	Superimposed on each scan is the best fit performed with the integral of a Landau function convolved with a Gaussian function.}
\end{figure}
\begin{figure}[t]
	\centering
	\subfloat[]{\includegraphics[width=0.45\textwidth]{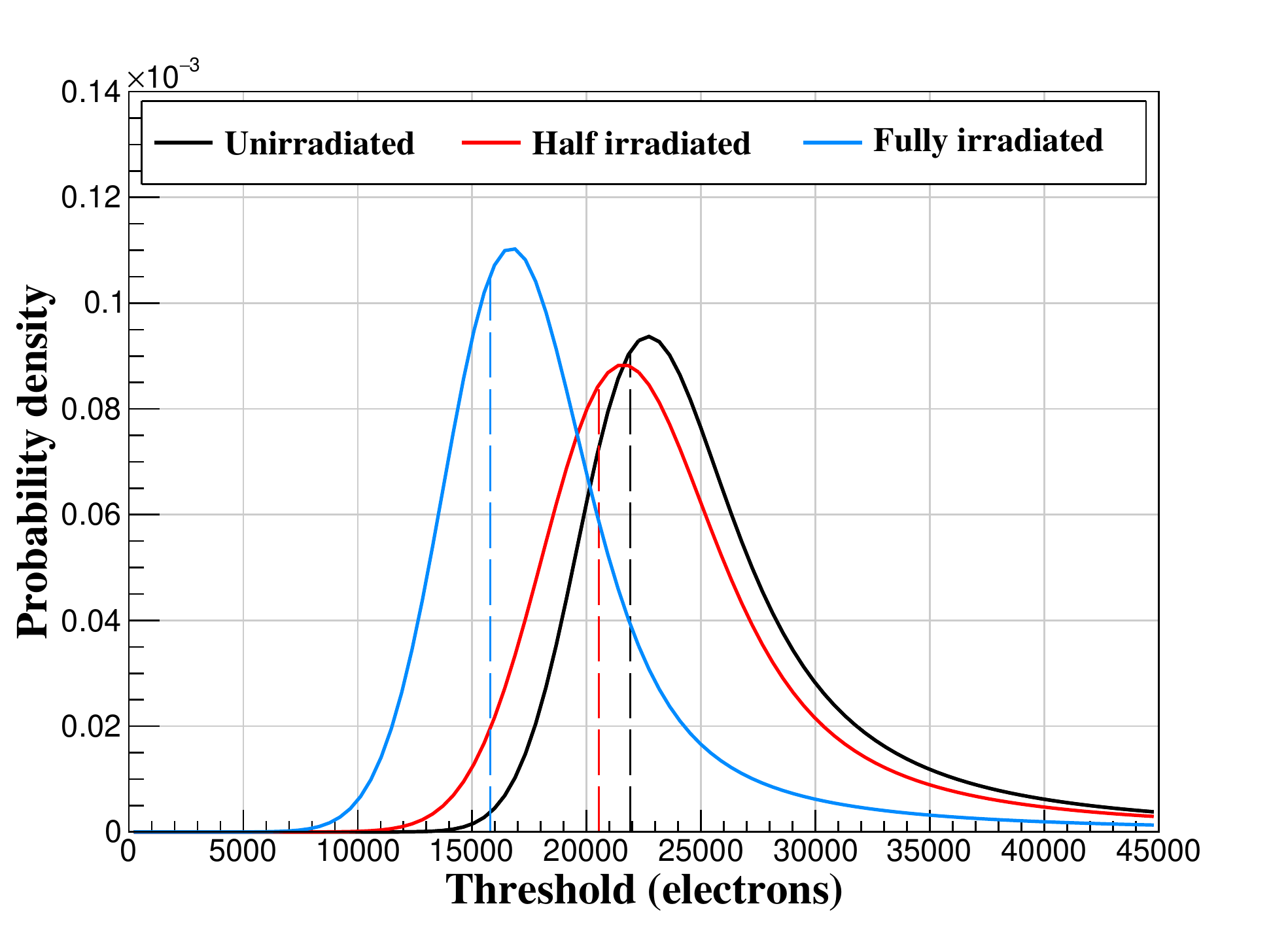}\label{fig:a}}%
	\subfloat[]{\includegraphics[width=0.45\textwidth]{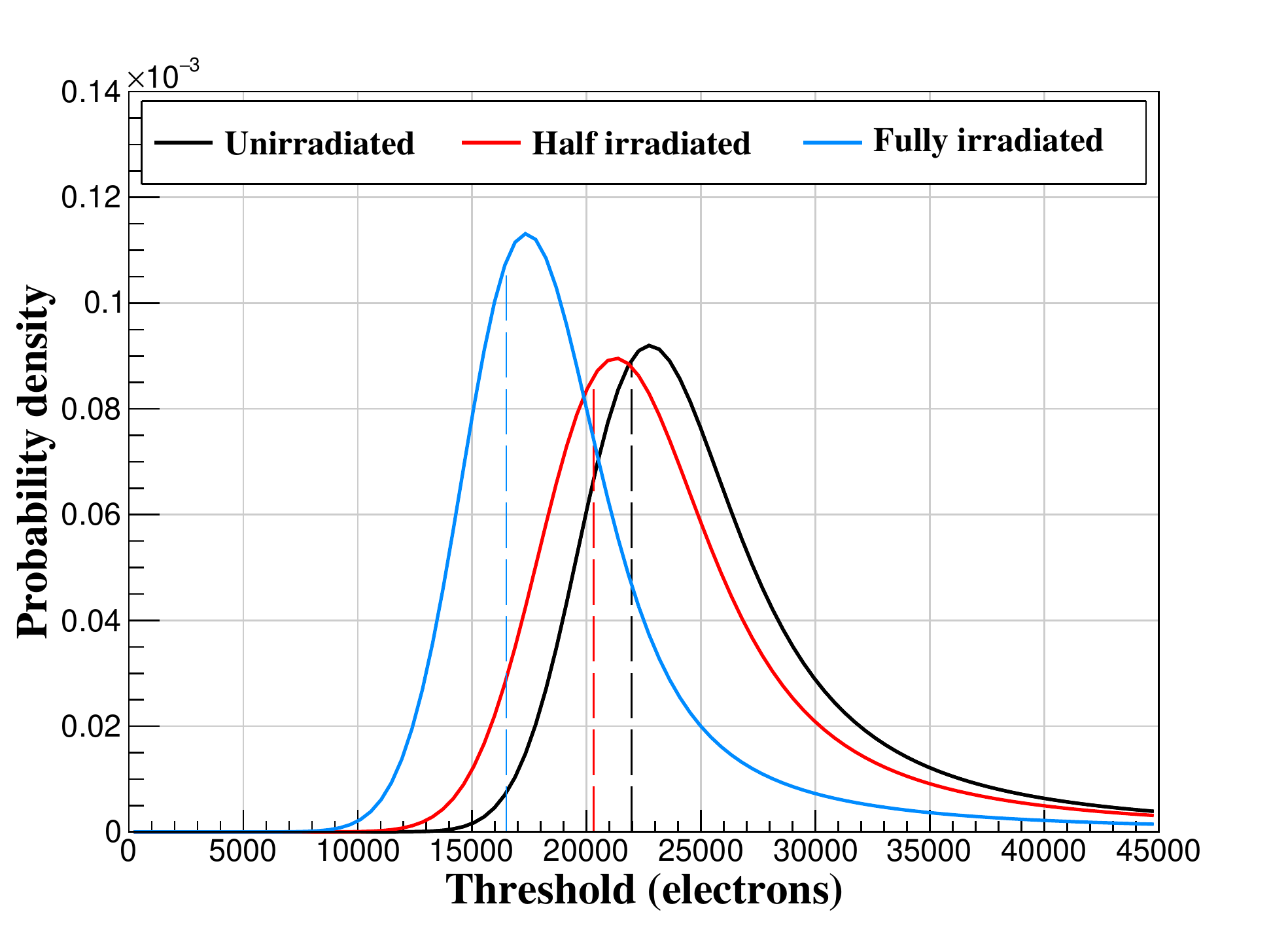}\label{fig:b}}%
	\caption{\label{fig:LangaussSummary}Distributions of the convolution of a Landau function with a Gaussian function, 
	obtained by differentiating the fitting functions to the threshold scans, for the
	(a) seed sensor and (b) correlated sensor. The dashed lines indicate the Landau functions' most probable value.}
\end{figure}
\begin{figure}[t]
	\centering
	\includegraphics[width=0.45\textwidth]{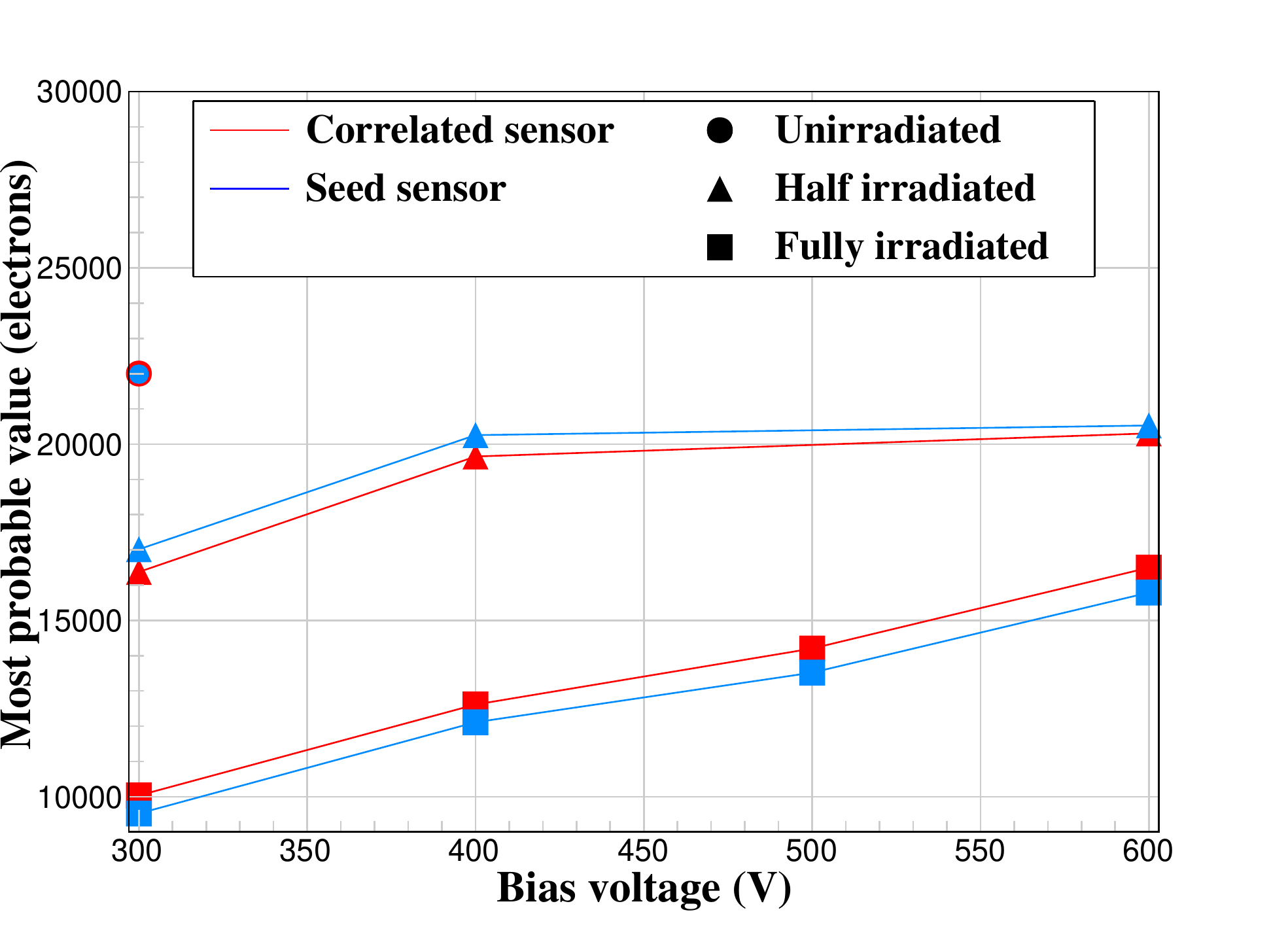}
	\caption{\label{fig:ThresholdScanSummary} Most probable value of the Landau distribution as a function of the bias voltage for different irradiation fluences.}
\end{figure}
When the detector is placed orthogonally to the beam, it is possible to measure the charge collected by the sensor by performing a threshold scan and 
measuring the efficiency using only tracks that point close to the center of the strip, where it is safe to assume 
that all charge released is collected by that strip. 
A cut of $\pm5\,\mu$m around the strip center is used for this analysis. 
Figure~\ref{fig:ThresholdSummaryCenter} shows the results of these scans from each test beam campaign, 
with the integral of a Landau function convolved with a Gaussian function superimposed on each scan.
Figure~\ref{fig:LangaussSummary} shows the Landau function convolved with a Gaussian function obtained by differentiating 
the fitting functions to the threshold scans. 
The dashed lines in the figure indicate the returned most probable value (MPV) of the Landau functions.
\\Figure~\ref{fig:ThresholdScanSummary}
shows the MPV of each Landau function, as a function of the bias voltage, for the unirradiated, 
half irradiated and fully irradiated mini-module (no bias scan was performed for the unirradiated mini-module).\\
The fits to the distributions return  22\,003 and 21\,985 electrons as MPVs for the 
correlated and seed sensor, respectively, in case of the unirradiated mini-module, 
values that are comparable with the expected value of 21\,600 electrons for a $300\mathrm{\mu m}$ thick silicon sensor. 
The values become 16\,496 and 15\,789 after the mini-module is fully irradiated. 
\\It is also notable that the two sensors behaved differently from one another after irradiation,
exhibiting a slightly different hit efficiency, as can be seen in Fig.~\ref{fig:ThresholdScanSummary}. 
This effect is visible after the detector was irradiated at half fluence and at full fluence.
The results, nevertheless, show a qualitative agreement with previous charge collection measurements~\cite{Sensor}.

%% file: tex/stubAnalysis.tex
\subsection{Stub reconstruction efficiency}
\begin{figure}[t]
	\centering
	\subfloat[]{\includegraphics[width=0.45\textwidth]{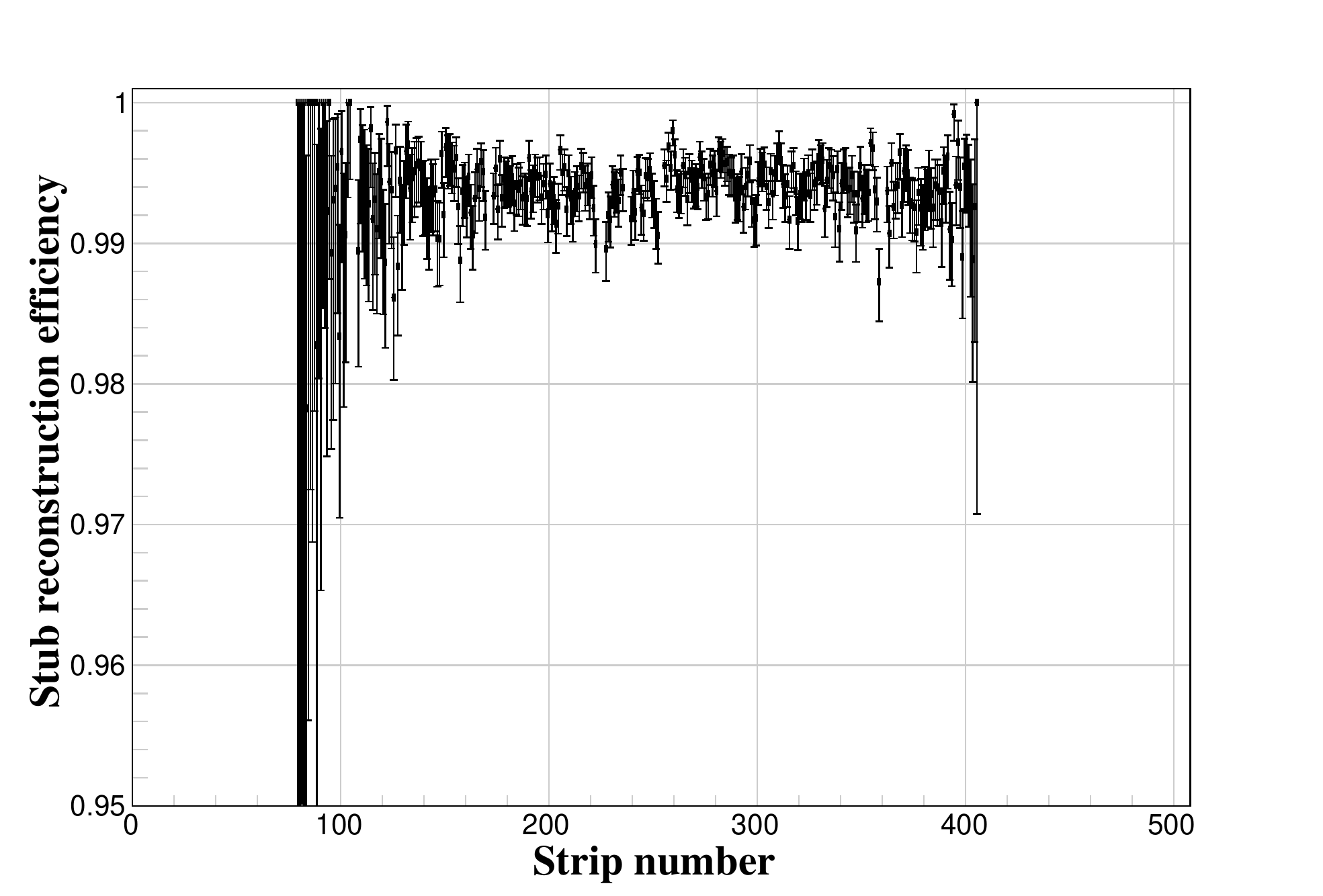}\label{fig:a}}%
	\subfloat[]{\includegraphics[width=0.45\textwidth]{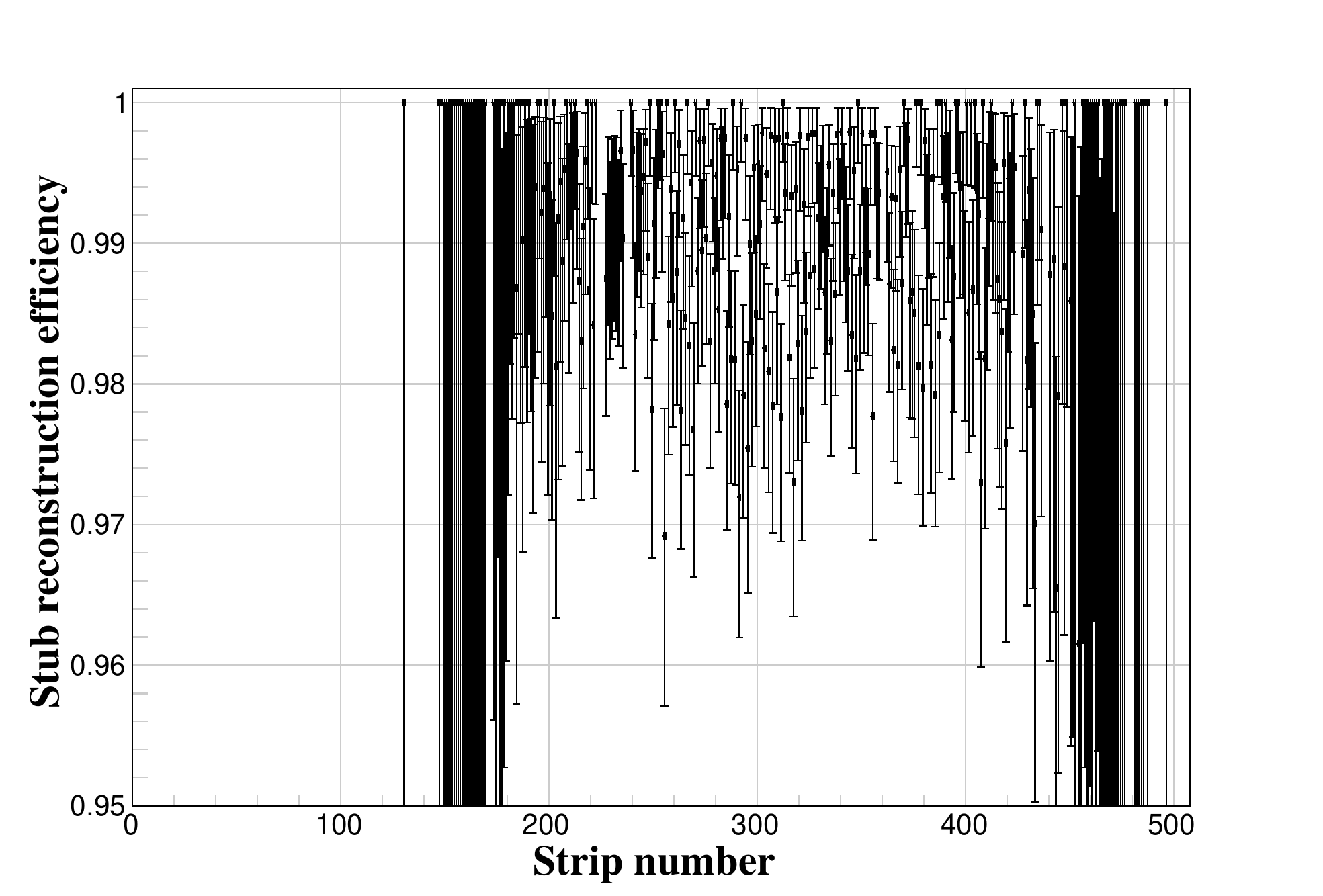}\label{fig:b}}%
	\caption{Stub reconstruction efficiency across the sensor for the unirradiated (a) and the fully irradiated (b) mini-module for normal incidence tracks.}
	\label{fig:stubRecoEff}
\end{figure}

\begin{figure}[t]
	\centering
	\subfloat[]{\includegraphics[width=0.45\textwidth]{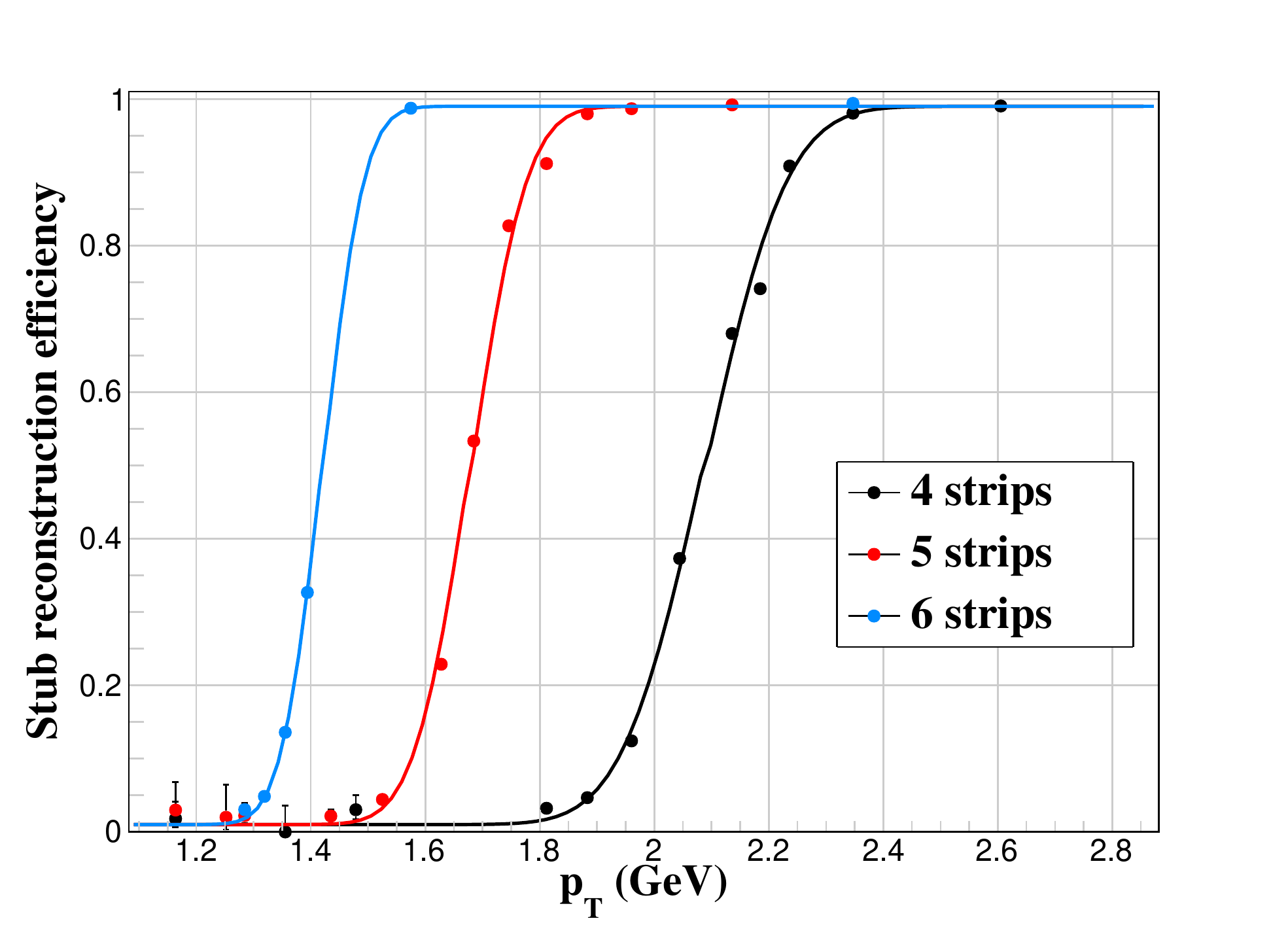}\label{fig:a}}%
	\subfloat[]{\includegraphics[width=0.45\textwidth]{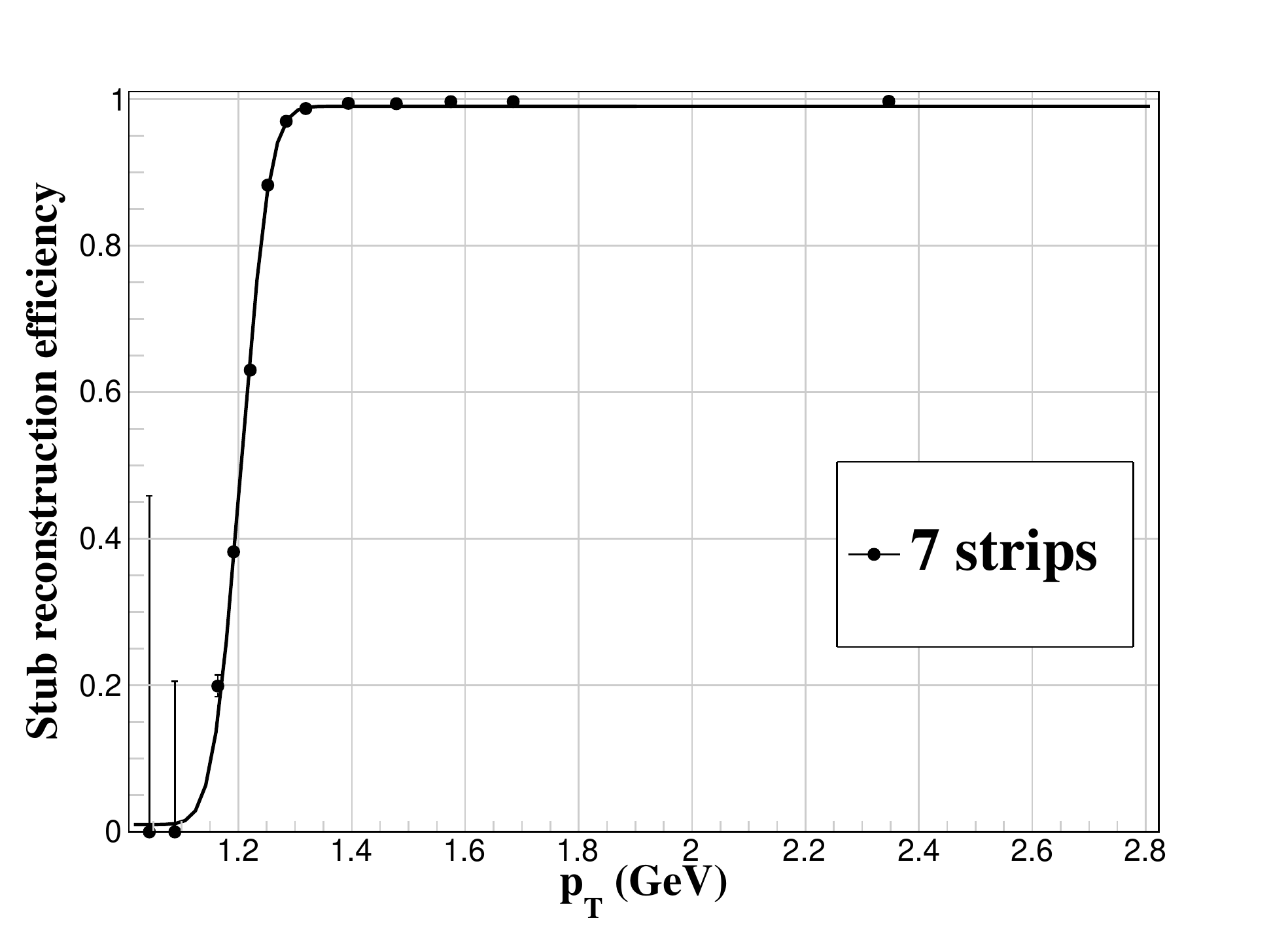}\label{fig:b}}%
	\caption{Stub efficiency vs. $p_\text{T}$ for the unirradiated (a) and fully irradiated (b) mini-module.}
	\label{fig:stubRecoEffVsPt}
\end{figure}

The CBC3 chip version used in the mini-module had the full logic implemented for the stub reconstruction. 
When hits on both sensors satisfied the requirements to create a stub, the chip was sending it to the FPGA where it was 
stored to be read out when a trigger was issued. 
The study of the stub reconstruction efficiency was one of the main goals of this beam test.
The same track criteria applied for the event selection described in Sec.~\ref{par:EfficiencyAnalysis} 
have been used for the \mbox{analysis} of the stub reconstruction efficiency.
For the stub efficiency, the denominator is \mbox{defined} as the number of tracks pointing to an active region of the sensor 
(excluding masked regions and edges, as already described in Sec.~\ref{par:EfficiencyAnalysis}) 
and there is no offline matching requiring clusters both on the seed and correlated sensors.
The numerator includes events in which the mini-module measured at least one stub within a window
of $\pm135\,\mu$m around the projected impact point.\\
Figure~\ref{fig:stubRecoEff} shows the uniformity of the stub reconstruction efficiency measured across the sensors for 
the unirradiated (a) and the fully irradiated (b) mini-module.
The efficiency is above $99\%$ everywhere along the sensor for the unirradiated mini-module.
For the fully irradiated mini-module the efficiency is still around $99\%$, within the errors.
\\Since there was no magnetic field during the beam tests, the correlation logic of the CBC3 has been tested by rotating the mini-module, 
thus emulating the bending of tracks in the CMS magnetic field. For incident angle $\beta$, the beam represents a charged particle track with bending radius 
in the transverse plane $r_\text{T}$. For a module in the tracker with radial position $R$, 
$\sin(\beta) = R/(2r_\text{T})$. The bending radius is related to the particle's electric charge ($q$) 
and transverse momentum ($p_\text{T}$), for a homogeneous magnetic
field of given strength ($B$), via the relation $r_\text{T} = p_\text{T}/(qB)$.\\
For the CMS field strength of $B = 3.8\,$T, the relationship between the beam incident angle ($\beta$)
and the transverse momentum ($p_\text{T}$) of a particle traversing a module at radial position 
$R$ is given by
\begin{equation}
	p_\text{T}[\text{GeV}] \approx \frac{0.57\cdot R[\text{m}]}{\sin{\beta}}.
	\label{eq:MomentumCut}
\end{equation}
The beam incident angle is related to the stub direction ($\Delta X = X_\text{seed} - X_\text{correlated}$), in strip units, the strip pitch ($p$) and the
sensor separation ($d$) via $\tan(\beta) = p\Delta X/d$. The CBC3s on the mini-module were configured to generate
stub triggers using a window size of $\pm$ 4, 5, 6, or 7 strips. From simple geometry, for the known 
$p = 90\,\mathrm{\mu m}$ and the estimated $d = 1.8\,\mathrm{mm}\,\pm\,40\,\mathrm{\mu m}$, the efficiency should be constant and high for small
angles. 
The efficiency is expected to start dropping when the incident angle is such that the $\Delta X$ is close to the selected window size.
If this module were to be placed in the first barrel layer of the CMS tracker at a radius $R = 71.5$\,cm, 
using Eq.~\ref{eq:MomentumCut}, the efficiency is expected to start dropping at $p_\text{T} \approx 2.36,~1.86,~1.54,~1.32$\,GeV, 
and to reach values around 0 for $p_\text{T} \approx 1.86,~1.54,~1.32,~1.16$\,GeV corresponding to a window size of 4, 5, 6, 7 strips. 
Unfortunately a direct comparison using the same window size cut between unirradiated and fully irradiated mini-module
stub reconstruction is not possible because of the lack of data. Nevertheless, 
the measured efficiencies as a function of the rotation angle~$\beta$ are in good agreement with these geometric expectations, as 
shown in Fig.~\ref{fig:stubRecoEffVsPt} both for the unirradiated (a) and the fully irradiated (b) mini-module. 
The efficiencies do not go exactly to zero, likely due to delta rays, stabilizing at around 3\%  
resulting in a rejection efficiency of about 97\% for low momentum track hits.
By fitting the efficiency curve with an error function, the effective $p_\text{T}$ threshold and resolution can be extracted. 
The effective $p_\text{T}$ threshold is defined as the value of $p_\text{T}$ for which the efficiency is $50\%$. 
For a window size of 5 strips, the effective $p_\text{T}$ is $1.68$\,GeV with a resolution of $6.5\%$, 
defined as the ratio of the sigma of the error function ($0.11$\,GeV) to the $p_\text{T}$ value at $50\%$.
The resolution calculated for the irradiated mini-module for a window of 7 strips is $4.6\%$, 
with a sigma of $0.055$\,GeV and a $p_\text{T}$ value at $50\%$ of $1.2$\,GeV.\\
The efficiency plateau for the unirradiated and fully irradiated module is above $99\%$. 
The high plateau efficiencies with sharp turn-on 
demonstrate that the module can reject hits at the radial distance from the interaction point of $R = 71.5$\,cm efficiently. 


%% file: tex/conclusions.tex
\section{Conclusions}
The future CMS Outer Tracker planned for the HL-LHC must provide information about tracks of
high transverse momentum to the first level trigger. 
The main building block of the new tracker will be the so-called \pt-module, equipped with two radiation-tolerant 
silicon sensors separated by a few millimeters and front-end ASICs that correlate hits on each sensor 
to reconstruct segments of tracks, 
called stubs. 
A fully functional mini-module, using the CMS binary chip version 3.0 for readout and equipped with two 
$300\,\mathrm{\mu m}$ n-in-p strip sensors, has been \mbox{analysed} in a test beam before and after being irradiated with neutrons. 
After commissioning, the performance was found to be in agreement with expectations. 
The particle detection efficiency is close to $100\%$ for the unirradiated mini-module and remains
above $99\%$ when the module is irradiated to a fluence that is even higher than the maximum 
fluence expected \mbox{during} the lifetime of the module when running in CMS, 
assuming that the HL-LHC upgrade will deliver 4000\,fb$^{-1}$ of proton-proton collisions.
The cluster width was in good agreement with geometric predictions, 
and the fraction of very broad clusters was below $1\%$. 
The \mbox{majority} of clusters are of width one or two, and the resolution for these clusters behaves as expected.
It has also been shown that the charge collected in the fully irradiated mini-module is consistent with the 
values found in literature. 
Finally, by emulating the $p_\text{T}$ dependent track bending in the magnetic field via a rotation of the mini-module,
it has been proven that the stub logic works as expected, rejecting 97\% of the the undesired low $p_\text{T}$ 
track points while accepting more than 99\% of high $p_\text{T}$ track points.
For example, it has been shown that particles with momentum below $1.2$\,GeV could be rejected with a 
resolution of about $0.055$\,GeV, when the module was irradiated at the highest fluence. 
Both before and after irradiation the stub reconstruction 
efficiency was above $99\%$, for high $p_\text{T}$ tracks.

%% file: tex/Acknowledgements_2022.tex
We thank the Fermilab accelerator and FTBF personnel for the excellent performance of the
accelerator and support of the test beam facility, in particular M. Kiburg, E. Niner, and E. Schmidt.\\
The tracker groups gratefully acknowledge financial support from the following funding agencies: 
BMWFW and FWF (Austria); FNRS and FWO (Belgium); CERN; MSE and CSF (Croatia); Academy of Finland, MEC, and 
HIP (Finland); CEA and CNRS/IN2P3 (France); BMBF, DFG, and HGF (Germany); GSRT (Greece); NKFIA K124850, 
and Bolyai Fellowship of the Hungarian Academy of Sciences (Hungary); DAE and DST (India); 
INFN (Italy); PAEC (Pakistan); SEIDI, CPAN, PCTI and FEDER (Spain); 
Swiss Funding Agencies (Switzerland); MST (Taipei); STFC (United Kingdom); DOE and NSF (U.S.A.). 
This project has received funding from the European Union’s Horizon 2020 research and innovation programme under the Marie Sk\l odowska-Curie grant agreement No 884104 (PSI-FELLOW-III-3i). 
Individuals have received support from HFRI (Greece).\\ 
This document was prepared using the resources of the Fermi National Accelerator Laboratory
(Fermilab), a U.S. Department of Energy, Office of Science, HEP User Facility. Fermilab is managed by Fermi Research Alliance, LLC (FRA), acting under Contract No. DE-AC02-07CH11359.

%% file: tex/TrackerAuthorList_2023.tex
\newcommand{\cmsAuthorMark}[1]
{\hbox{\textsuperscript{\normalfont#1}}}







\newpage

\section*{The Tracker Group of the CMS Collaboration}
\addcontentsline{toc}{section}{The Tracker Group of the CMS Collaboration}

{\setlength{\parindent}{0cm}
{\setlength{\parskip}{0.4\baselineskip}

\textcolor{black}{\textbf{Institut~f\"{u}r~Hochenergiephysik, Wien, Austria}\\*[0pt]
W.~Adam, T.~Bergauer, K.~Damanakis, M.~Dragicevic, R.~Fr\"{u}hwirth\cmsAuthorMark{1}, H.~Steininger}

\textcolor{black}{\textbf{Universiteit~Antwerpen, Antwerpen, Belgium}\\*[0pt]
W.~Beaumont, M.R.~Darwish\cmsAuthorMark{2}, T.~Janssen, T.~Kello\cmsAuthorMark{3}, H.~Rejeb~Sfar, P.~Van~Mechelen
% ...
}

\textcolor{black}{\textbf{Texas~A\&M~University, College~Station, USA}\\*[0pt]
R.~Eusebi}

\textcolor{black}{\textbf{Vanderbilt~University, Nashville, USA}\\*[0pt]
P.~D'Angelo, W.~Johns}

\textcolor{black}{\textbf{Wayne State University, Detroit, USA}\\*[0pt]
R.~Harr, N.~Poudyal\cmsAuthorMark{22}}
}
\newpage
% ...
1: Also at Vienna University of Technology, Vienna, Austria \\
2: Also at Institute of Basic and Applied Sciences, Faculty of Engineering, Arab Academy for Science, Technology and Maritime Transport, Alexandria, Egypt \\
3: Also at Universit\'{e}~Libre~de~Bruxelles, Bruxelles, Belgium \\
% ...
22: Now at University of South Dakota, Vermillion, USA

}